\pdfoutput=1
\newcommand*{\ATLASLATEXPATH}{}
\documentclass[cernpreprint, texlive=2016, UKenglish]{\ATLASLATEXPATH atlasdoc}
 
\usepackage{\ATLASLATEXPATH atlaspackage}
\usepackage{\ATLASLATEXPATH atlasbiblatex}
 
\usepackage{\ATLASLATEXPATH atlasphysics}
\usepackage{\ATLASLATEXPATH atlasjetetmiss}
 
\graphicspath{{logos/}{figures/}}
\usepackage{graphicx}
\usepackage{enumerate}
\usepackage{multirow}
 
\usepackage{ANA-PERF-2017-04-PAPER-defs}

\AtlasTitle{Identification of boosted Higgs bosons decaying into $b$-quark pairs with the ATLAS detector at 13~\TeV}
 
\AtlasAbstract{
 
This paper describes a study of techniques for identifying Higgs bosons at high transverse momenta decaying into
bottom-quark pairs, $H \rightarrow b\bar{b}$, for proton--proton collision data collected by the ATLAS detector
at the Large Hadron Collider at a centre-of-mass energy $\sqrt{s}=13$~\TeV. These decays are reconstructed
from calorimeter jets found with the anti-$k_{t}$ $R = 1.0$ jet algorithm. To tag Higgs bosons, a combination of requirements
is used: $b$-tagging of $R = 0.2$ track-jets matched to the large-$R$ calorimeter jet, and requirements on
the jet mass and other jet substructure variables. The Higgs boson tagging efficiency and corresponding multijet
and hadronic top-quark background rejections are evaluated using Monte Carlo simulation. Several benchmark tagging
selections are defined for different signal efficiency targets. The modelling of the relevant input distributions used
to tag Higgs bosons is studied in 36~\ifb\ of data collected in 2015 and 2016 using $g\to b\bar{b}$ and
$Z(\to b\bar{b})\gamma$ event selections in data. Both processes are found to be well modelled within
the statistical and systematic uncertainties. }
 
\author{The ATLAS Collaboration}
 
\AtlasRefCode{PERF-2017-04}
 
\PreprintIdNumber{CERN-EP-2019-085}

\AtlasJournalRef{Eur. Phys. J. C 79 (2019) 836}

\AtlasDOI{DOI:10.1140/epjc/s10052-019-7335-x}

\hypersetup{pdftitle={ATLAS document},pdfauthor={The ATLAS Collaboration}}
 
\begin{document}
 
\maketitle

\clearpage
 
\section{Introduction}
\label{sec:intro}
 
The Large Hadron Collider (LHC) centre-of-mass energy of 13~\TeV\ greatly extends the sensitivity of the ATLAS
experiment~\cite{PERF-2007-01} to heavy new particles. In several new physics scenarios~\cite{Randall:1999ee,SUSYh1,2HDM1}, these
heavy new particles may have decay chains including the Higgs boson~\cite{HIGG-2012-27,CMS-HIG-12-028}. The large mass-splitting
between these resonances and their decay products results in a high-momentum Higgs boson, causing its decay products to be collimated.
The decay of the Higgs boson into a $b\bar{b}$ pair has the largest branching fraction within the Standard Model (SM), and thus is a major
decay mode to use when searching for  resonances involving high-momentum Higgs bosons (see e.g.\ Ref.~\cite{PhenoBBBB}), as well as
for measuring the SM Higgs boson properties. The signature of a boosted Higgs boson decaying into a $b\bar{b}$ pair is a collimated
flow of particles, in this document called a `\HiggsJet', having an energy and angular distribution of the jet constituents consistent
with a two-body decay and containing two $b$-hadrons.
The techniques described in this paper to identify Higgs bosons decaying into bottom-quark pairs have been used
successfully in several analyses~\cite{EXOT-2016-31,EXOT-2016-10,EXOT-2016-12} of 13~\TeV\ proton--proton collision data recorded by ATLAS.
 
In order to identify, or tag, boosted Higgs bosons it is paramount to understand the details of $b$-hadron identification and the internal
structure of jets, or jet substructure, in such an environment~\cite{ref:boostedhiggsprl}. The approach to tagging presented in this paper is
built on studies from LHC runs at $\sqrt{s} = 7$ and $8$~\TeV, including extensive studies of jet reconstruction and grooming
algorithms~\cite{PERF-2012-02}, detailed investigations of track-jet-based $b$-tagging in boosted topologies~\cite{ATL-PHYS-PUB-2014-013},
and the combination of substructure and $b$-tagging techniques applied in the Higgs boson pair search in the four-$b$-quark final state~\cite{EXOT-2014-11}
and for discrimination of $Z$ bosons from $W$ bosons~\cite{PERF-2015-02}. Gluon splitting into $b$-quark pairs at small opening angles has been
studied at $\sqrt{s} = 13$~\TeV\ by ATLAS~\cite{STDM-2017-17}. The identification of Higgs bosons at high transverse momenta through
the use of jet substructure has also been studied by the CMS Collaboration and their techniques are described in Refs.~\cite{CMS-BTV-16-002,CMS-HIG-17-010}.
 
The Higgs boson tagging efficiency and background rejection for the two most common background processes, the multijet and hadronic top-quark
backgrounds, are evaluated using Monte Carlo simulation. In addition, two processes with a topology similar to the signal,
$Z \to b \bar{b}$ decays and $g\to b\bar{b}$ splitting, are used to validate \HiggsJet\ tagging techniques in data at $\sqrt{s}=13$~\TeV. In particular
the modelling of relevant \HiggsJet\ properties in Monte Carlo simulation is compared with data. The $g\to b\bar{b}$ process allows the
modelling of one of the main backgrounds to be validated. The $Z \to b \bar{b}$ process is a colour-singlet resonance with a mass close to the Higgs boson mass
and thus very similar to the $H \to b\bar{b}$ signal.
 
After a brief description of the ATLAS detector in Section~\ref{sec:detector} and of the data and simulated samples in Section~\ref{sec:samples},
the object reconstruction, selection and labelling is discussed in Section~\ref{sec:objects}. Section~\ref{sec:systematics} describes relevant
systematic uncertainties. The \HiggsJet\ tagging algorithm and its performance are presented in Section~\ref{sec:Baseline_tagger}.
Sections~\ref{sec:performance_in_gbb} and \ref{sec:performance_in_Zbb} discuss a comparison between relevant distributions in data
control samples dominated by $g\to b\bar{b}$ and $Z(\to b\bar{b})\gamma$ and the corresponding simulated events, respectively. Finally,
conclusions are presented in Section~\ref{sec:conclusions}.

\newcommand{\AtlasCoordFootnote}{
ATLAS uses a right-handed coordinate system with its origin at the nominal interaction point (IP)
in the centre of the detector and the $z$-axis along the beam pipe.
The $x$-axis points from the IP to the centre of the LHC ring,
and the $y$-axis points upwards.
Cylindrical coordinates $(r,\phi)$ are used in the transverse plane,
$\phi$ being the azimuthal angle around the $z$-axis.
The pseudorapidity is defined in terms of the polar angle $\theta$ as $\eta = -\ln \tan(\theta/2)$.
Angular distance is measured in units of $\Delta R \equiv \sqrt{(\Delta\eta)^{2} + (\Delta\phi)^{2}}$.}

\section{ATLAS detector}
\label{sec:detector}
 
The ATLAS detector~\cite{PERF-2007-01} at the LHC covers nearly the entire solid angle around the collision point.\footnote{\AtlasCoordFootnote}
It consists of an inner tracking detector surrounded by a thin superconducting solenoid, electromagnetic and hadronic calorimeters,
and a muon spectrometer incorporating three large superconducting toroid magnets.
The inner-detector system (ID) is immersed in a \SI{2}{\tesla} axial magnetic field
and provides charged-particle tracking in the range $|\eta| < 2.5$.
 
Preceding data-taking at a centre-of-mass energy of 13~\TeV, the
high-granularity silicon pixel detector was equipped with a new barrel layer,
located at a smaller radius (of about 34 mm) than the other layers~\cite{ATLAS-TDR-19,Abbott:2018ikt}. The
upgraded pixel detector covers the vertex region and typically provides four measurements for tracks
originating from the luminous region.
It is followed by a silicon microstrip tracker, which usually provides four space points per track.
These silicon detectors are complemented by a transition radiation tracker,
which enables radially extended track reconstruction up to $|\eta| = 2.0$.
The transition radiation tracker also provides electron identification information
based on the fraction of hits above a certain energy deposit threshold corresponding to transition radiation.
 
The calorimeter system covers the pseudorapidity range $|\eta| < 4.9$.
Within the region $|\eta|< 3.2$, electromagnetic calorimetry is provided by barrel and
endcap high-granularity lead/liquid-argon (LAr) calorimeters,
with an additional thin LAr presampler covering $|\eta| < 1.8$
to correct for energy loss in material upstream of the calorimeters.
Hadronic calorimeter within $|\eta| < 1.7$ is provided by a steel/scintillating-tile calorimeter,
segmented into three barrel structures, and two copper/LAr hadronic endcap calorimeters covering $1.5<|\eta| < 3.2$.
The solid angle coverage is completed with forward copper/LAr and tungsten/LAr calorimeter modules
optimised for electromagnetic and hadronic measurements respectively.
 
The muon spectrometer (MS) comprises separate triggering and
high-precision tracking chambers measuring the deflection of muons in a magnetic field generated by superconducting air-core toroids.
The precision chamber system covers the region $|\eta| < 2.7$ with three layers of monitored drift tubes,
complemented by cathode strip chambers in the forward region, where the background is highest.
The muon trigger system covers the range $|\eta| < 2.4$ with resistive plate chambers in the barrel, and thin gap chambers in the endcap regions.
 
A two-level trigger system is used to select interesting events~\cite{TRIG-2016-01}.
The level-1 trigger is implemented in hardware and uses a subset of detector information
to reduce the event rate to a design value of at most \SI{100}{\kHz}.
This is followed by a software-based high-level trigger, which reduces the event rate further to an average of \SI{1}{\kHz}.
 

\section{Data and simulated event samples}
\label{sec:samples}
 
The data used in this paper were recorded with the ATLAS detector during the 2015 and 2016 LHC proton--proton $(pp)$ collision runs, and correspond to a total integrated luminosity of
$36.1$~fb$^{-1}$ at $\sqrt{s} = 13$~\TeV. This integrated luminosity is calculated after the imposition of data quality requirements, which ensure
that the ATLAS detector was in good operating condition.
 
Several Monte Carlo (MC) simulated event samples were used for the optimisation of the Higgs boson tagger, estimation of its performance, and
the comparisons between data and simulation.
 
Simulated events with a broad transverse momentum (\pt) spectrum of Higgs bosons were generated as decay products of Randall--Sundrum
gravitons $G^{*}$ in a benchmark model with a warped extra dimension~\cite{Randall:1999ee}, $G^{*} \rightarrow HH \rightarrow b\bar{b}b\bar{b}$,
over a range of graviton masses between 300 and 6000~\GeV. The events were simulated using the \MGMCatNLO\,generator~\cite{Alwall:2014hca}.
Parton showering, hadronisation and the underlying event were simulated with \PythiaEight\,\cite{Sjostrand:2014zea} using the leading-order (LO)
NNPDF2.3 parton distribution function (PDF) set~\cite{Ball:2012cx} and the ATLAS \textsc{A14}~\cite{ATL-PHYS-PUB-2014-021} set of tuned parameters.
 
Events containing the $Z(\rightarrow b\bar{b})\gamma$ and \Gammajets processes were simulated with the
\SHERPA~v2.1.1~\cite{Gleisberg:2008ta, Hoeche:2009rj, Gleisberg:2008fv, Schumann:2007mg} LO generator. The matrix elements were configured to allow up
to three partons in the final state in addition to the $Z$ boson or the photon. The $Z$ boson was produced on-shell and required to decay hadronically.
The CT10 next-to-leading-order (NLO) PDF set~\cite{Lai:2010vv,PhysRevD.89.033009} was used. The $t\bar{t}\gamma$ MC events were modelled by \MADGRAPH
interfaced with \PythiaEight\ for showering, hadronisation and the underlying event with the LO NNPDF2.3 PDF set and the \textsc{A14} underlying-event tune.
Simulated events of hadronically decaying $W\gamma$ were generated using \SHERPA~v2.1.1, with the same configuration as the one used for the $Z\gamma$ sample.
 
To cover a large range of top-quark transverse momenta, hadronically decaying top quarks were generated using $Z'$ bosons decaying into \ttbar\ pairs
over a range of $Z'$ boson masses between 400 and 5000~\GeV. These samples were simulated using \PythiaEight\ with the LO NNPDF2.3 PDF set
and the \textsc{A14} underlying-event tune.
 
Finally, inclusive multijet events were generated using \PythiaEight, with the LO NNPDF2.3 PDF set and the \textsc{A14} underlying-event tune;
and with \Herwigpp~\cite{Bahr:2008pv}, with the CTEQ~\cite{Pumplin:2002vw} PDF set and the UEEE~\cite{Gieseke:2012ft} underlying event tune.
To increase the number of simulated events with semimuonically decaying hadrons for the $g\to b\bar{b}$ analysis,
samples of multijet events filtered to have at least one muon with \pT\ above 3~\GeV\ and $|\eta| < 2.8$ were produced
with \PythiaEight\ and \Herwigpp\ using the same PDF set and underlying-event tunes as the unfiltered multijet samples.
 
In all cases except events generated using \SHERPA, EvtGen~\cite{EvtGen} was used to model the decays of $b$- and $c$-hadrons. All simulated
event samples included the effect of multiple $pp$ interactions in the same and neighbouring bunch crossings (`pile-up') by overlaying simulated minimum-bias events on each
simulated hard-scatter event. The minimum-bias events were simulated with the single-, double- and non-diffractive $pp$ processes of
\PythiaEight\ using the A2 tune~\cite{ATL-PHYS-PUB-2012-003} and the MSTW2008 LO PDF~\cite{Martin:2009iq,Martin:2009bu,Martin2010}.
The detector response to the generated events was simulated with \GEANT4~\cite{Agostinelli:2002hh,SOFT-2010-01}.


\section{Object and event reconstruction}
\label{sec:objects}
 
In this section the object reconstruction, associations among the objects, jet labelling, and the procedure to determine
the heavy-flavour content of jets are described.
 
\textbf{Calorimeter jets:} Calorimeter-based jets are built from noise-suppressed topological clusters and are reconstructed
using \Fastjet~\cite{Cacciari:2011ma} with the \antikt\ algorithm~\cite{Cacciari:2008gp} with a radius parameter of $R=1.0$ (\largeR\ jets) or
$R=0.4$ (\smallR\ jets). The topological clusters of the \largeR\ jets are brought to the hadronic energy
scale using the local hadronic cell weighting scheme~\cite{PERF-2014-07}. The \largeR\ jets are groomed using trimming~\cite{jetTrimming,PERF-2012-02} to discard the softer components of jets
that originate from initial-state radiation, pile-up interactions or the underlying event. This is done by reclustering the constituents
of the initial jet, using the \kt\ algorithm~\cite{Catani:1993hr,Ellis:1993tq}, into subjets of radius parameter $\Rsub=0.2$ and
removing any subjet that has a \pT\ less than 5\% of the parent jet \pT. The simulation-based calibration of the trimmed jet \pT\ and mass
is described in Ref.~\cite{JETM-2018-02}.  \LargeR\ jets are required to
have $\pt > 250$~\GeV\ and $|\eta| < 2.0$. \SmallR\ jets are calibrated with a series of simulation-based
corrections and in situ techniques, including corrections to account for pile-up energy entering the jet area, as described in
Ref.~\cite{PERF-2016-04}. They are required to have $\pT > 20$~\GeV\ and $|\eta| < 2.5$. To reduce the number of \smallR\ jets originating
from pile-up interactions, these jets are required to pass the jet vertex tagger (JVT)~\cite{PERF-2014-03} requirement if the jets are in the
range $\pT < 60$~\GeV\ and $|\eta| < 2.4$. The JVT requirement has an inclusive hard-scatter efficiency of about 97\% in that kinematic region.
 
\textbf{Truth jets:} Truth jets are built in simulated events by using `truth' information from MC generator's event record to cluster stable particles
with a lifetime $\tau_0$ in the rest frame such that $c\tau_0 > 10$~mm. Particles such as muons and
neutrinos which do not leave significant energy deposits in the calorimeter are excluded. The same jet-clustering algorithm and trimming procedure
as for calorimeter jets are used to reconstruct truth jets.
 
\textbf{Track-jets:} Track-jets are built with the \antikt\ algorithm with a radius parameter of $R=0.2$~\cite{ATL-PHYS-PUB-2014-013}
from at least two ID tracks with $\pt>0.4$~\GeV\ and $|\eta| < 2.5$ that are either associated with the primary vertex or have a longitudinal impact parameter
$|z_{0}\sin(\theta)|<3$~mm. Such requirements greatly reduce the number of tracks from pile-up vertices whilst being highly efficient
for tracks from the hard-scatter vertex. Once the track-jet's axis is determined, tracks selected with looser impact parameter requirements are matched to the jet in order to collect the tracks needed to effectively run the jet flavour tagging
algorithms. The tracks are matched to the jet by using the angular separation $\Delta R$ between the track and the track-jet's axis. The $\Delta R$
requirement varies as a function of jet \pT, being wide for low-\pT\ jets and narrower for high-\pT\ jets as described in
Ref.~\cite{PERF-2012-04}. Only track-jets with $\pt>10$~\GeV\ and $|\eta|<2.5$ are used for the analysis.
 
\textbf{Muons:} Muons are reconstructed from a combination of measurements from the ID and the MS. They are required to
pass identification requirements based on quality criteria applied to the ID and MS tracks. The `Loose' 
identification working point defined in Ref.~\cite{PERF-2015-10} is used. Muons selected for this analysis are required to have
$\pt>5$~\GeV\ and $|\eta|<2.4$.
 
\textbf{Photons:} Photons are reconstructed from clusters of energy deposits in the electromagnetic calorimeter.  Clusters without
matching tracks are classified as unconverted photon candidates.  A photon candidate that can be matched to a reconstructed vertex
or track consistent with a photon conversion is considered as a converted photon candidate~\cite{PERF-2017-02}. The photon energy
estimate is described in Ref.~\cite{PERF-2017-03}. Requirements on the shower shape in the electromagnetic calorimeter and on
the energy fraction measured in the hadronic calorimeter are used to identify photons; the `Tight' identification working point is
applied in the analysis~\cite{PERF-2017-02}. In order to select prompt photons, the photons are required to fulfil the `Tight'
isolation criteria. The photons are required to have $|\eta|<1.37$ or $1.52<|\eta|<2.37$ and $\ET > 175$~\GeV. The latter requirement is
applied to insure efficient triggering.
 
\textbf{Track-jet ghost association:} In events with a dense hadronic environment an ambiguity often exists when matching track-jets
to calorimeter jets. The track-jet matching to \largeR\ jets is performed by applying ghost association~\cite{GhostMatching,jetareas,PERF-2012-02}:
the \largeR\ jet clustering process using the \antikt\ algorithm with $R = 1.0$ is repeated with the addition of `ghost' versions of the track-jets that
have the same direction but infinitesimally small \pT, so that they do not change the properties of the \largeR\ calorimeter jets. A track-jet is associated
with the \largeR\ jets if its ghost version is contained in the jet after reclustering. The reclustering is applied to the untrimmed \largeR\ jets.
The reclustered jets are identical to the jets before the reclustering, with the addition of the matched track-jets retained as associated objects.
This provides a robust matching procedure, and matching to jets with irregular boundaries can be achieved in a way that is less ambiguous
than a simple geometric matching.
 
\textbf{Jet labelling:} The performance of the tagger is evaluated on the basis of labelled \largeR\ jets. \HiggsJets\ are defined
as calorimeter-based \largeR\ jets with a Higgs boson and the corresponding two $b$-hadrons from the
Higgs boson decay found in the MC event record within $\Delta R = 1$ of the \largeR\ jet. Only the Higgs boson with the
highest \pT\ in the event is considered and it is required to have $\pt > 250$~\GeV\ and $|\eta| < 2.0$. The $b$-hadron must
have \pt\ above 5~\GeV\ and $|\eta| < 2.5$. Configurations where more than one Higgs boson is found within the \largeR\ jet
are excluded. Top-jets are defined as \largeR\ jets in which exactly one top quark is found in the MC event record within
$\Delta R = 1$ of the \largeR\ jet.
 
\textbf{Jet flavour labelling:} The labelling of the flavour of the track-jets in simulation is done by geometrically matching the
jet with truth hadrons. If a weakly decaying $b$-hadron with \pT\ above 5~\GeV\ is found within $\Delta R = 0.2$ of the track-jet's direction, the track-jet
is labelled as a $b$-jet.  In the case that the $b$-hadron could match more than one track-jet, only the closest track-jet is labelled
as a $b$-jet. If no $b$-hadron is found, the procedure is repeated for weakly decaying $c$-hadrons to label $c$-jets. If no $c$-hadron
is found, the procedure is repeated for $\tau$-leptons to label $\tau$-jets. A jet for which no such matching can be made is labelled
as a light-flavour jet.
 
\textbf{{\textit b}-jet identification:} Track-jets containing $b$-hadrons are identified using a multivariate MV2c10
algorithm~\cite{PERF-2012-04,ATL-PHYS-PUB-2016-012}, which exploits the information about the jet kinematics, the
impact parameters of tracks within jets, and the presence of displaced vertices. The training is performed on jets from
\ttbar\ events with $b$-jets as signal, and a mix of approximately 93\% light-flavour jets and 7\% $c$-jets as background.
A particular $b$-tagging requirement on MV2c10 results in a given efficiency, known as an efficiency working point (WP).
The efficiency WP is calculated from the inclusive \pt and $\eta$ spectra of jets from an inclusive \ttbar\
sample. For example a WP with 70\% efficiency corresponds to a factor of 120 in the light-quark/gluon-track-jet rejection and a
factor of seven in the $c$-track-jet rejection. Different WPs (60\%, 70\%, 77\% and 85\%) are
studied in the analyses presented in this paper and jets satisfying a particular MV2c10 criterion WP are referred to as `$b$-tagged jets'.
 
\textbf{\LargeR\ jet mass:}  To overcome the limited angular resolution for the energy deposits used to reconstruct the calorimeter-based
jet mass (\mcalo), an independent jet mass estimate using tracking information is developed, the `track-assisted jet mass',
\mTA~\cite{JETM-2018-02}. A weighted combination of calorimeter-based and track-assisted jet masses, \mcomb~\cite{JETM-2018-02}, is used in
the analysis. The \mcomb\ resolution is very similar to the \mcalo\ resolution at \HiggsJet\ \pt
below 700~\GeV\ and improves with increasing \pt. Muons from semileptonic $b$-hadron decays do not leave significant energy
deposits in the calorimeter, so they are considered separately in the calculation of the \mcomb\ observable. The resulting neutrinos
are not taken into account because they are not measured by the detector directly. The four-momentum of the closest muon candidate
within $\Delta R=0.2$ of the $b$-tagged track-jet is added to the four-momentum of the \largeR-jet after subtraction of the muon energy
loss in the calorimeter. Only the calorimeter-based component of the \mcomb\ observable is corrected~\cite{ATLAS-CONF-2016-039}.
The resolution of the muon-corrected \HiggsJet\ mass, \masscorr, is improved by about 10\% at transverse momenta below 500~\GeV,
while the improvement is not as pronounced at higher \pT, as was shown in Ref.~\cite{ATL-PHYS-PUB-2015-035}.


\section{Systematic uncertainties}
\label{sec:systematics}
 
\textbf{\LargeR\ jets:} The uncertainties in the jet energy, mass, and substructure scales are evaluated by comparing the ratio of calorimeter-based
to track-based measurements in dijet data and simulation~\cite{JETM-2018-02}. The sources of uncertainty in these measurements are treated as
fully correlated among \pT, mass, and substructure scales. The resolution uncertainty of the \largeR\ jet observables is evaluated in measurements
documented in Ref.~\cite{JETM-2018-02} and is assessed by applying an additional smearing to these observables.
The jet energy resolution uncertainty is estimated by degrading the nominal resolution by an absolute 2\%.
Similarly, the jet mass resolution is degraded by a relative 20\% to estimate the jet mass resolution uncertainty. The parton-shower-related
uncertainty for the $g \rightarrow b\bar{b}$ analysis is estimated by comparing the
nominal \PythiaEight multijet sample with \Herwigpp samples.
 
\textbf{Flavour tagging:} The flavour-tagging efficiency and its uncertainty for $b$- and $c$-jets is estimated in $t\bar{t}$ events, while the light-flavour-jet
misidentification rate and uncertainty is determined using dijet events~\cite{ATLAS-CONF-2018-001,PERF-2016-05,
ATLAS-CONF-2018-006}. Correction factors are applied to the simulated event samples to compensate for differences between data
and simulation in the $b$-tagging efficiency for track-jets with $\pT < 250$~\GeV. Correction factors and uncertainties for $c$-jets and light-flavour jets are derived for calorimeter-based jets and extrapolated to track-jets using MC simulation. An additional term is included to extrapolate the measured
uncertainties to \pT\ above 250~\GeV. This term is estimated from simulated events by varying the quantities affecting
the flavour-tagging performance such as the impact parameter resolution, percentage of poorly measured tracks, description of the detector
material, and track multiplicity per jet. The total uncertainties are 1--10\%, 15--50\%, and 50--100\% for $b$-jets, $c$-jets, and light-flavour jets respectively.
 
\textbf{Muon:} The uncertainties in the muon momentum scale and resolution are derived from data events with dimuon decays of $J/\psi$ and $Z$
bosons. In total, there are three independent components: one corresponding to the uncertainty in the inner detector track \pT\ resolution, one
corresponding to the uncertainty in the muon spectrometer \pT\ resolution, and one corresponding to the momentum scale uncertainty~\cite{PERF-2015-10}.
 
\textbf{Photon:}  The uncertainties in the reconstruction, identification, and isolation efficiency for photons are determined from data samples of
$Z\rightarrow \ell\ell\gamma$, $Z\rightarrow ee$, and inclusive photon events~\cite{PERF-2017-02}. Uncertainties in the electromagnetic shower energy
scale and resolution are taken into account as well~\cite{PERF-2017-03}.
 
\textbf{Background modelling uncertainties for $t\bar{t}\gamma$, $\gamma+$jets and $W(\rightarrow q\bar{q})\gamma$:} These correspond to the main
backgrounds in the  $Z(\rightarrow b\bar{b})\gamma$ studies presented in Section~\ref{sec:performance_in_Zbb}. The background modelling uncertainty
for the $\gamma+$jets sample was estimated with the alternative MC generator, \PythiaEight using the LO NNPDF2.3 PDF set and the \textsc{A14} underlying
event tune. The alternative sample includes LO photon plus jet events from the hard process and photon bremsstrahlung in dijet events.
 
In the case of the $W(\rightarrow q\bar{q})\gamma$ background, the nominal samples were compared
with samples produced using the \MGMCatNLO\ generator interfaced with \PythiaEight. For the $t\bar{t}\gamma$ background three different
sources of modelling uncertainty were considered: uncertainty due to the parton shower and hadronisation estimated by comparing the nominal samples produced
using \MADGRAPH interfaced with \PythiaEight, with samples from \MADGRAPH interfaced with \textsc{Herwig7}~\cite{Bahr:2008pv,Bellm:2015jjp}; uncertainty due to
different initial- and final-state radiation conditions from \PythiaEight tunes with high or low QCD radiation activity;
and uncertainty due to the choice of renormalisation and factorisation scales.

Uncertainties related to the photons and the $\gamma$+jets, $W(\rightarrow q\bar{q})\gamma$, and $t\bar{t}\gamma$ background modelling are applied only in the $Z(\to b\bar{b})\gamma$
analysis.

\section{\HiggsTagger}
\label{sec:Baseline_tagger}
 
The \HiggsTagger\ algorithm consists of several reconstruction steps. First, the Higgs boson candidate is reconstructed
as a \largeR\ jet. Second, the $b$-tagging requirement is applied to track-jets associated with the \largeR\ jet in order to select candidates corresponding
to $H \rightarrow b\bar{b}$ decays. Third, the $b$-tagged \largeR\ jet mass can be required to be around the SM Higgs boson mass of
125~\GeV. Finally, a requirement on other \largeR\ jet substructure variables can be applied depending on the \HiggsTagger\ working point.
 
\begin{figure}
\centering
\includegraphics[width=0.60\textwidth]{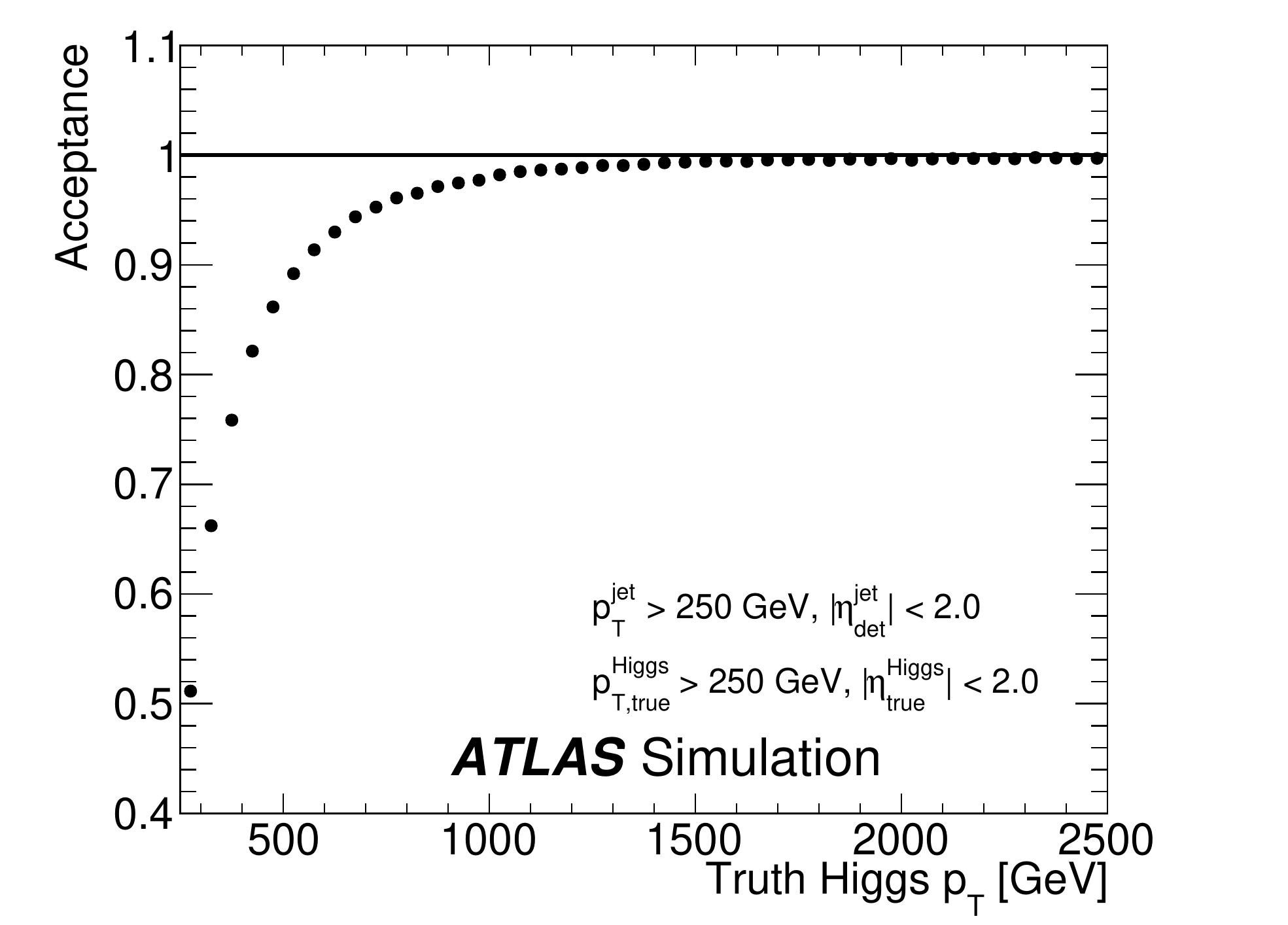}
\caption{Fraction of Higgs bosons in simulation which are reconstructed and labelled as a \HiggsJet\ following the definition in
Section~\ref{sec:objects}, as a function of Higgs boson \pT. Only Higgs bosons with $\pT > 250$~\GeV, $|\eta| < 2.0$ and
with associated $b$-hadrons from its decay are considered. Same $\pT$ and $\eta$ requirements are applied to the \HiggsJet s.}
\label{fig:signal_acceptance}
\end{figure}

The signal acceptance for the first reconstruction step where the Higgs boson candidate is reconstructed as a \largeR\ jet depends
strongly on its transverse momentum. The angular separation between Higgs boson decay products can be approximated as
$\Delta R \approx 2\mH / \pT$. Therefore, in most of the cases the Higgs boson decay products will fall within a single \largeR\ jet with a radius
parameter of $R=1.0$ if the Higgs boson \pT\ is at least 250~\GeV. The signal acceptance shown in Figure~\ref{fig:signal_acceptance}
is determined as the fraction of Higgs bosons in simulation which are reconstructed and labelled as a \HiggsJet\ following the definition in Section~\ref{sec:objects}.
Only Higgs bosons with $\pT > 250$~\GeV, $|\eta| < 2.0$, and associated $b$-hadrons from its decay that have $\pt > 5$~\GeV\ and $|\eta| < 2.5$ are considered.
The Higgs boson acceptance is around 50\% at 250~\GeV, where the jet \pT\ resolution have a significant impact as well, and increases to 95\% for transverse momenta above 750~\GeV.
 
The \HiggsJet\ tagging efficiency is defined as the number of \HiggsJets\ passing a given selection requirement divided by
the total number of \HiggsJets. The background rejection is defined as the inverse of the efficiency for a background jet to
pass the given selection requirement.
 
\subsection{Two-step sample reweighting}
\label{sec:reweighing}
 
To construct the signal sample, all graviton samples are combined. To allow a valid comparison between the signal
efficiency and the background rejection, the \largeR\ jet \pt\ spectrum of the combined graviton sample  is
reweighted to the reconstructed multijet \pt\ spectrum for the Higgs boson tagger performance studies in a two-step
procedure. The same two-step reweighting procedure is also applied to the $Z'\rightarrow t\bar{t}$ background sample.
The multijet spectrum is chosen as a reference because of its smoothly falling \pT\ spectrum being representative for
many analyses. During the first step of the reweighting the highest-\pt\ truth \HiggsJet\ is used, whereas for the second
reweighting step the highest-\pt\ reconstructed \HiggsJet\ is used. The reconstructed \HiggsJet\ and the truth \HiggsJet\
must both contain the highest-\pt\ Higgs boson to mitigate effects from initial-state radiation (ISR).
 
In the first step, the  \pT\ spectrum  of the truth \HiggsJet\ in the combined signal sample is reweighted to the \pT\ spectrum of the
reconstructed \largeR\ jet in the multijet sample. In the second step, the reconstructed
\HiggsJet\ \pT\ spectrum is reweighted to the reconstructed \largeR\ jet \pT\ spectrum in the multijet sample. A one-step
reweighting using the reconstructed \HiggsJet\ \pT\ spectrum results in large weights for jets with \pT\ much
larger or smaller than half of the graviton mass. Furthermore, the reconstructed \HiggsJet\ can contain additional
energy which does not stem from the Higgs boson decay, such as ISR, energy missing due to neutrinos, `out-of-cone'
effects, or trimming. The frequency of these effects depends on the Higgs boson boost, i.e.\ on the graviton mass, introducing
a dependence on the choice of simulated graviton masses used in the combined signal sample. The second step is
needed to account for a residual difference between reconstructed and truth \HiggsJet\ transverse momenta.
 
\subsection{Flavour-tagging working points}
\label{sec:ftag_wp}
 
To apply $b$-tagging to identify $H\rightarrow b\bar{b}$ decays, the track-jets are matched to the \largeR\ jets by ghost association as described
in Section~\ref{sec:objects}. At least two track-jets must be matched to the \largeR\ jet for the double-$b$-tagging
benchmarks, and at least one track-jet in the case of single-$b$-tagging benchmarks. The track-jet is considered to be $b$-tagged if
its MV2c10 $b$-tagging discriminant value is larger than a given threshold value. These threshold
values are defined for several $b$-tagging working points: 60\%, 70\%, 77\% and 85\% $b$-jet tagging efficiencies.
 
The following $b$-tagging benchmarks are studied:
 
\begin{itemize}
\item double $b$-tagging: the two highest-\pT\ track-jets must both pass a given $b$-tagging requirement;
\item asymmetric $b$-tagging: the track-jet which is more consistent with the interpretation of being a $b$-jet must
pass a given fixed 60\%, 70\%, 77\%, or 85\% working point, while the $b$-tagging requirement on the second track-jet is varied;
\item single $b$-tagging: at least one of the two highest-\pT\ track-jets must pass the $b$-tagging requirement;
\item leading single $b$-tagging: the highest-\pT\ track-jet must pass the $b$-tagging requirement.
\end{itemize}

\begin{figure}
\centering
\includegraphics[width=0.49\textwidth]{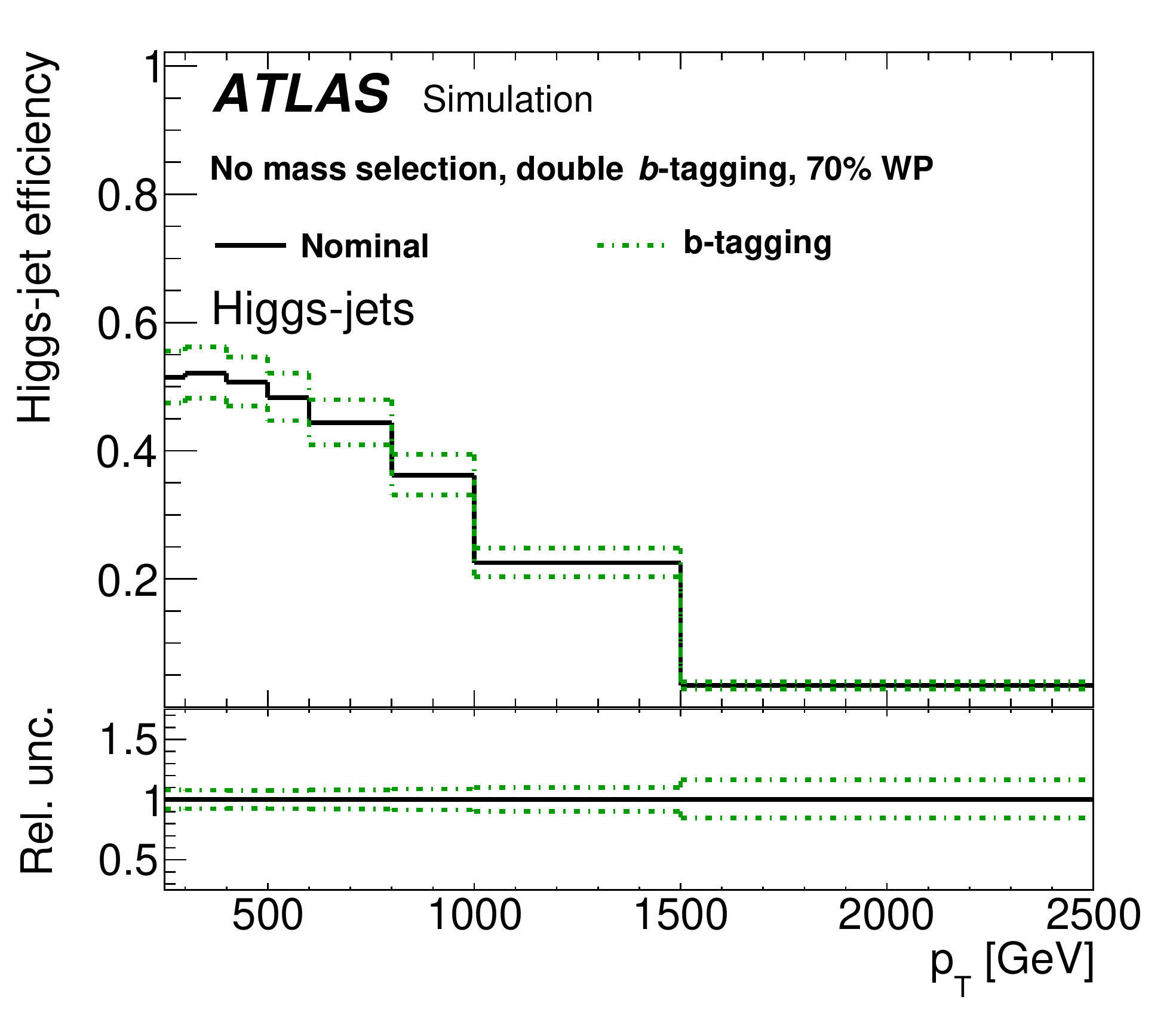}
\includegraphics[width=0.49\textwidth]{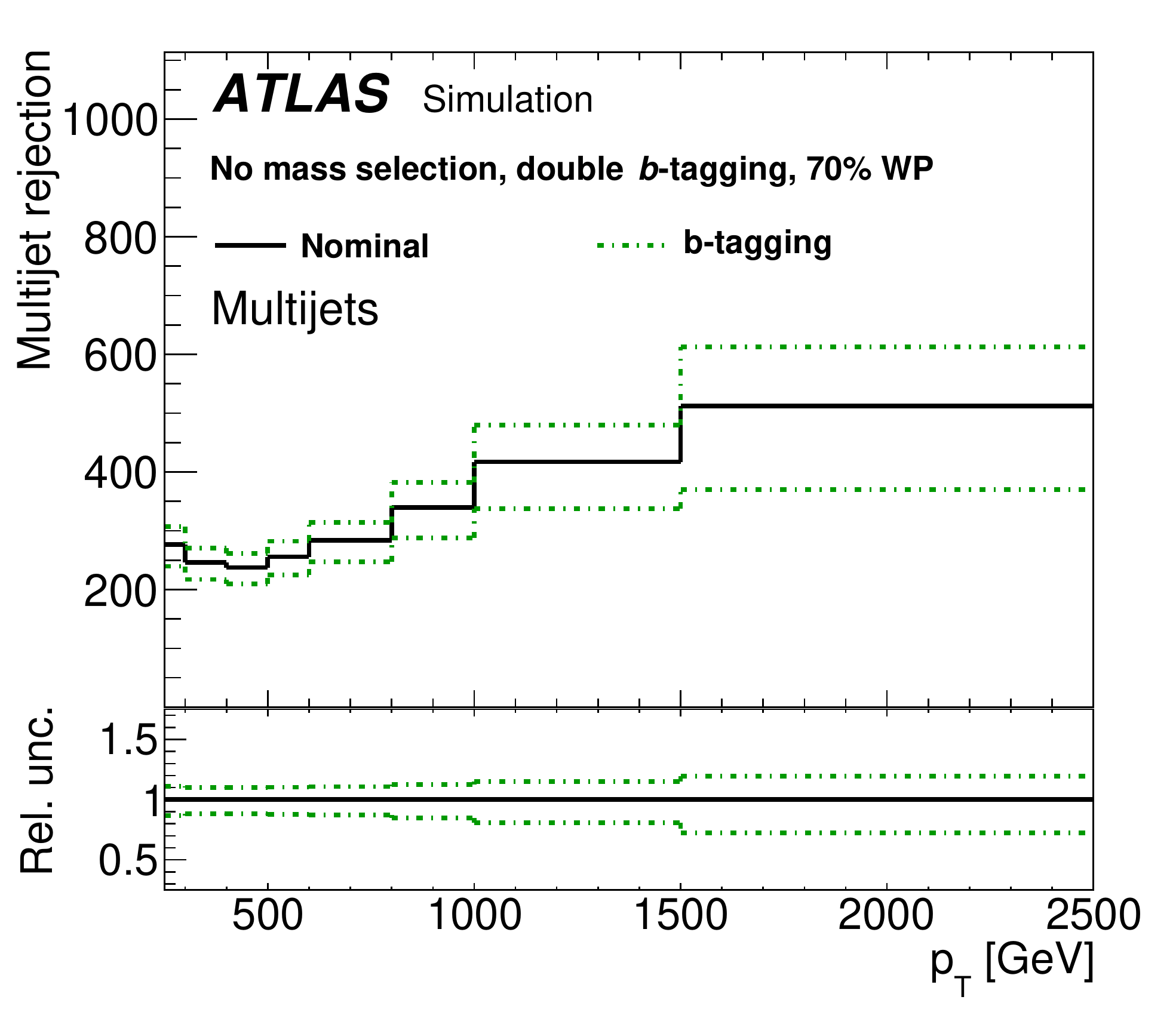} \\
\includegraphics[width=0.49\textwidth]{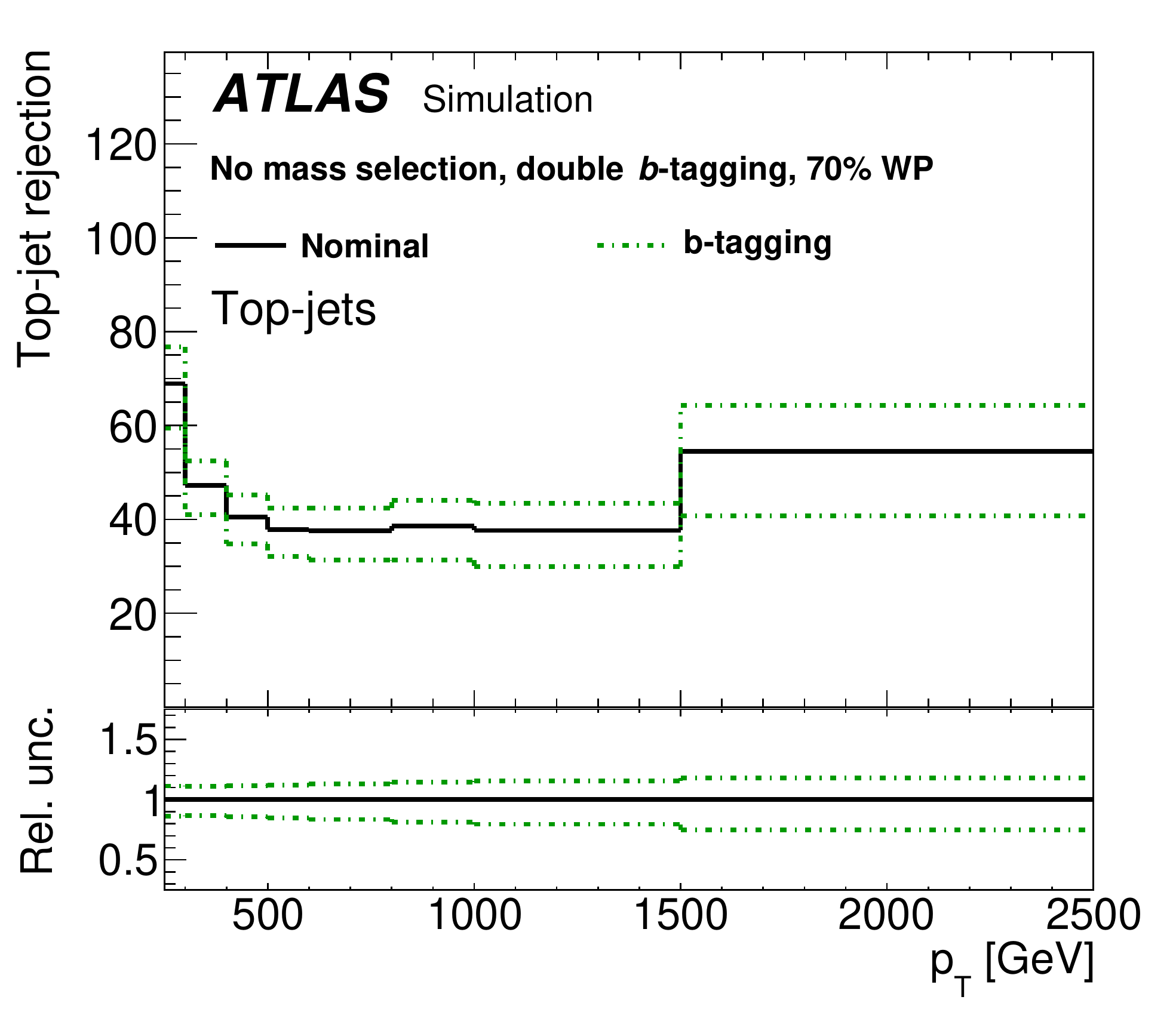}
\caption{The \HiggsJet\ efficiency (top left) and rejection against multijet (top right) and top-jet backgrounds (bottom) as
a function of the jet \pT\ for the 70\% double-$b$-tagging working point. The nominal curves correspond to the requirement
on the MV2c10 discriminant described in Section~\ref{sec:ftag_wp}. The $b$-tagging-related uncertainties
defined in Section~\ref{sec:systematics} are shown. }
\label{fig:btag_efficiency}
\end{figure}

The \HiggsJet\ efficiencies and background rejections as a function of the jet \pT\ for the 70\% double-$b$-tagging benchmark
are shown in Figure~\ref{fig:btag_efficiency}. The signal efficiency varies from 52\% at low \pt to about 5\% for $1500 < \pt < 2500$~\GeV.
The drop in efficiency at high transverse momenta due to the increasing collimation and eventual merging of the two $b$-jets can be
partially recovered using single-$b$-tagging working points as indicated in Figure~\ref{fig:acceptanceXeff}. The multijet (top-jet) rejection
is relatively constant over the whole \pT\ range and is about 250 (60) at low \pT and 500 (50) at high \pT.
 
\begin{figure}
\centering
\includegraphics[width=0.49\textwidth]{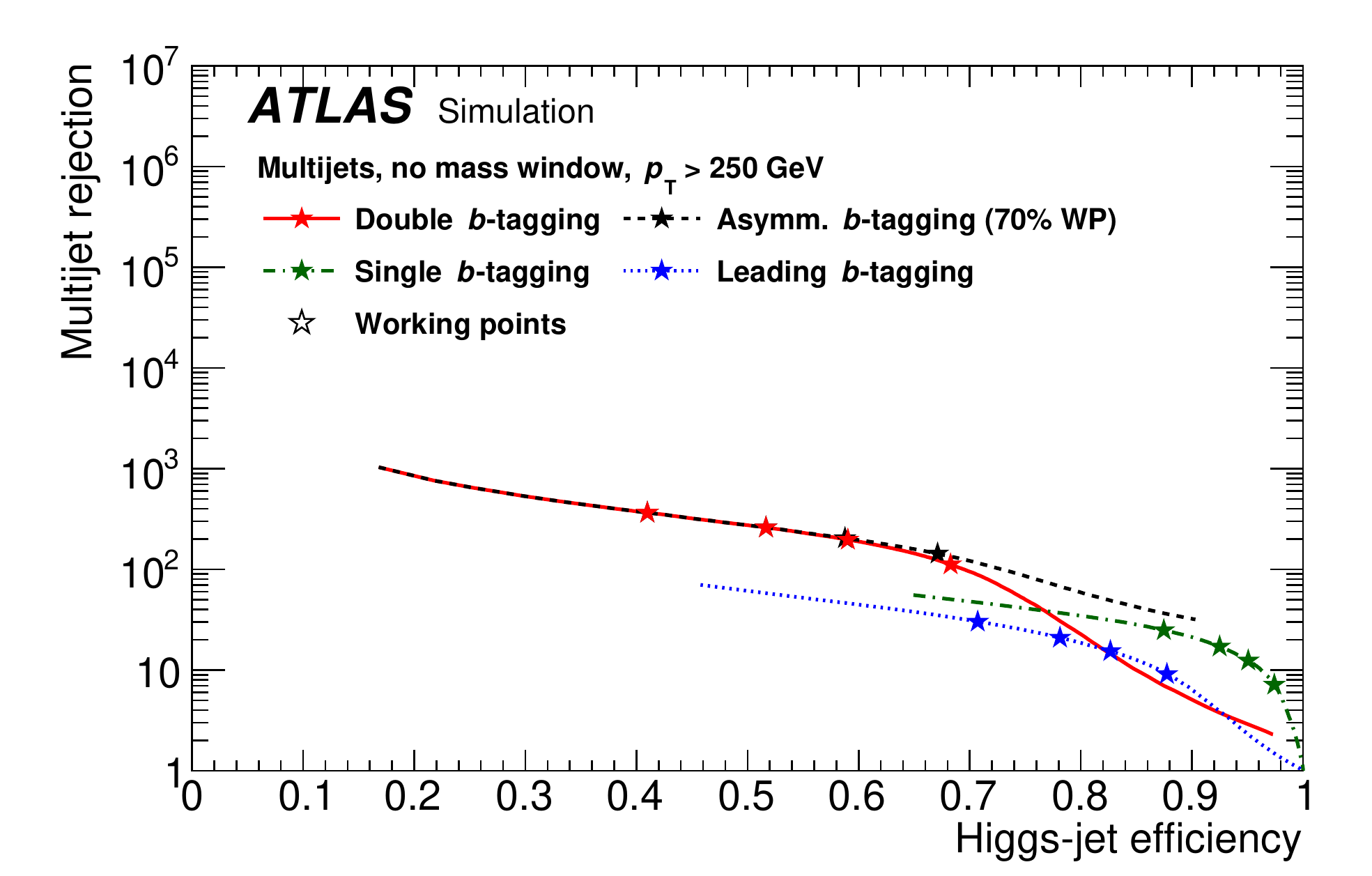}
\includegraphics[width=0.49\textwidth]{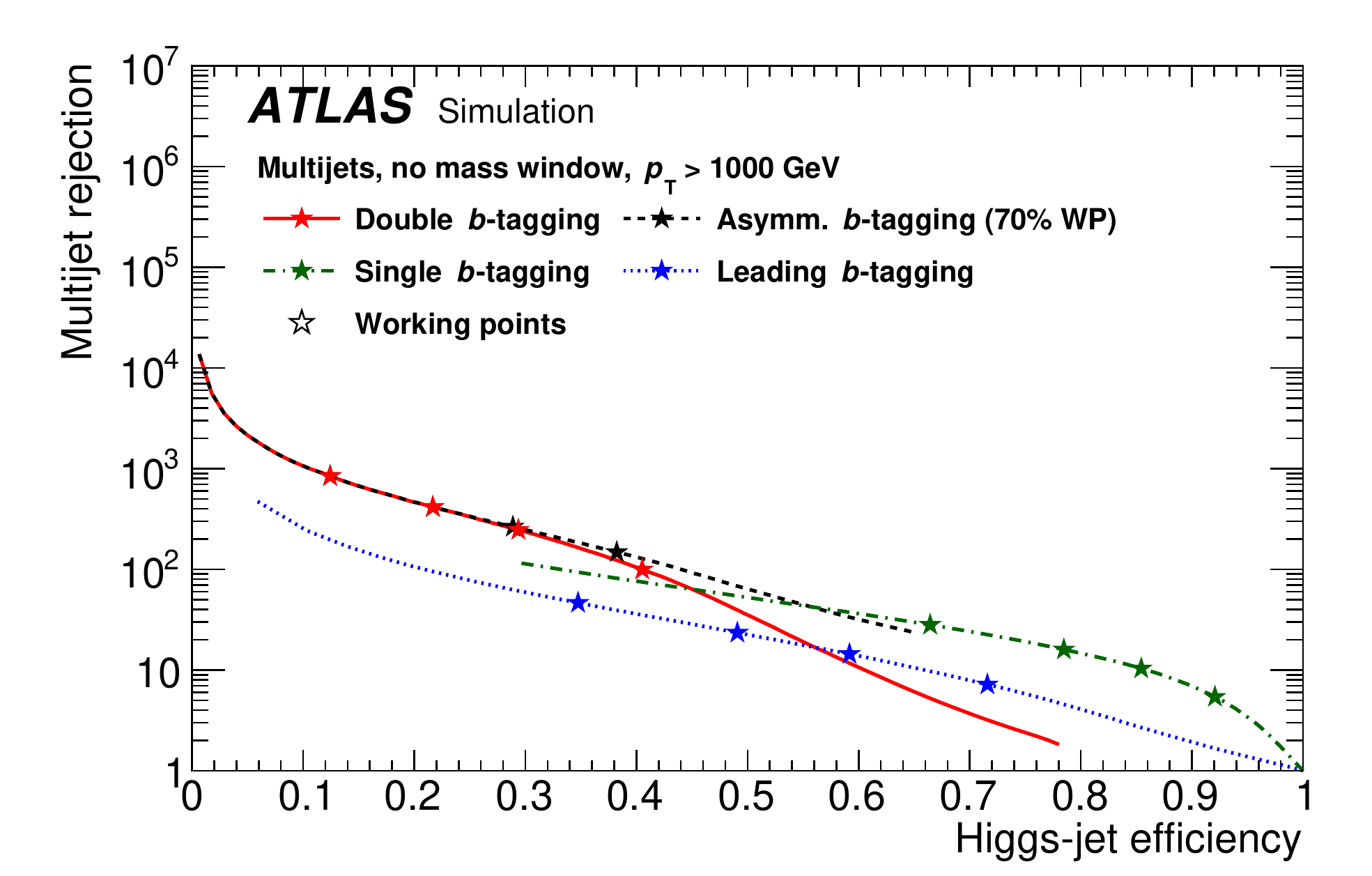} \\
\includegraphics[width=0.49\textwidth]{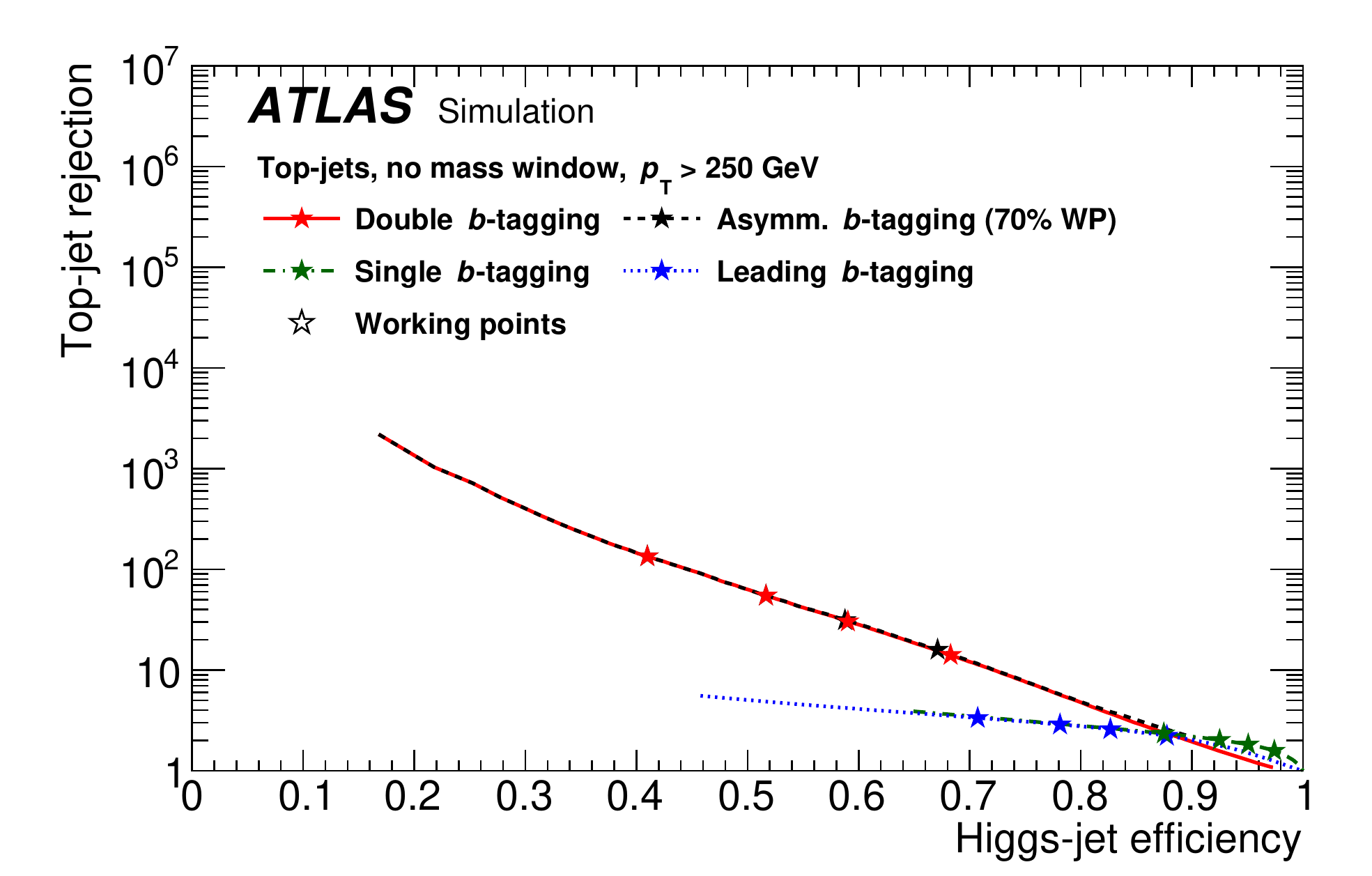}
\includegraphics[width=0.49\textwidth]{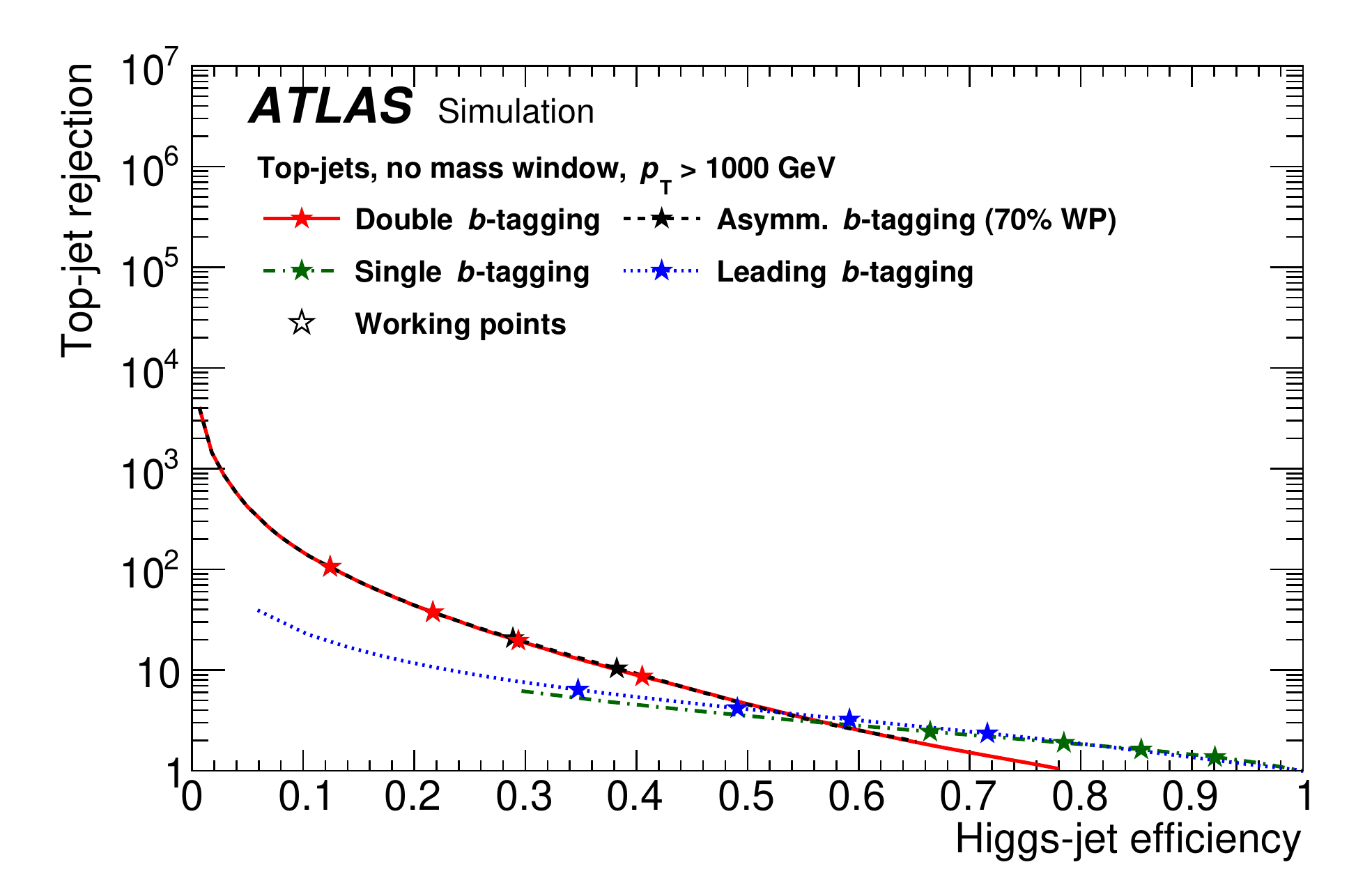}
\caption{The multijet (top) and the top-jet (bottom) rejection as a function of the Higgs  tagging
efficiency for \largeR\ jet \pT\ above 250~\GeV\ (left) and above 1000~\GeV\ (right) for various $b$-tagging benchmarks
defined in Section~\ref{sec:ftag_wp}. The stars correspond to the 60\%, 70\%, 77\% and 85\% $b$-tagging WPs (from left to right).
The curves for the double-$b$-tagging and asymmetric-$b$-tagging working points coincide over a large range of \HiggsJet\ efficiency.}
\label{fig:background_rejection}
\end{figure}

The multijet and top-quark background rejections as a function of the Higgs  tagging efficiency for various $b$-tagging
benchmarks are shown in Figure~\ref{fig:background_rejection}. Plots on the left show the performance for \HiggsJet\ \pT\ above
250~\GeV\ and plots on the right show the performance for \HiggsJet\ \pT\ above 1000~\GeV. The double-$b$-tagging and asymmetric-$b$-tagging selections
give the best background rejection in a large range of Higgs  tagging efficiencies. At high \HiggsJet\ efficiencies above
$\sim$90\% ($\sim$55\%) for \HiggsJet\ transverse momenta above 250 (1000)~\GeV\ the single-$b$-tagging benchmark shows a higher multijet
and top-quark background rejection. To achieve such a high \HiggsJet\ efficiency, a very loose double-$b$-tagging or asymmetric-$b$-tagging requirement is
needed, which results in a low light-flavour jet rejection. The double-$b$-tagging and asymmetric $b$-tagging working points do not reach an efficiency of 100\% due
to a requirement of at least two track-jets. In the case of asymmetric $b$-tagging, Higgs  tagging efficiencies are below 100\%
because of the fixed $b$-tagging working point requirement on one of the track-jets. The drop in performance is pronounced at
high jet transverse momenta due to the lower efficiency to reconstruct two subjets and the decrease in the MV2c10 $b$-tagging
performance~\cite{ATL-PHYS-PUB-2017-010}.

\subsection{Mass window optimisation}
\label{sec:mass_window}
 
The reconstructed Higgs boson mass distribution provides a powerful way to distinguish the Higgs boson signal from background processes.
The muon-corrected combined mass described in Section~\ref{sec:objects} is used to impose the Higgs boson mass requirement and
select \largeR\ jets with a mass around the SM Higgs boson mass. The Higgs boson mass resolution, $\sigma_{m}$, varies as a function
of the reconstructed \largeR\ jet \pT, so the mass window is optimised and parameterised as a function of \HiggsJet\ \pT. Two
working points are defined:
 
\begin{itemize}
\item tight mass window, containing 68\% of \HiggsJets;
\item loose mass window, containing 80\% of \HiggsJets.
\end{itemize}
 
The mass window is defined as the smallest window containing the given fraction of \HiggsJets. The out-of-cone effects, ISR and the missing
neutrinos from semileptonic $b$-hadron decays have an impact on the mass resolution that is similar to their impact on the \pt\ response; therefore, the mass window
optimisation depends on the applied \HiggsJet\ selection and on the \HiggsJet\ \pT\ spectrum.
 
\begin{figure}
\centering
\includegraphics[width=0.60\textwidth]{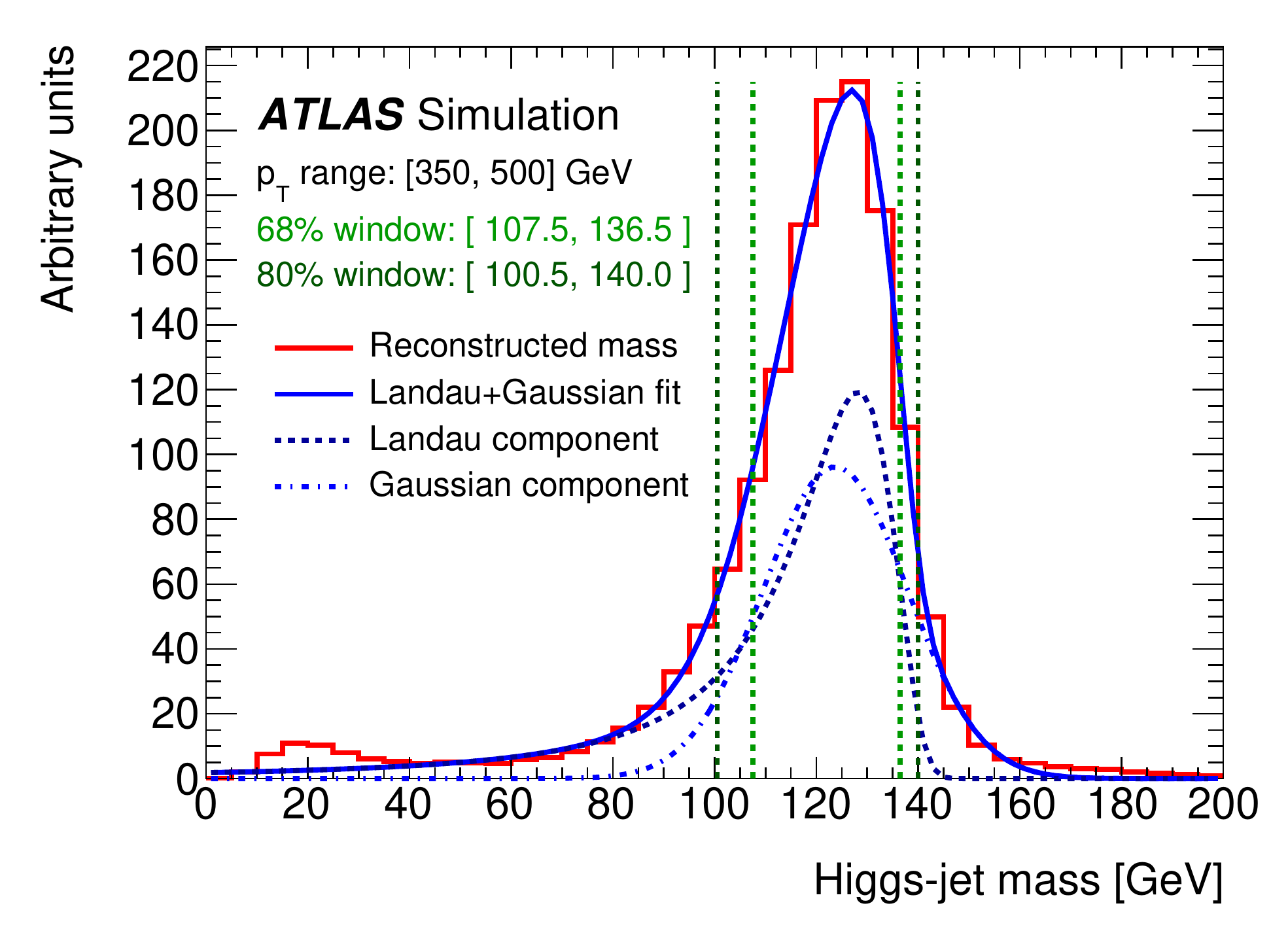}
\caption{The \HiggsJet\ mass distribution for jet transverse momenta in the range 350 to 500~\GeV\ after reweighting the \pT\ spectrum.
The dotted and dash-dotted blue curves correspond to the two components of the fit function, while the solid blue curve shows the combination thereof.
The vertical lines indicate the boundaries of the mass ranges for 68\% (light green) and 80\% (dark green) containment.}
\label{fig:mass_dist_Pt1}
\end{figure}

Figure~\ref{fig:mass_dist_Pt1} shows the reconstructed Higgs boson mass distribution for \HiggsJet s with a \pT\ in the range
350 to 500~\GeV. The mass region below 50~\GeV\ is affected by grooming and out-of-cone effects. In the case of
asymmetric $H \rightarrow b \bar{b}$ decays, where one of the $b$-hadrons carries a large fraction of the Higgs boson \pT,
the \largeR\ jet's axis is close to the direction of the higher-\pT\ $b$-hadron. The decay products of the lower-\pT\ $b$-hadron could be removed by grooming or not
fully captured in the \largeR\ jet. That leads to smaller \HiggsJet\ masses. The mass region above 150~\GeV\ suffers from additional
contributions from initial-state radiation. A large fraction of the ISR is suppressed by selecting the reconstructed \HiggsJet\ containing the
highest-\pT\ Higgs boson candidate. However, the high mass tails are still substantial in high \HiggsJet\ \pT\ regions and affect the Higgs boson
mass window definition.
 
In order to suppress the impact of the tails on the mass window definition, a fit of the mass distribution is performed.
The fit function is chosen empirically to describe the core of the mass distribution, while mitigating the tails. The chosen function is a linear
combination of a Landau function to describe the low mass part of the distribution and a Gaussian function to describe the high mass part.
 
The fit is performed in 12 \HiggsJet\ \pT\ bins across the entire range of transverse momentum from 250 to 2500~\GeV.
 
\begin{figure}
\centering
\includegraphics[width=0.49\textwidth]{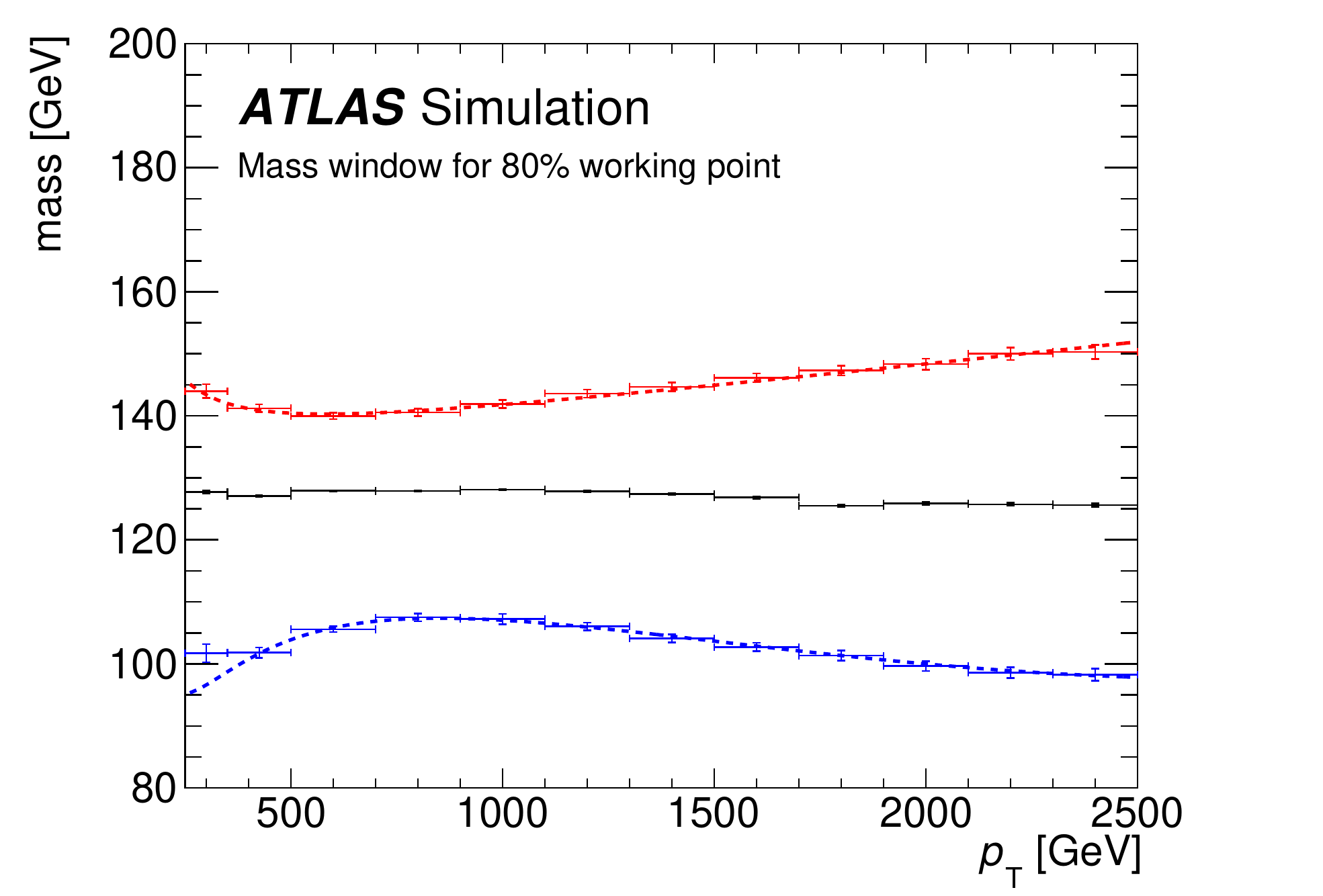}
\includegraphics[width=0.49\textwidth]{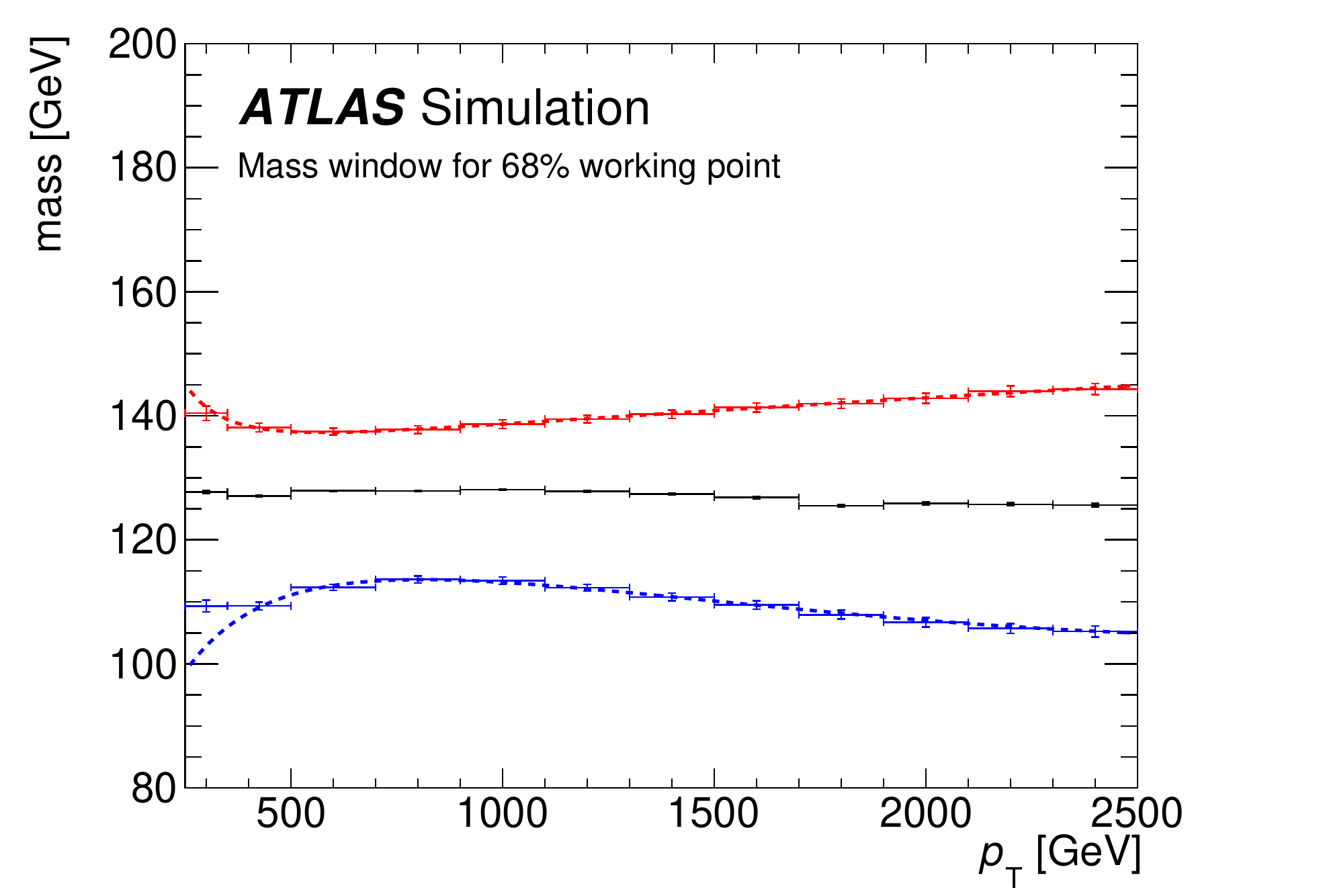}
\caption{The \HiggsJet\ mass window interval for a loose (left) and a tight (right) working point. The dashed lines show a fit
to the derived intervals (blue and red markers) as a function of the \HiggsJet\ \pT. The black markers show the position of the
maximum of the \HiggsJet\ mass distribution.}
\label{fig:fit_masswindow}
\end{figure}

A toy MC simulation is used as input to model the mass window and to estimate the statistical uncertainty on the mass window determination. This toy MC
simulation samples the fit functions mentioned
above and is performed many times in each \pT\ slice.
For each toy MC sample, the mass window is calculated by selecting the smallest window containing the required signal fraction. The final upper and lower
boundaries for a given \pT\ slice are found by averaging over the upper and lower boundaries from the corresponding toy MC samples. The mean defines the position and the RMS the uncertainty
of the window boundaries in each \pT\ slice. Using the mean and RMS from the toy MC samples as input, the mass window is parameterised as a function of the
\HiggsJet\ \pT\ using the fit function:
$f(\pT) = \sqrt{ {\left( a + b / \pT \right)}^{2} + {\left( c \cdot \pT + d \right)}^2}$. The jet mass depends primarily on the energies of the
jet constituents and their angular separations.  Consequently, there are two competing effects: the improving precision of the calorimeter
energy scale with increasing jet \pT\ and the decreasing ability of the calorimeter granularity to resolve individual energy deposits due to increasing
decay collimation with increasing jet \pT. Fit results are shown in Figure~\ref{fig:fit_masswindow} for tight and loose mass window working points.
 
\begin{figure}
\centering
\includegraphics[width=0.60\textwidth]{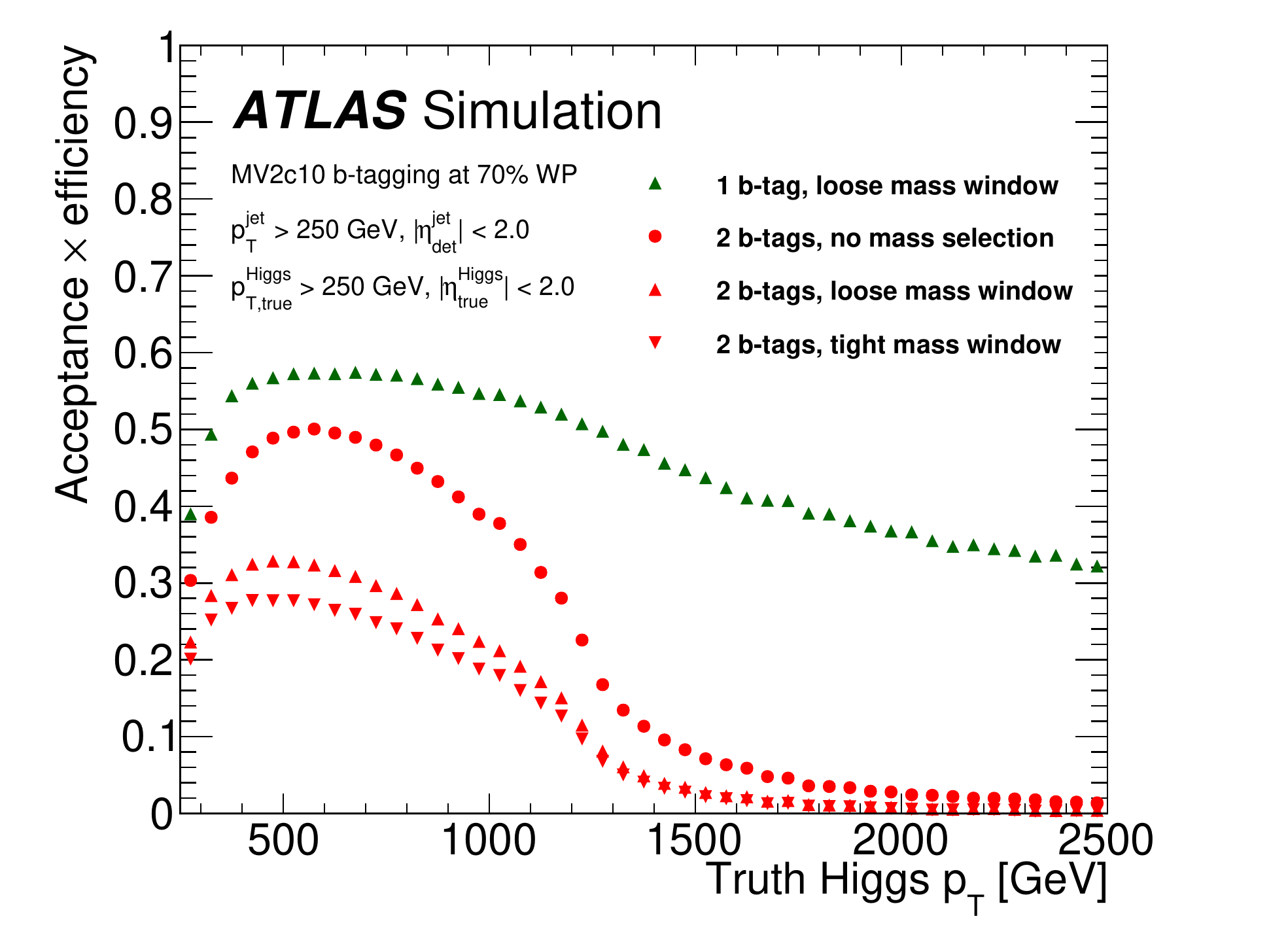}
\caption{The Higgs boson acceptance times efficiency is shown for a few working points: the double and single $b$-tagging with the loose
mass window requirement and the double $b$-tagging with the tight, loose and no mass window requirements.}
\label{fig:acceptanceXeff}
\end{figure}

The Higgs boson acceptance times efficiency is presented in Figure~\ref{fig:acceptanceXeff}. In addition to the truth-matching requirements defined
for Figure~\ref{fig:signal_acceptance}, the double- and single-$b$-tagging, tight, loose and no mass window working points are applied. The double-$b$-tagging
requirement in particular leads to a significant drop in the Higgs boson acceptance times efficiency at high Higgs boson transverse momenta, where the efficiency
to reconstruct two track-jets and the double-$b$-tagging efficiency decrease quickly.
 
\begin{figure}
\centering
\includegraphics[width=0.49\textwidth]{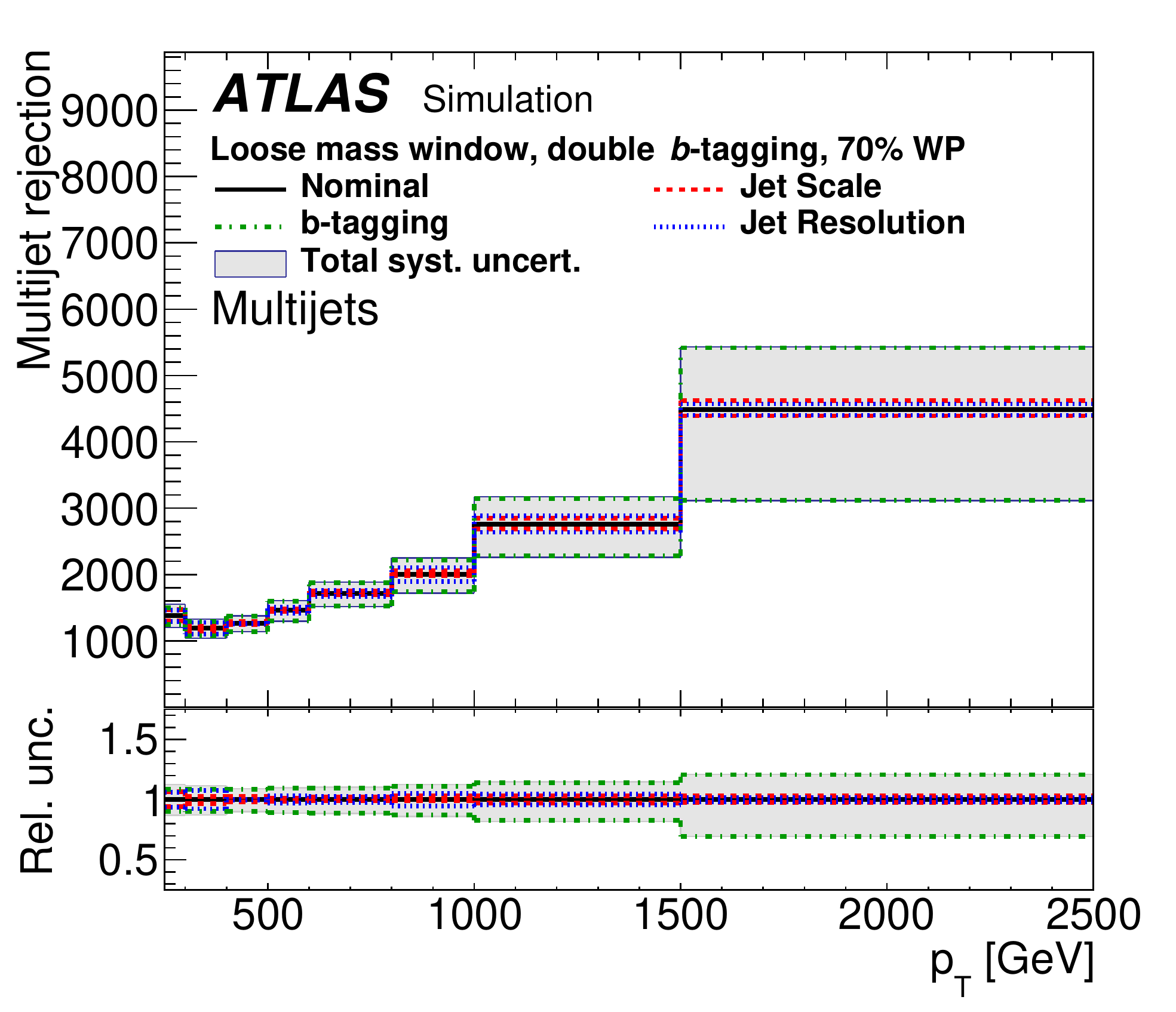}
\includegraphics[width=0.49\textwidth]{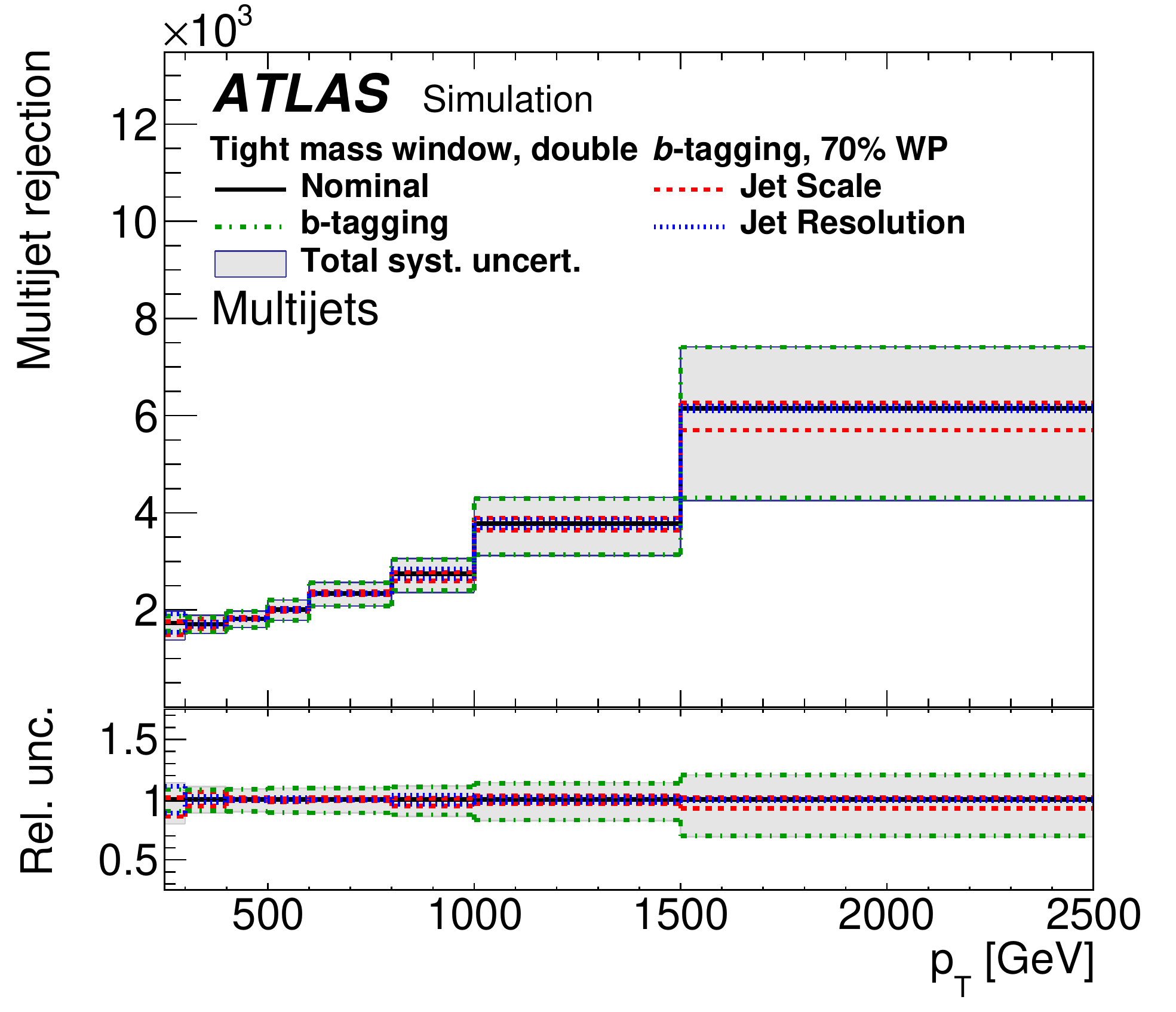}
\caption{Rejection of multijet background as a function of the \HiggsJet\ \pT\ for the loose (left) and tight (right)
mass window requirements, in combination with the 70\% double-$b$-tagging working point. The nominal curves
correspond to the requirement on the MV2c10 discriminant described in Section~\ref{sec:ftag_wp}. Systematic uncertainties
defined in Section~\ref{sec:systematics} as well as their sum in quadrature (total uncertainty) are shown. `Jet Scale' refers to
the sum in quadrature of the jet energy and mass scale uncertainties and `Jet Resolution' refers to the sum in quadrature of
the jet energy and mass resolution uncertainties.}
\label{fig:qcd_mass_efficiency}
\end{figure}

Figure~\ref{fig:qcd_mass_efficiency} shows the rejection of the multijet background as a function of the \HiggsJet\ \pt. Applying a
combination of loose mass window and double-$b$-tagging requirements improves the rejection by a factor of about four relative to the corresponding benchmark without the mass requirement shown in Figure~\ref{fig:btag_efficiency}. The tight mass window
requirement leads to an additional improvement of about 30--50\% in the background rejection. The efficiency of the mass window requirements changes by a few percent after
the application of the double $b$-tagging-requirement due to the dependence of the $b$-tagging efficiency on the jet kinematics.
 
\begin{figure}
\centering
\includegraphics[width=0.49\textwidth]{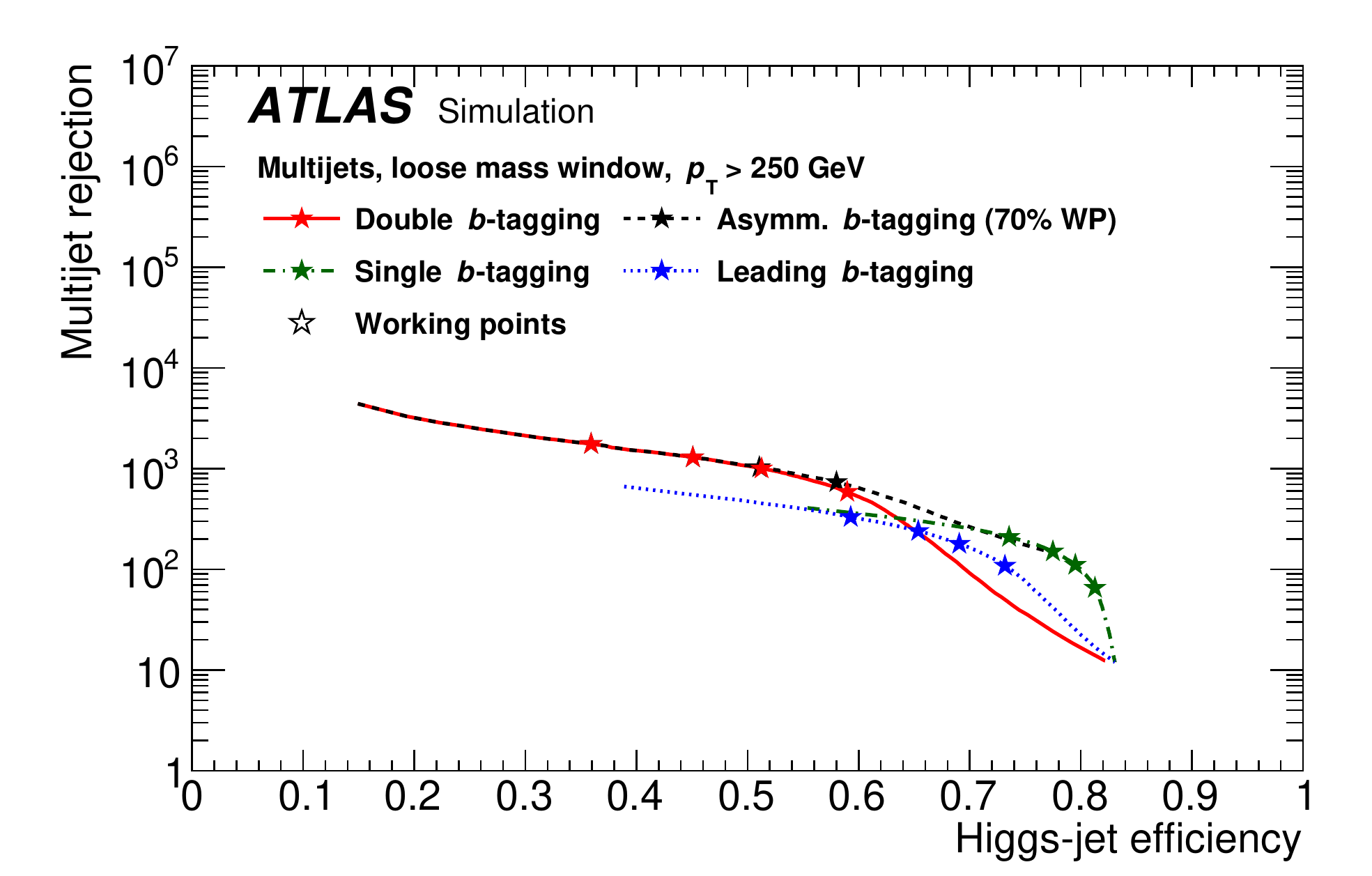}
\includegraphics[width=0.49\textwidth]{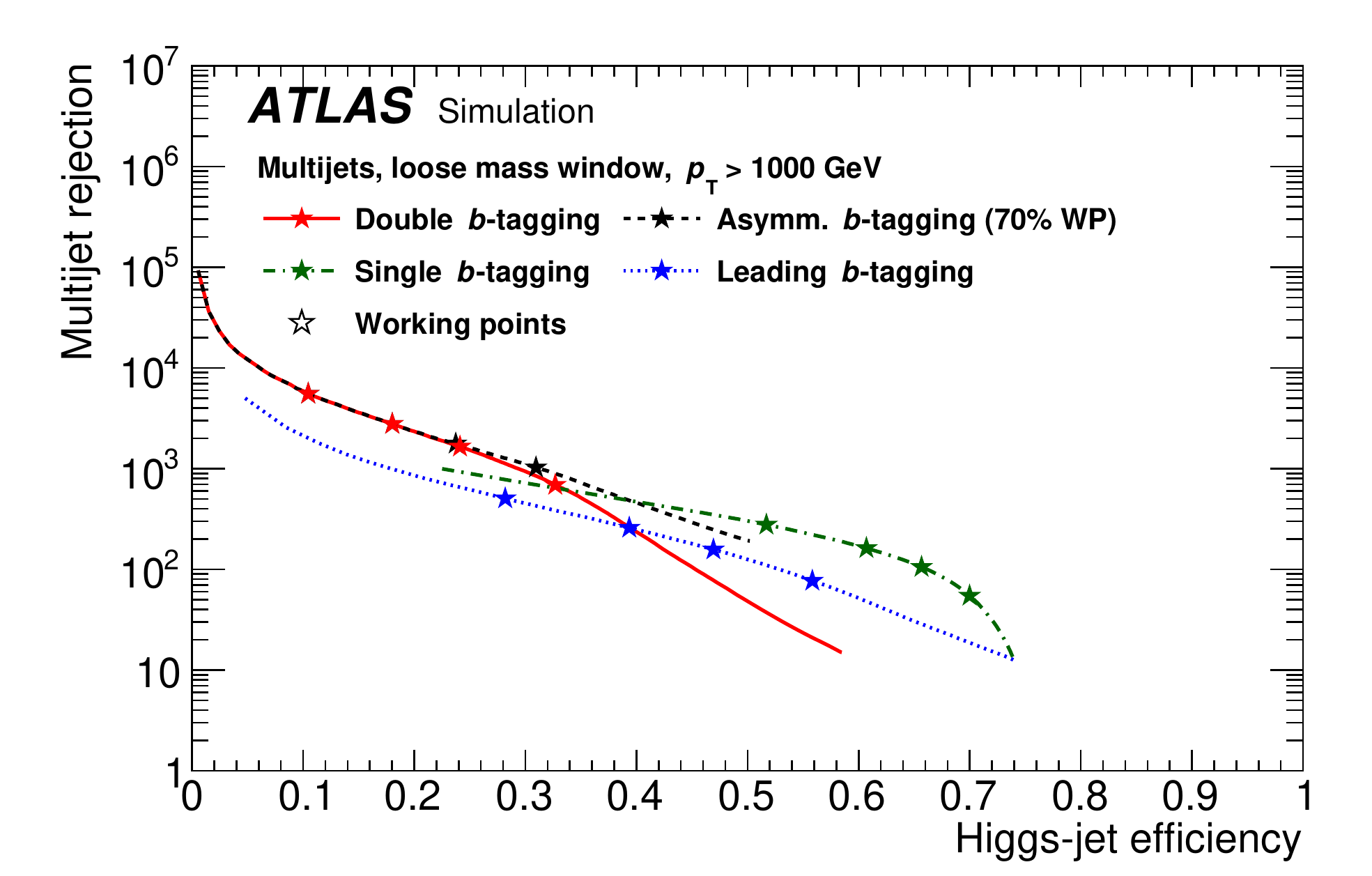} \\
\includegraphics[width=0.49\textwidth]{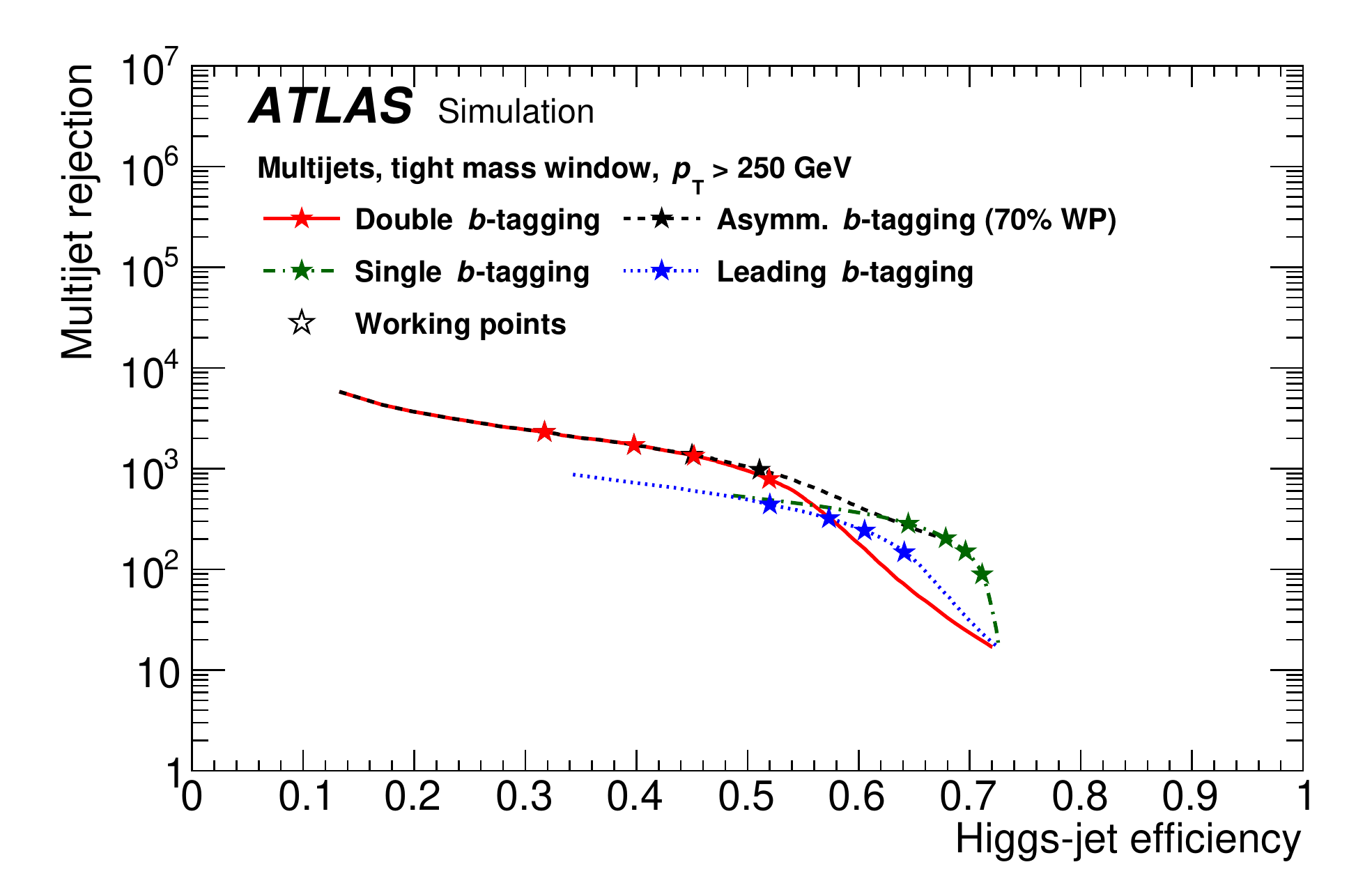}
\includegraphics[width=0.49\textwidth]{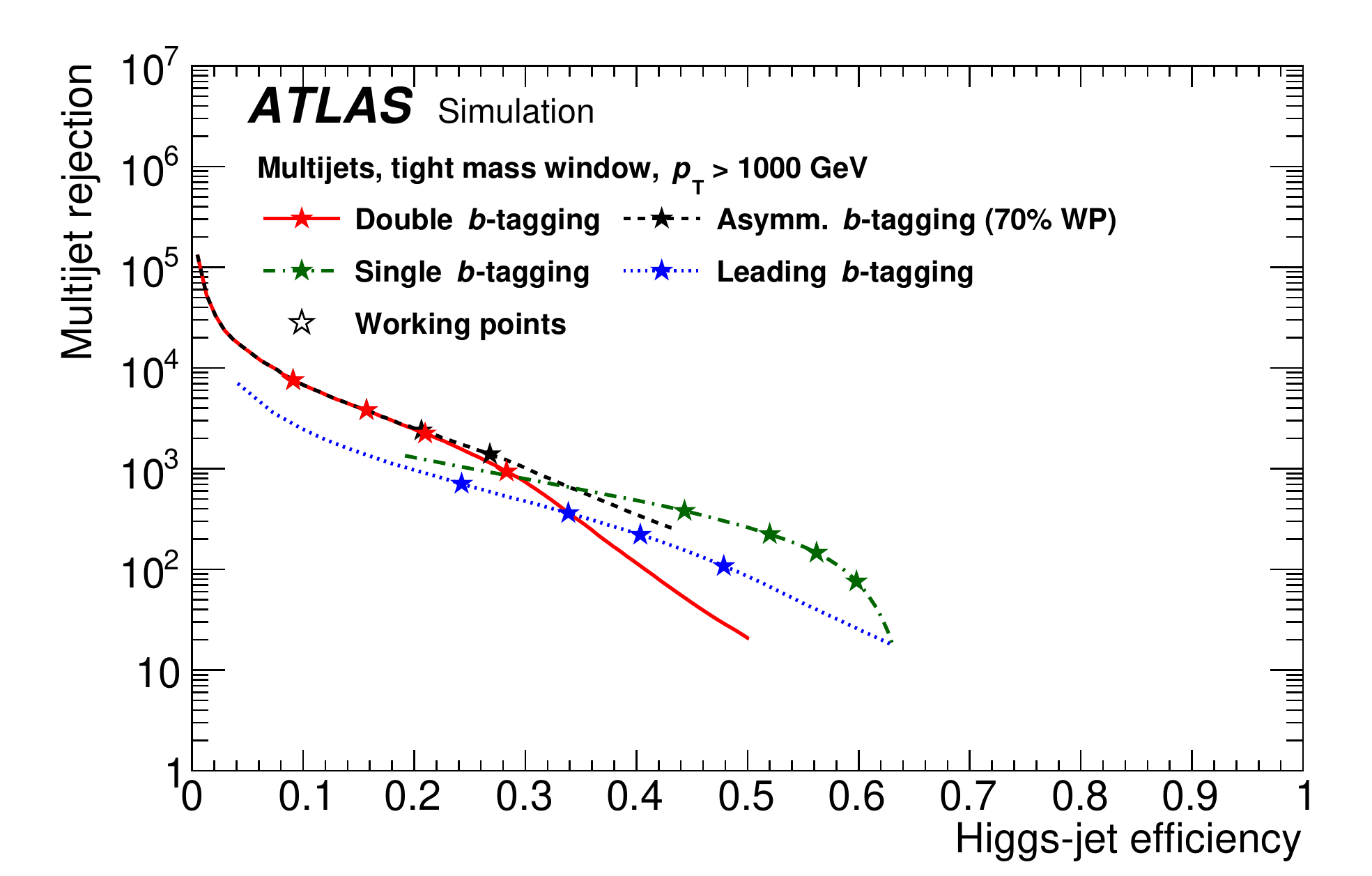}
\caption{Rejection of multijet background as a function of the Higgs boson tagging efficiency for loose (top) and
tight (bottom) mass window requirements for \largeR\ jet \pT\ above 250~\GeV\ (left) and above 1000~\GeV\ (right)
for various $b$-tagging benchmarks. The stars correspond to the 60\%, 70\%, 77\% and 85\%
$b$-tagging WPs (from left to right). The curves for the double- and asymmetric-$b$-tagging working points coincide
over a large range of \HiggsJet\ efficiency.}
\label{fig:qcd_mass_rejection}
\end{figure}

The corresponding rejection of the multijet background as a function of the \HiggsJet\ efficiency is shown in Figure~\ref{fig:qcd_mass_rejection}
for different \HiggsJet\ \pt ranges, $b$-tagging benchmarks, and mass window requirements. Application of the mass window
requirement improves the performance of the tagger substantially. For a fixed signal efficiency of 40\% and \largeR\ jet \pT\ above 250~\GeV,
the multijet rejection rises from roughly 360 after applying the double-$b$-tagging requirement to about 1480 (1670) for the combination
of the double-$b$-tagging and loose (tight) mass window requirements.
 
\begin{figure}
\centering
\includegraphics[width=0.49\textwidth]{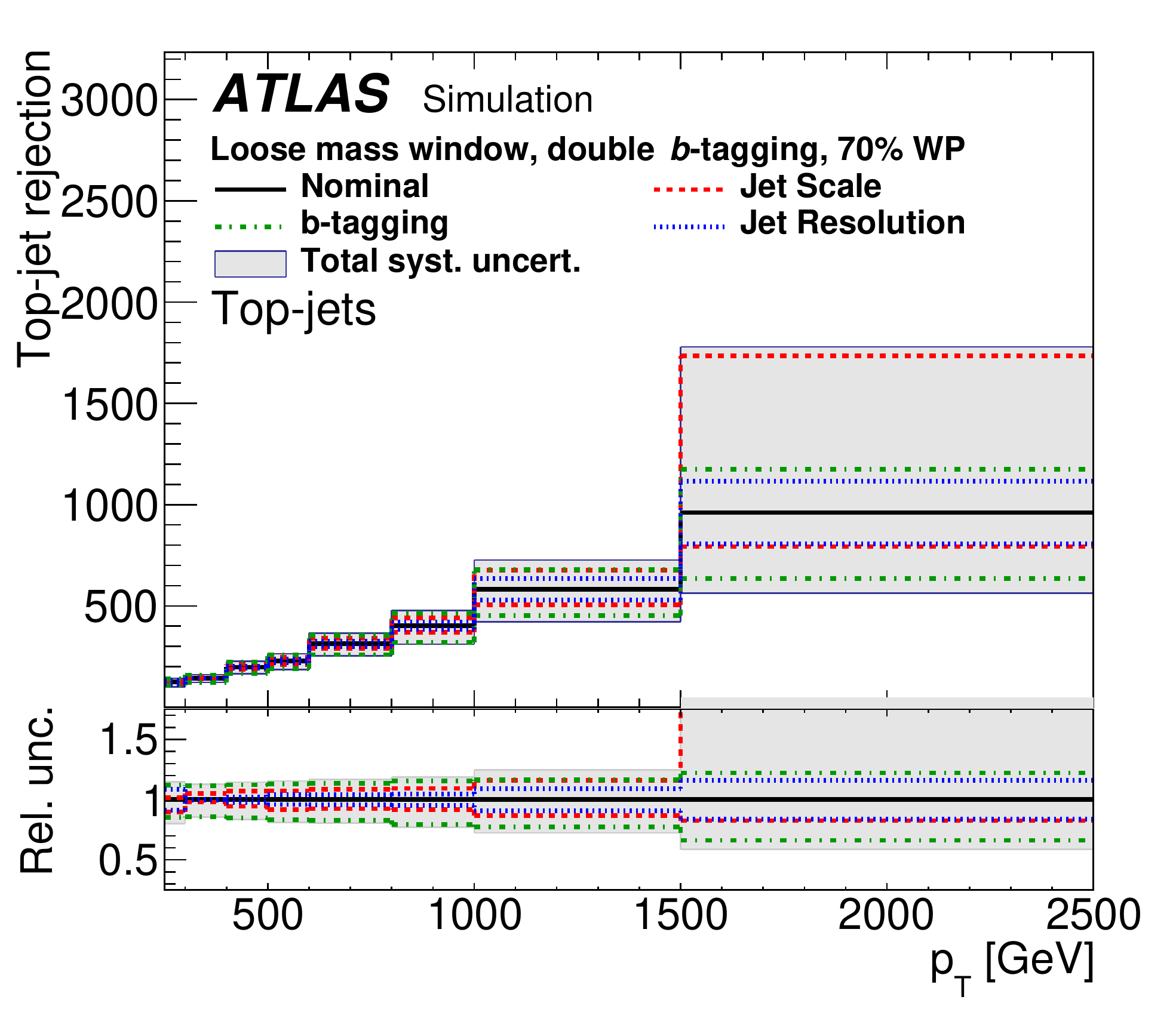}
\includegraphics[width=0.49\textwidth]{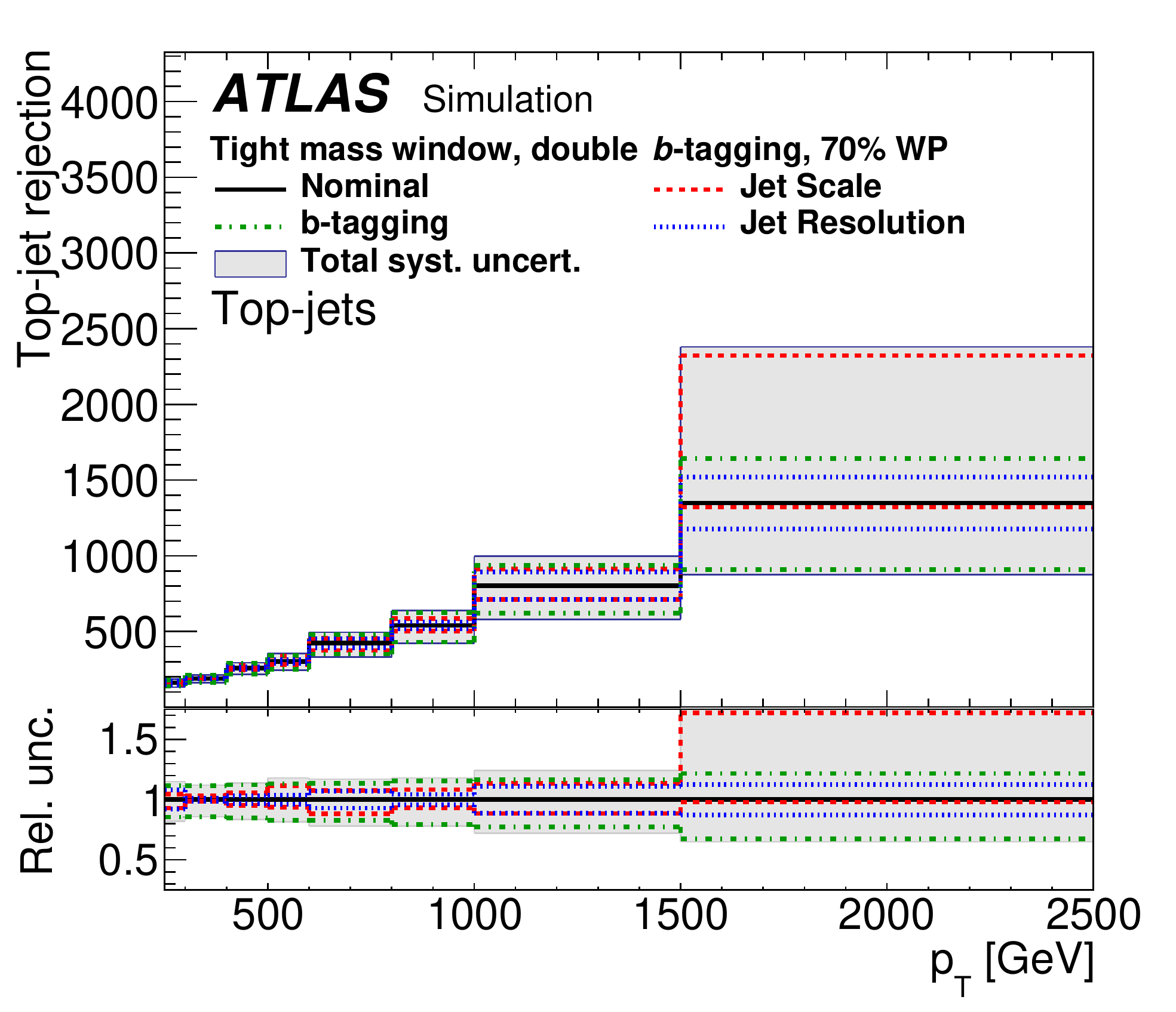}
\caption{Rejection of the top-jet background as a function of the Higgs-jet \pT\ for the loose (left) and tight (right) mass window requirements,
in combination with the 70\% double-$b$-tagging working point. The nominal curves correspond to the requirement
on the MV2c10 discriminant described in Section~\ref{sec:ftag_wp}. Systematic uncertainties
defined in Section~\ref{sec:systematics} as well as their sum in quadrature (total uncertainty) are shown. `Jet Scale' refers to
the sum in quadrature of the jet energy and mass scale uncertainties and `Jet Resolution' refers to the sum in quadrature of
the jet energy and mass resolution uncertainties.}
\label{fig:mtt_mass_efficiency}
\end{figure}

Figure~\ref{fig:mtt_mass_efficiency} shows the hadronic top-quark background rejection as a function of the \HiggsJet\ \pt for
combinations of mass window and $b$-tagging benchmarks. The background rejection is higher
for multijets than for hadronically decaying top quarks. The rejection varies between 120 (170)  at low \pt and 1000 (1300) at
high \pt for the loose (tight) mass window and double-$b$-tagging benchmark. In comparison with the benchmarks without the mass
window requirement, the rejection is improved by about one order of magnitude, but the shape as function of \pt\ is fundamentally
different. At low \pt, not all decay products of the top quark are contained in the \largeR\ jet. Thus the reconstructed jet mass has
a long tail towards low jet masses with a substantial fraction of jets within the mass window of the tagger. Hence, the rejection
at low jet \pt\ is not improved as much as at high jet \pt. The tight mass window requirement further
improves the background rejection by 15--40\% as function of \pt.
 
\begin{figure}
\centering
\includegraphics[width=0.49\textwidth]{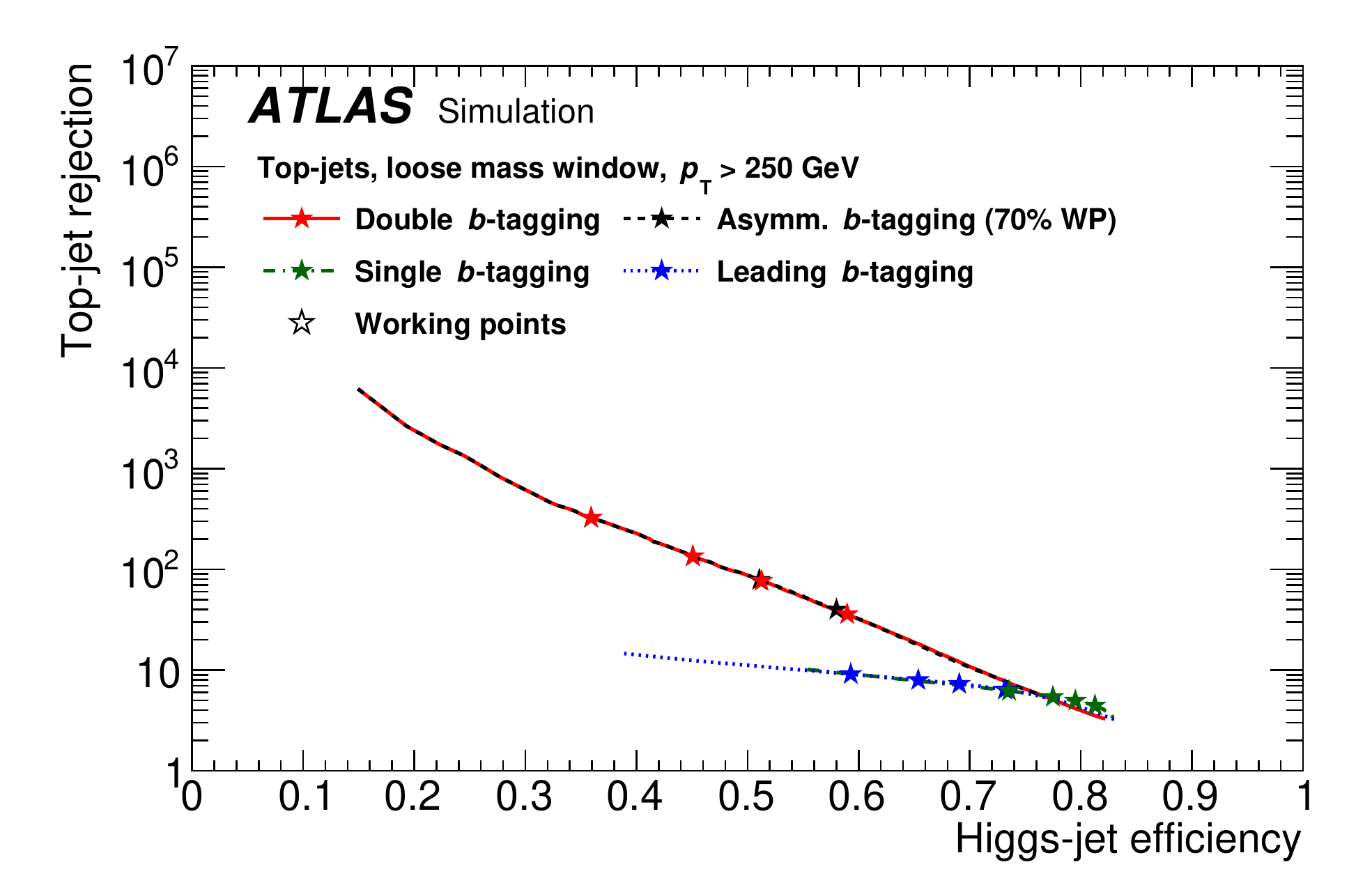}
\includegraphics[width=0.49\textwidth]{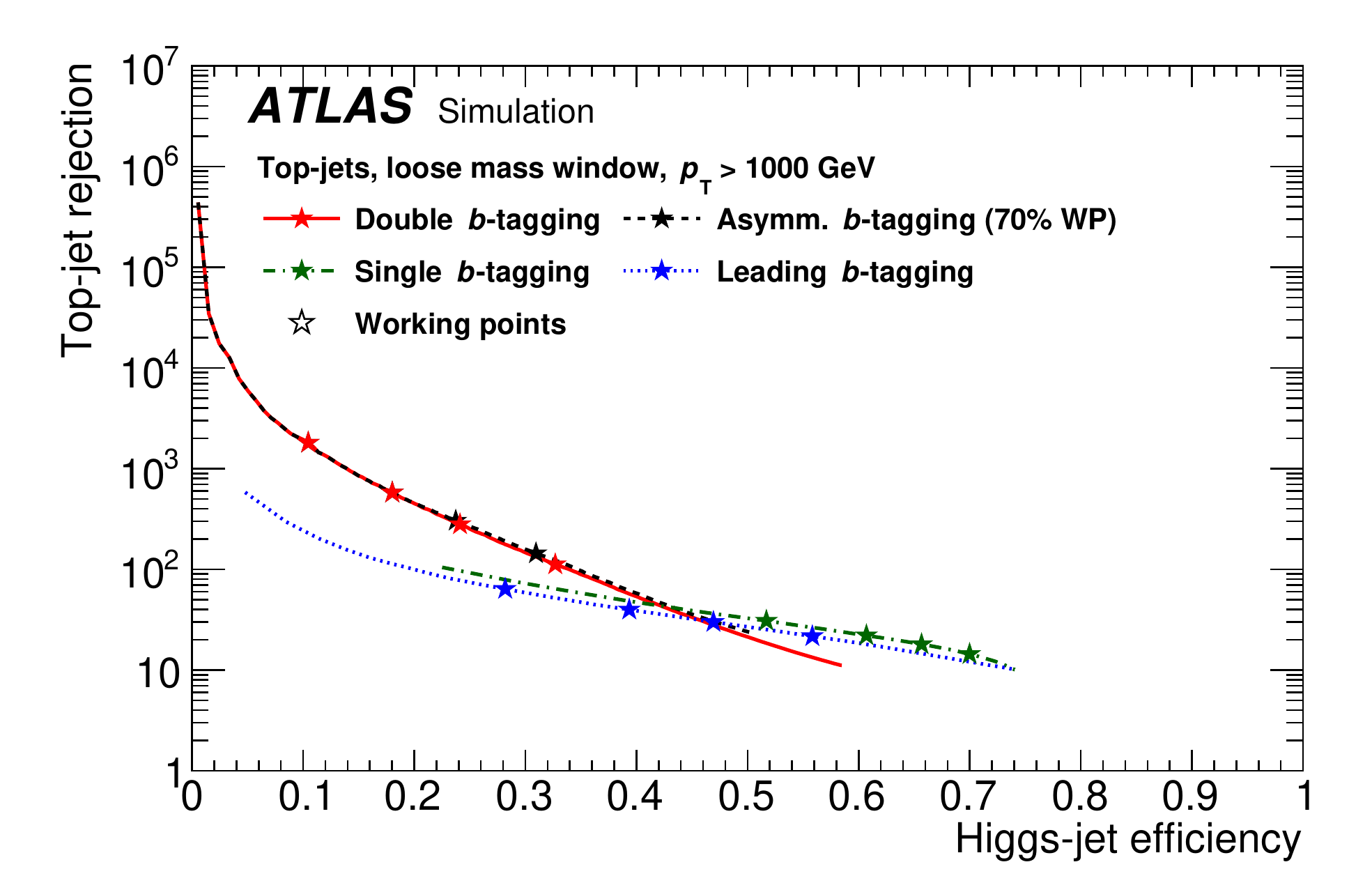} \\
\includegraphics[width=0.49\textwidth]{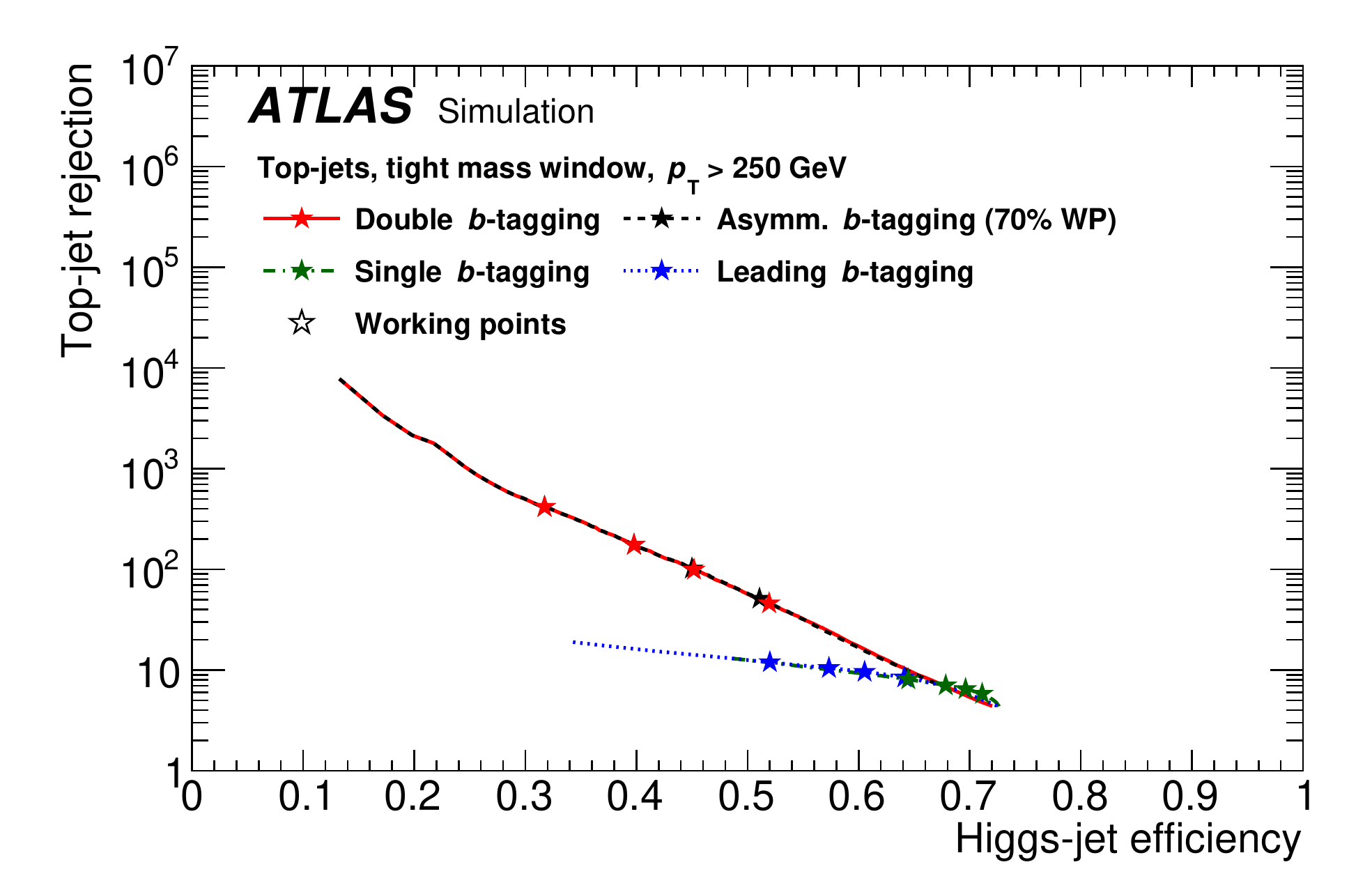}
\includegraphics[width=0.49\textwidth]{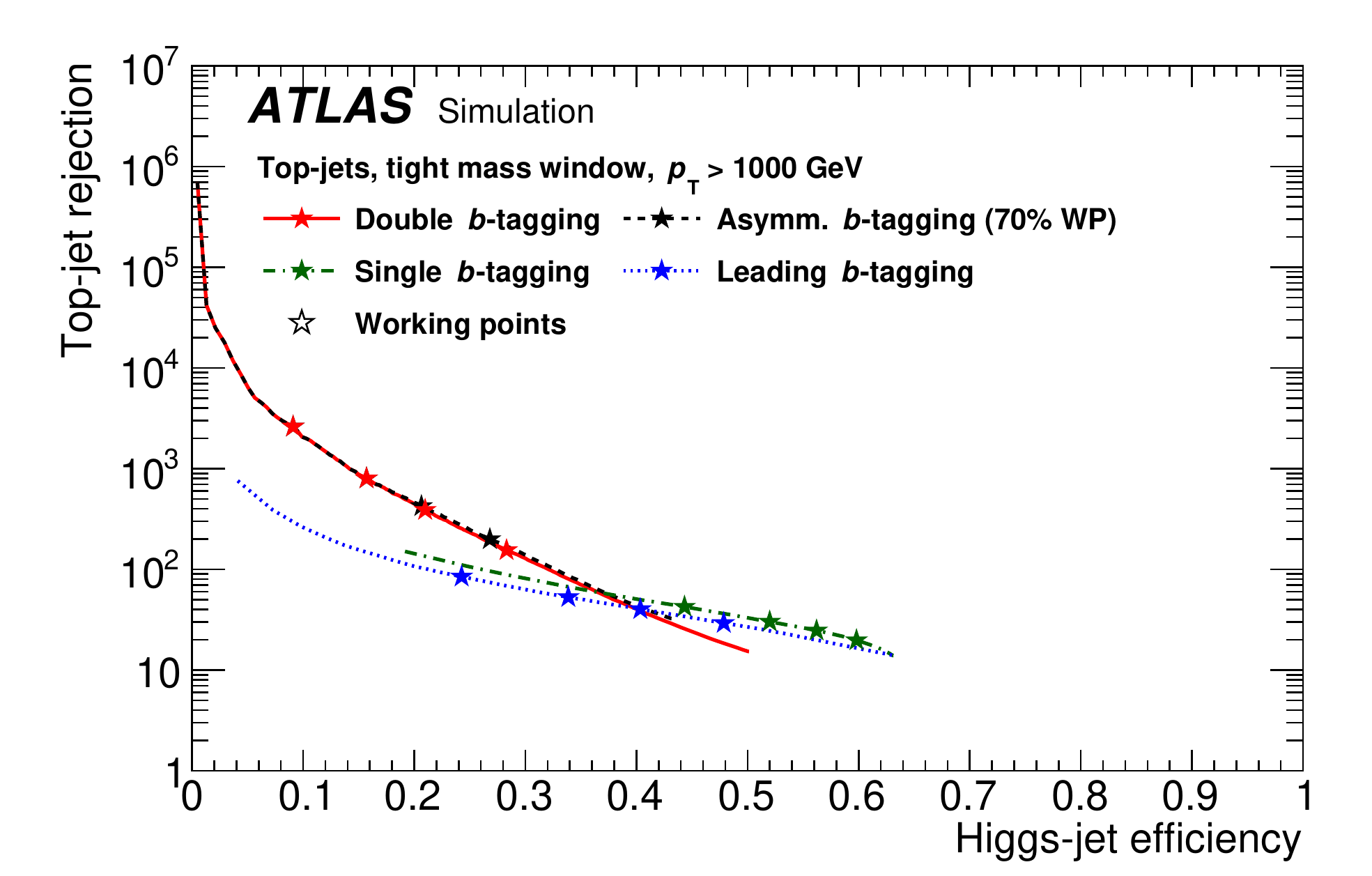}
\caption{Rejection of the top-jet background as a function of the Higgs  tagging efficiency for loose (top) and tight (bottom) mass window
requirements for \largeR\ jet \pT\ above 250~\GeV\ (left) and above 1000~\GeV\ (right) for various $b$-tagging benchmarks. The stars correspond to the
60\%, 70\%, 77\% and 85\% $b$-tagging WPs (from left to right). The curves for the double- and asymmetric-$b$-tagging working
points coincide over a large range of \HiggsJet\ efficiency.}
\label{fig:mtt_mass_rejection}
\end{figure}

The rejection of the hadronic top-quark background as a function of the Higgs  tagging efficiency is shown in Figure~\ref{fig:mtt_mass_rejection}. For the loose mass
window requirement, an improvement from 140 to 200 is found at a fixed \HiggsJet\ efficiency of 40\%, whereas for the  tight mass window a smaller improvement from 140 to
160 is observed relative to no mass requirement for \largeR\ jet \pT\ above 250~\GeV. The rejection values are lower for double $b$-tagging and asymmetric $b$-tagging
for \largeR\ jet \pT\ above 1~\TeV, and for high Higgs tagging efficiency single and single leading $b$-tagging are better options.
 
\subsection{Jet substructure}
\label{sec:jet_substructure}
 
Sections~\ref{sec:ftag_wp} and~\ref{sec:mass_window} present the performance of the \HiggsJet\ tagger based on
the $b$-tagging and jet mass requirements designed to distinguish \largeR\ jets produced by Higgs boson decays from
backgrounds. This section discusses the possibility of improving the background rejection with the help of
other jet substructure variables and tighter selections on jet mass and $b$-tagging applied on top of the
previously defined jet mass window and $b$-tagging benchmark working points. These additional
selections are referred to as secondary selections.
 
\begin{table}[tb]
\caption{Overview of jet substructure variables. A short description of these substructure variables can be found in Refs.~\cite{PERF-2015-03,PERF-2015-04}.
$^{(\ast)}$ Exclusive dipolarity forces the jet to have exactly two subjets from the \kt\ algorithm to begin with, which is different from
the dipolarity, which runs \kt\ clustering and then takes all jets with \pT\ above 5~\GeV.
}
\medskip
\centering
\begin{tabular}{lll}
\hline \hline
Symbol                    & Description & Reference\\
\hline \hline
\multicolumn{3}{c}{Energy correlation functions} \\
\hline
$E_{\textrm{CF}i}$                                  & $i$-th energy correlation function                                                               & \cite{ECF,D2}\\
\CTwoBetaOne                                              & $E_\textrm{CF3} \cdot E_\textrm{CF1} / E_\textrm{CF2}$                        & \\
\DTwoBetaOne                                              & $E_\textrm{CF3} \cdot \left( E_\textrm{CF1} / E_\textrm{CF2} \right)^3 $  & \\
\hline \hline
\multicolumn{3}{c}{$n$-subjettiness} \\
\hline
$\tau_{n}$                                                            &  $n$-subjettiness                                                                                                & \cite{Thaler:2010tr,tau2} \\
$\tau_{n}^{\mathrm{wta}}$                                  & $n$-subjettiness variant \textit{winner takes all} (wta)                                         & \\
$\tau_{ji}$, $\tau_{ji}^{\mathrm{wta}}$                & $\tau_{j} / \tau_{i}$ or $\tau_{j}^\mathrm{wta} / \tau_{i}^\mathrm{wta}$, $j > i$     & \\
\hline \hline
\multicolumn{3}{c}{Centre-of-mass observables} \\
\hline
$\FoxWolfram_{i}$                   & $i$-th Fox--Wolfram moment                 & \cite{foxwolfram} \\
\hline \hline
\multicolumn{3}{c}{Dipolarity} \\
\hline
 
$\Dipolarity^{\mathrm{excl}}$  & Exclusive dipolarity$^{(\ast)}$ & \cite{Dipolarity} \\
\hline \hline
\multicolumn{3}{c}{Cluster sequence} \\
\hline
$\kt \Delta R$            & $\Delta R$ of two subjets within the \largeR\ jet & \cite{Catani:1993hr} \\
\MassDrop                & \kt mass drop & \cite{ref:boostedhiggsprl} \\
\hline \hline
\multicolumn{3}{c}{Splitting measures} \\
\hline
$\sqrt{d_{ij}}$           & \kt splitting scale from $i\rightarrow j$ splitting &  \cite{Butterworth:2002tt,STDM-2012-12} \\
 
\hline \hline
\multicolumn{3}{c}{Thrust} \\
\hline
\ThrustMin, \ThrustMax  & Thrust & \cite{Thrust} \\
 
\hline \hline
\multicolumn{3}{c}{Shape} \\
\hline
\Aplanarity          & Aplanarity & \cite{SphericityAplanarity} \\
\PlanarFlow        & Planar flow & \cite{PlanarFlow}  \\
\Sphericity          & Sphericity & \cite{SphericityAplanarity} \\
\hline \hline
\multicolumn{3}{c}{Other} \\
\hline
\Angularity         & Angularity & \cite{STDM-2011-38} \\
 
\hline \hline
\end{tabular}
\label{tab:jssvars}
\end{table}

Many jet substructure variables exist that can capture features of a jet's internal structure and  can potentially
give additional discrimination power against backgrounds from multijet production and top-quark decays. They are based
on the jet constituents and exploit quantities such as transverse momentum and angular distance between the constituents. They
give information about different jet attributes such as shape (e.g.\ sphericity, aplanarity) or number of axes (e.g. two-subjettiness $\tau_2$).
Ratios are often used to avoid scale dependence of substructure variables. Table~\ref{tab:jssvars} lists the jet substructure variables
that are investigated in this study, together with a short description and references. Secondary selections on jet mass and the
flavour-tagging discriminant for the track-jets, MV2c10, are also considered relative to the previously defined mass window and
$b$-tagging benchmark working points and their performance is compared with that achieved by the application of additional jet
substructure variables to these benchmarks. Two categories of secondary selections are used for the $b$-tagging discriminant MV2c10,
and these exploit the potential of tighter $b$-tagging working points where the criteria are tightened for both track-jets (double $b$-tagging) or
for only one track-jet (single $b$-tagging).
 
\begin{figure}
\center
\includegraphics[width=\columnwidth]{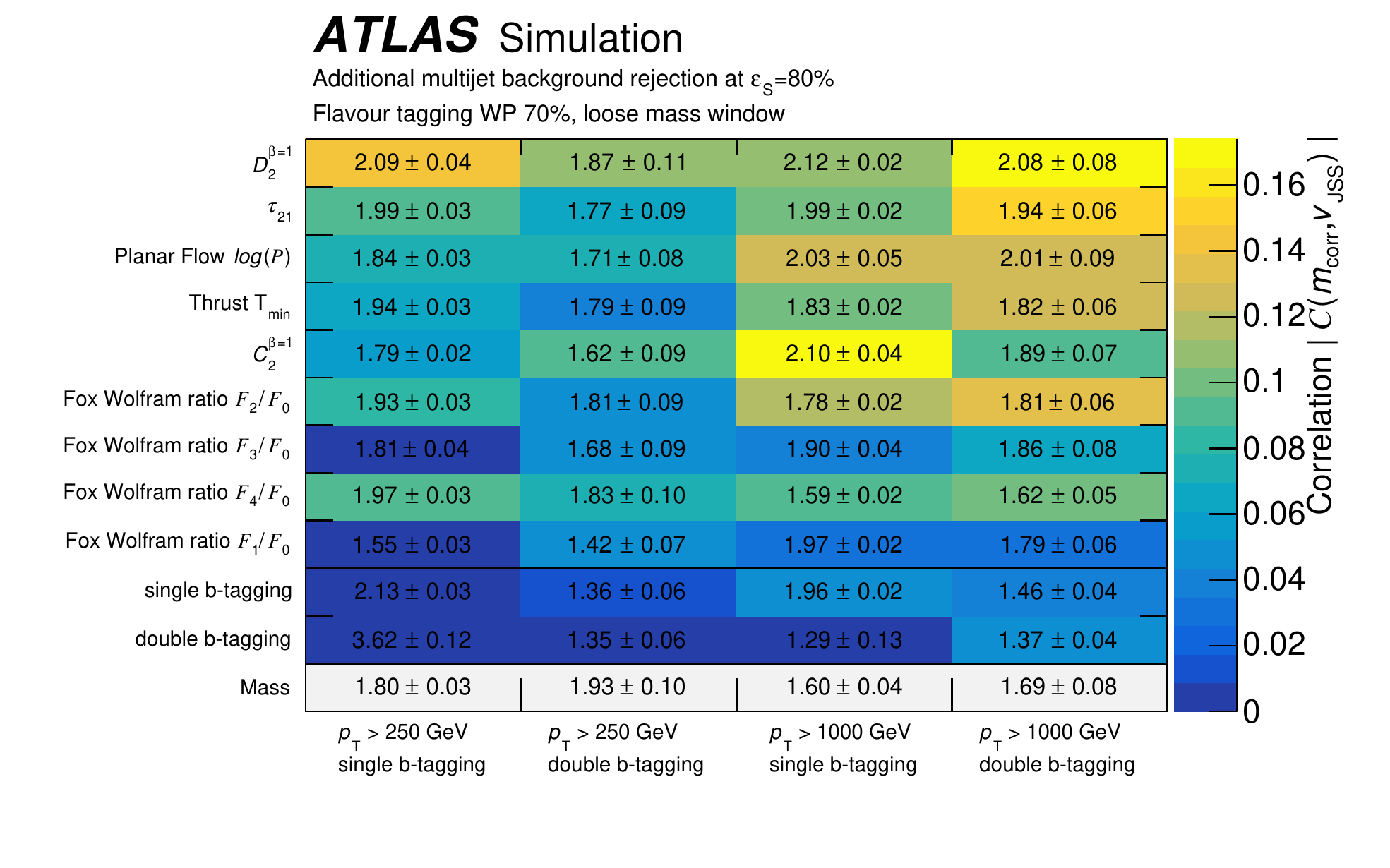}
\caption{Multijet background rejection at 80\% signal efficiency ($\varepsilon_\text{S} = 80\%$) for a variety of substructure variables using different benchmarks
in terms of $b$-tagging strategy and  transverse momentum range. The $z$-axis colour scale represents the absolute value of the linear-correlation coefficient,
$|C(m_\text{corr},v_\text{JSS})|$, between the jet mass and the jet substructure variables. The selection efficiency is determined relative to the mass window
and $b$-tagging benchmark working points defined in Sections~\ref{sec:mass_window} and~\ref{sec:ftag_wp} respectively. }
\label{fig:jss_qcd_scan_z}
\end{figure}
 
\begin{figure}
\center
\includegraphics[width=\columnwidth]{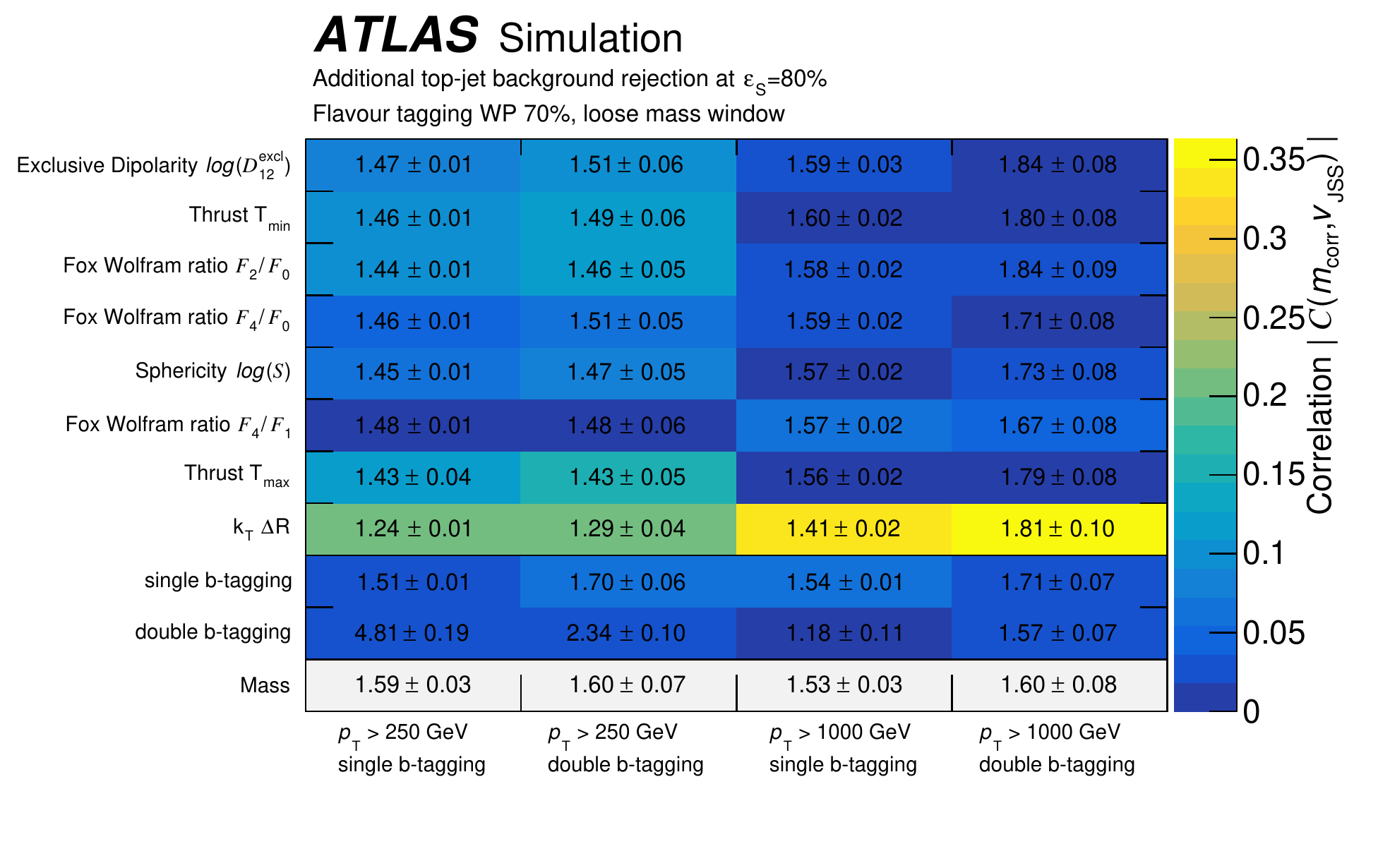}
\caption{Hadronic top-quark background rejection at 80\% signal efficiency ($\varepsilon_\text{S} = 80\%$) for a variety of substructure variables using different
benchmarks in terms of $b$-tagging strategy and  transverse momentum range. The $z$-axis colour scale represents the absolute value of the
linear-correlation coefficient, $|C(m_\text{corr},v_\text{JSS})|$, between the jet mass and the jet substructure variables. The selection efficiency is determined
relative to the mass window and $b$-tagging benchmark working points defined in Sections~\ref{sec:mass_window}
and~\ref{sec:ftag_wp} respectively. }
\label{fig:jss_top_scan_z}
\end{figure}
 
For all secondary selection variables an optimal two-sided range is chosen for each variable and each benchmark working point. Searches
of new-physics resonances typically use tagging definitions with relatively high signal efficiency, around 40\% (75\%) for Higgs-jets with
$\pT = 500$~\GeV\ for double (single) $b$-tagging and a mass requirement. Hence, the two-sided range for a secondary variable which contains
the smallest fraction of background but at least 80\% of signal events is determined. Figures~\ref{fig:jss_qcd_scan_z} and~\ref{fig:jss_top_scan_z}
show the background rejection for a 80\% retention of signal efficiency relative to the jet mass and $b$-tagging benchmark working points for multijet and
hadronic top-quark backgrounds, respectively. The matrices in Figures~\ref{fig:jss_qcd_scan_z} and~\ref{fig:jss_top_scan_z} show the background
rejection for substructure variables, secondary jet mass, and MV2c10 $b$-tagging discriminant on the $y$-axis for the four benchmark points of
the \HiggsJet\ tagger on the $x$-axis. The $z$-axis colour scale represents the absolute value of the linear-correlation coefficient of the substructure variable
and the jet mass for the corresponding background. For each benchmark, five variables with the largest background rejection are selected and all selected variables
for every benchmark are shown.
 
In general, there are improvements across the various benchmark points. The background rejection is often higher for the multijet background
than for the hadronically decaying top quarks. The secondary $b$-tagging discriminant is very powerful, and there are only a few areas of phase
space where substructure yields larger improvements than an optimised $b$-tagging working point. However, substructure variables are an
interesting alternative to tighter $b$-tagging working points for \largeR\ jet \pT\ above 1~\TeV. For the multijet background (Figure~\ref{fig:jss_qcd_scan_z}),
a tighter requirement on the double $b$-tagging achieves a background rejection of 3.62 (1.35) in the inclusive range $\pT > 250$~\GeV\ for the single-$b$-tagging (double-$b$-tagging) working point. In contrast, the improvement from the double-$b$-tagging discriminant is small for working points for $\pT > 1000$~\GeV, achieving
a background rejection of 1.29 (1.37) for the single-$b$-tagging (double-$b$-tagging) working point. At large \pT\ the background rejection for substructure variables
varies between 2.12 (\DTwoBetaOne) and 1.55 (Fox--Wolfram ratio $\FoxWolfram_1 / \FoxWolfram_0$) for a signal efficiency of 80\%. In general,
correlations with the jet mass greater than 10\% are observed for most of the jet substructure variables. The Fox--Wolfram ratios
$\FoxWolfram_3 / \FoxWolfram_0$ and $\FoxWolfram_1 / \FoxWolfram_0$ show the lowest correlations: less than 1\% for most of the benchmarks.
 
The room for improvement is smaller if  secondary jet substructure selections on top of the jet mass window and $b$-tagging benchmark working  points are used in the case of the hadronic top-quark background (Figure~\ref{fig:jss_top_scan_z}). A tighter double-$b$-tagging working point reaches
a factor of 4.81 (2.34) background rejection in the inclusive range $\pT > 250$~\GeV\ for the single-$b$-tagging (double-$b$-tagging) working point. In contrast,
the improvement from the double-$b$-tagging discriminant is small at large \pT, achieving a background rejection of 1.18 (1.57) for the single-$b$-tagging (double-$b$-tagging) working point. The background rejection for other variables varies between 1.84 (Fox--Wolfram ratio $\FoxWolfram_2 / \FoxWolfram_0$
and exclusive dipolarity) and 1.24 ($\kt \Delta R$) for a signal efficiency of 80\%. Compared with the multijet background the correlations
between the jet mass and the jet substructure variables are smaller in the case of the top-quark background, especially for
$\pT > 1000$~\GeV. The Fox--Wolfram ratio $\FoxWolfram_4 / \FoxWolfram_1$ shows the lowest correlation: less than 1\% for most of the benchmarks.
 
In conclusion, the application of jet substructure variables improves the background rejection moderately, while better improvements are observed
for high transverse momenta. Furthermore, it is important to take into account the correlation between the \largeR\ jet mass and the substructure
variables since requirements on the substructure variables sculpt the jet mass distribution~\cite{JETM-2018-03, ATL-PHYS-PUB-2018-014}.

\section{Modelling tests in $g \rightarrow b\bar{b}$ data}
\label{sec:performance_in_gbb}
 
Multijet events enriched in $b$-jets, which predominately originate from gluon to \bb\ production, are used to evaluate the $b$-tagging
efficiency in data and simulation as well as the modelling of jet substructure variables.  The multijet background is one of the main
backgrounds for searches in fully hadronic final states, for example the Higgs boson pair search in the four-$b$-quark final state~\cite{EXOT-2015-11}.
This background also provides  a unique opportunity to validate the modelling of the double-$b$-jets in a large data sample.
Events with one \largeR\ jet with two ghost-associated track-jets (`$g \to b\bar{b}$ candidate jet') and one recoiling ISR \smallR\ jet (`recoil jet', $j_\text{recoil}$)
are used for this study.
 
\subsection{Event selection}
 
Events are required to have a primary vertex that has at least two tracks, each with $\pt > 500$~\MeV~\cite{ATL-PHYS-PUB-2015-026}. The primary
vertex with the highest $\pt^2$ sum of associated tracks is selected.
A single-\smallR-jet trigger with an online \ET\ threshold
of 380~\GeV\ was used to collect the data. An offline $R = 0.4$ recoil jet with \pt above 500~\GeV\ is matched to the jet which fired the trigger.
 
Non-collision backgrounds originating from calorimeter noise, beam-halo interactions, or cosmic rays can lead to spurious calorimeter signals.
This effect is suppressed by applying the criteria described in Ref.~\cite{DAPR-2012-01}.

Selected events are required to have at least one \largeR\ jet with $\pt>500$~\GeV\ and $|\eta|<2.0$, for which the \smallR\ jet trigger
is fully efficient and unbiased. The \largeR\ jet must have at least two ghost-associated $R=0.2$ track-jets. To enrich the event sample in
jets containing $b$-hadrons, it is required that at least one of the ghost-associated track-jets be matched to a muon. The highest-\pt\ track-jet
matched to a muon is called the muon-tagged jet, \muonjet . The matching is performed using a geometric $\Delta R < 0.2$ requirement
between the track-jet's axis and the muon.
The highest-\pt\ jet among the remaining track-jets matched by ghost association to the \largeR\ jet is called the non-muon jet, \nonmuonjet. The
highest-\pt\ \largeR\ jet satisfying these criteria is selected as gluon-jet candidate. Furthermore, the event must satisfy
$\Delta R(j_\text{recoil},\muonjet)>1.5$. This requirement ensures that the triggering jet and the gluon-jet candidate are well
separated.

\subsection{Flavour fraction corrections}
 
To reduce discrepancies between data and MC simulation in the flavour composition of the \largeR\ jet,
the flavour fractions of the sample are determined from the data before applying $b$-tagging. Each \largeR\ jet carries two flavours,
that of  \muonjet\ and \nonmuonjet, leaving nine possible flavour combinations for the \largeR\ jet (each track-jet can be a $b$-jet, $c$-jet, or
light-flavour jet; B, C and L abbreviations are used in the following). The long decay length of $b$- and $c$-hadrons makes the signed
impact parameter significance, \sdzero, of tracks
associated with a jet a good discriminating variable for different jet flavours. The \sdzero\ of a track is defined as:
\begin{linenomath}
\begin{align*}
\sdzero = \frac{d_0}{\sigma(d_0)}s_j,
\end{align*}
\end{linenomath}
where $d_0$ is the track's transverse impact parameter relative to the primary vertex, $\sigma(d_0)$ is the uncertainty
in the $d_0$ measurement, and $s_j$ is the sign of $d_0$ relative to the track-jet's axis, depending on whether the track
crosses the track-jet's axis in front of or behind the primary vertex. For a given track-jet, the average $\left< \sdzero \right>$ is built from
the three highest-\pt\ tracks associated with the track-jet. The tracks from $b$- and $c$-hadron decays
are expected to have higher \pt\ than tracks in light-flavour jets, because the heavy-flavour hadrons carry on average
a larger fraction of the jet energy. The requirement that $\left< \sdzero \right>$ is built from the three highest-\pT\ tracks helps
to distinguish them from light-flavour jets, which may have tracks with large
\sdzero\ values, e.g.\ from $\Lambda$ and $K_\text{s}$ decays.
 
The impact parameter resolution depends on the intrinsic track resolution, the traversed detector material, the detector alignment,
and other effects. To determine the impact parameter resolution in data, minimum-bias, dijet, and $Z$+jets events are used. The impact parameter
resolution is extracted in fine bins of track \pt\ and $\eta$ with an iterative method described in Ref.~\cite{PERF-2012-04}. The simulation
is corrected to match the measured impact parameter resolution as a function of track \pt\ and $\eta$ by using a Gaussian function to smear the
impact parameter resolution in the simulation.
 
\begin{figure}
\begin{center}
\includegraphics[width=0.49\textwidth]{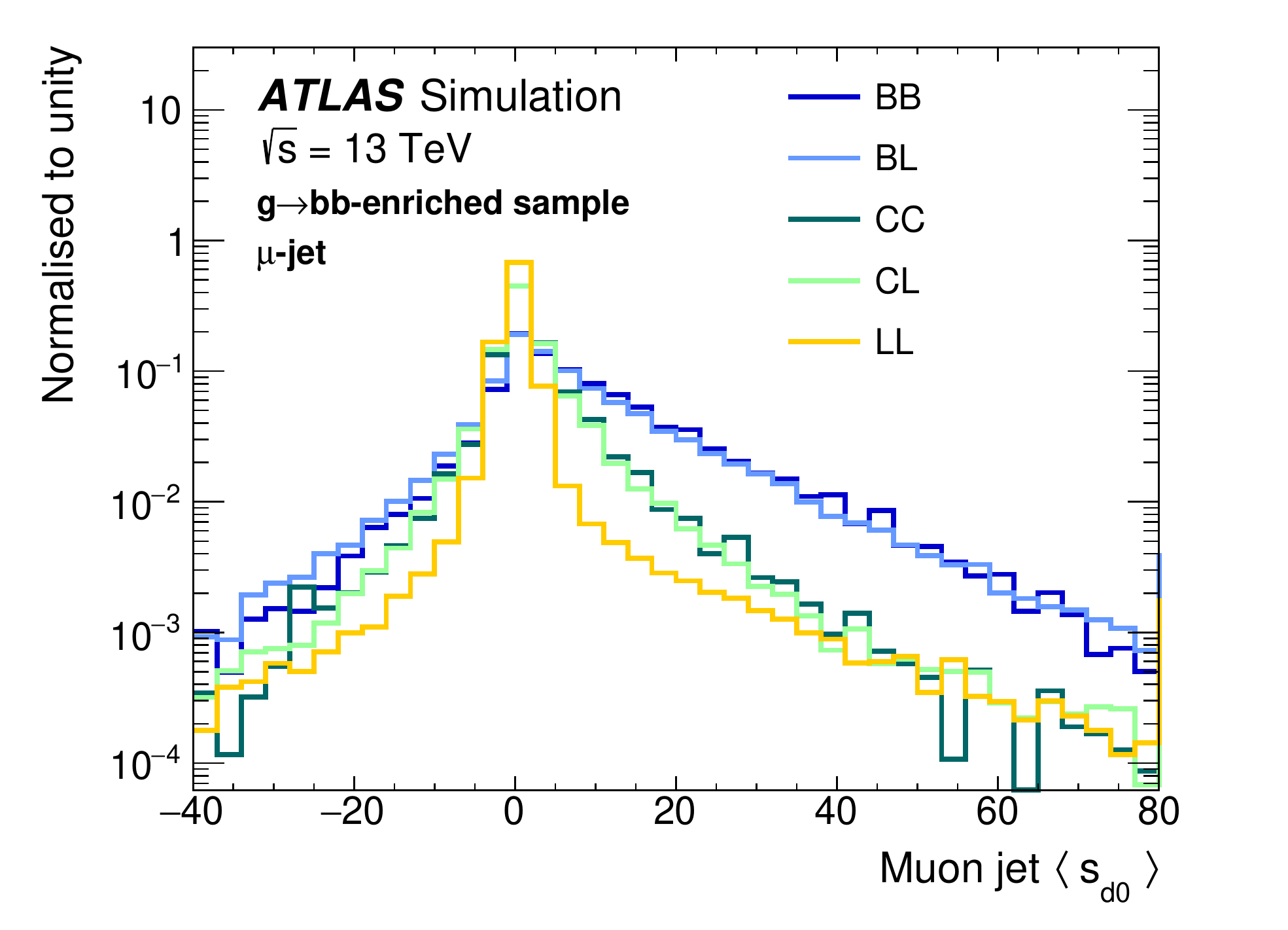}
\includegraphics[width=0.49\textwidth]{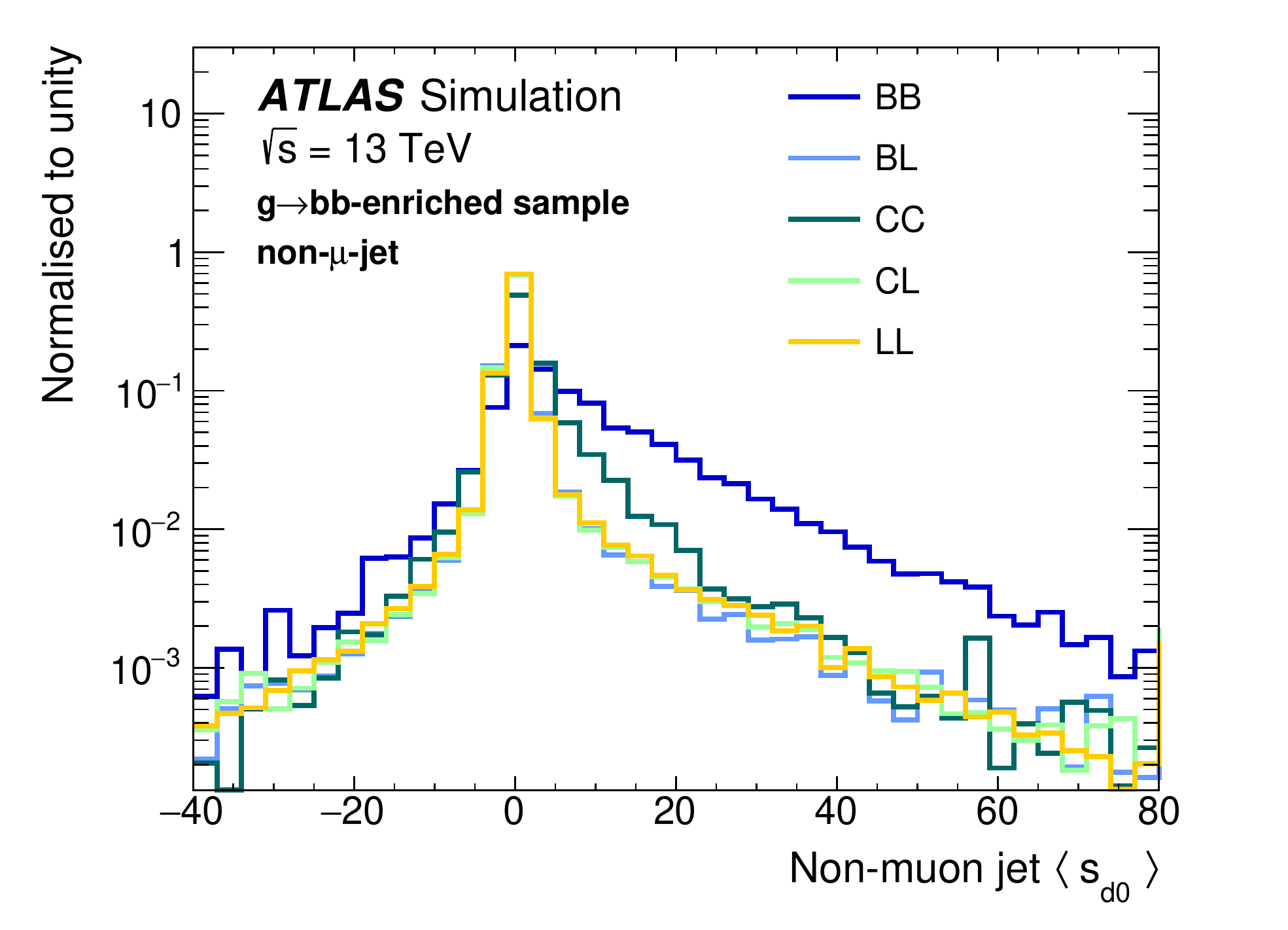}
\end{center}
\caption{Averaged impact parameter significance, $\left< \sdzero \right>$, distributions for the muon (left) and non-muon jets (right) inclusive in \muonjet\ and \nonmuonjet\ transverse momenta. The double flavour labels denote the true flavour of the jet pair, with the \muonjet\ given first.}
\label{fig:Sd0_template}
\end{figure}

The $\left< \sdzero \right>$ values of \muonjet\ and \nonmuonjet\ are found to be uncorrelated and thus the one-dimensional distributions of
each jet's $\left< \sdzero \right>$ are fit simultaneously.  Furthermore, the flavour combinations of (\muonjet,\nonmuonjet) = \{(B,C), (C,B), (L,C), (L,B)\}
are predicted to be less than 1\% of the total, so they are merged with other flavour categories which have the closest shape.
The shape similarity is determined using the $\chi^2$-statistic. Thus a total of five flavour categories are used,
(\muonjet,\nonmuonjet) = \{(B,B), (B,L), (C,C), (C,L), (L,L)\}. Figure~\ref{fig:Sd0_template} shows the templates inclusive in \pt .
 
\begin{figure}
\begin{center}
\includegraphics[width=0.49\textwidth]{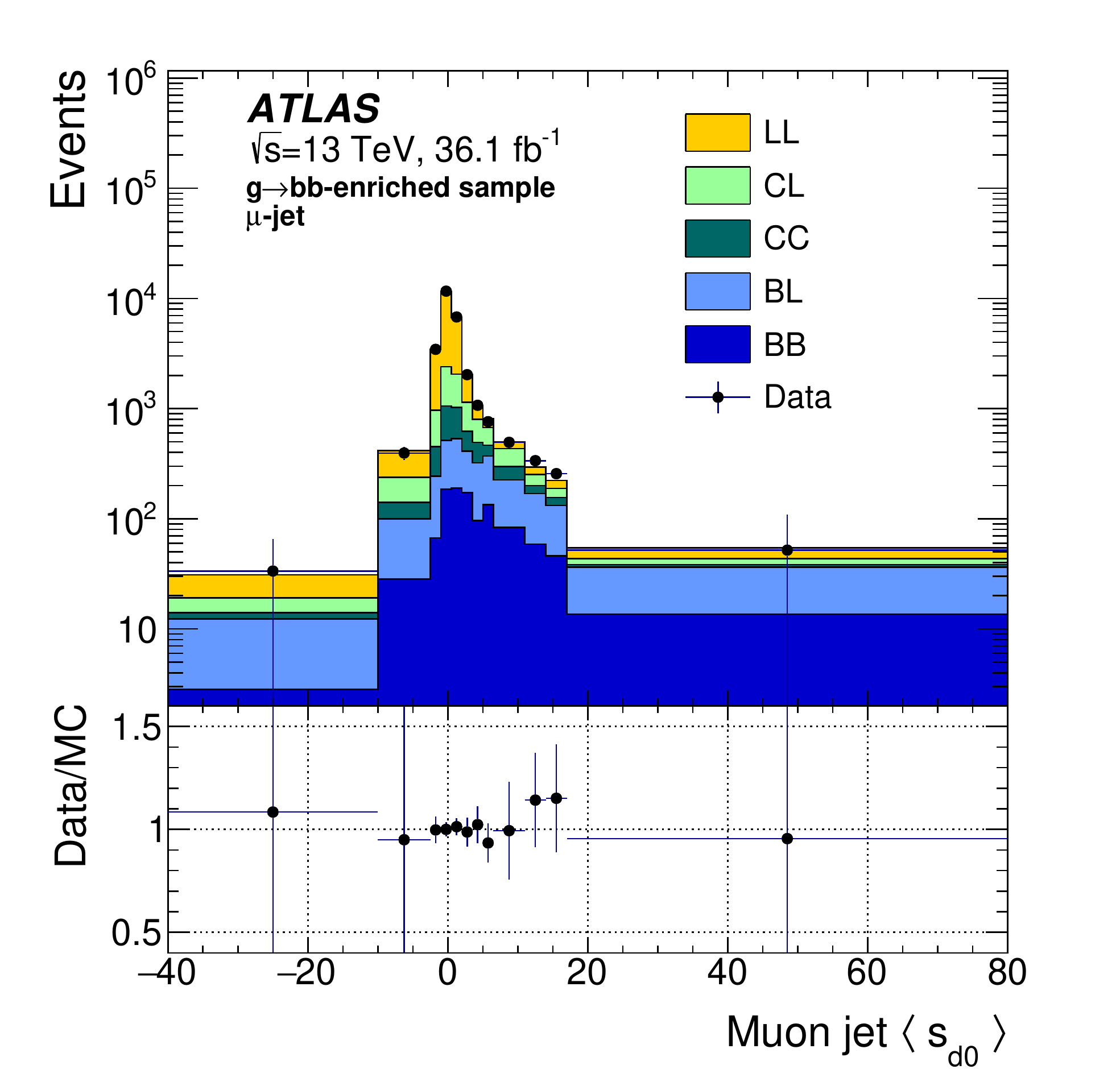}
\includegraphics[width=0.49\textwidth]{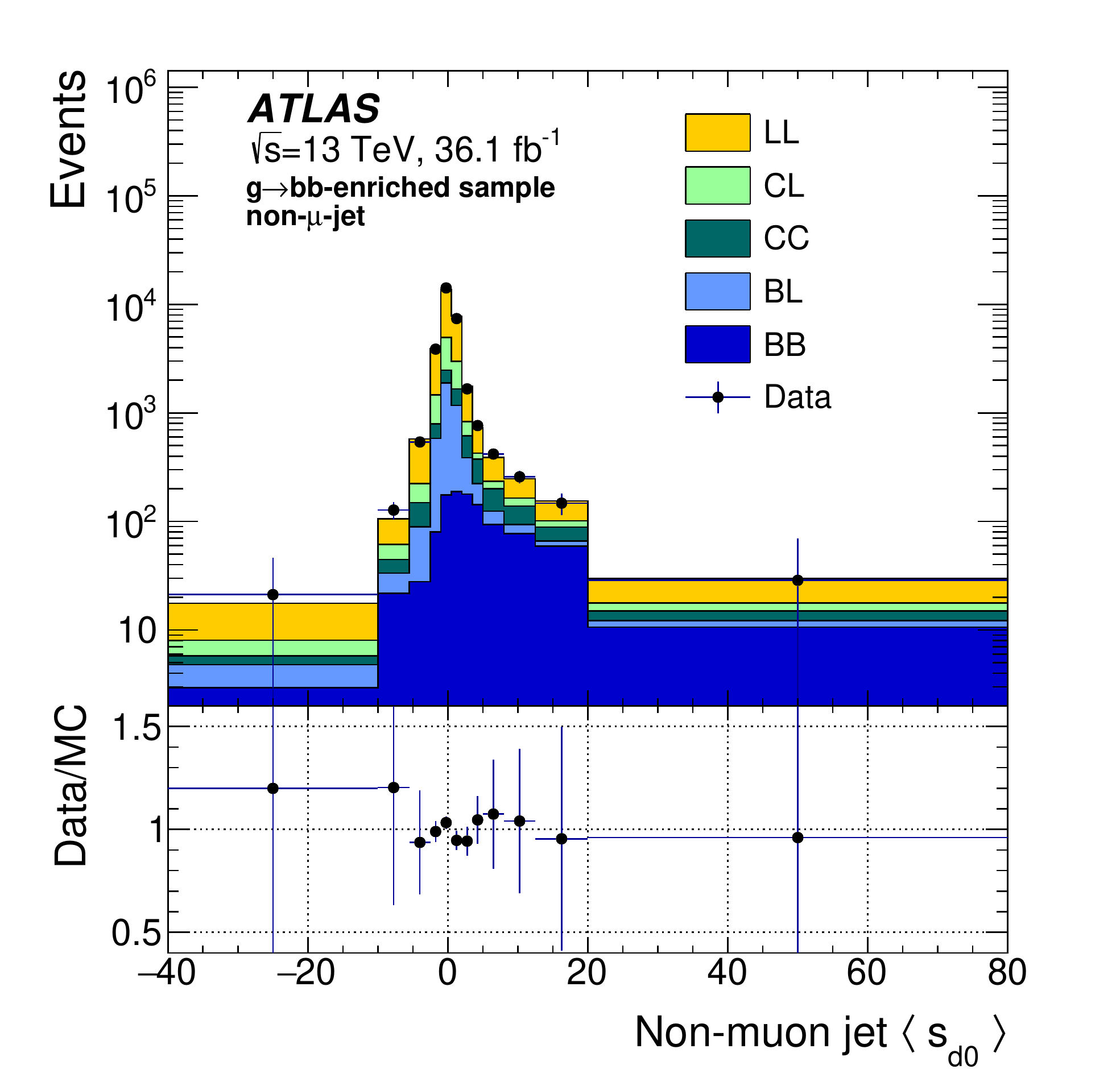}
\end{center}
\caption{Averaged impact parameter significance, $\left< \sdzero \right>$, distributions of the muon (left) and non-muon jet (right) in the (100--200)~\GeV\
bin of the \muonjet\ and \nonmuonjet\ transverse momenta.}
\label{fig:Sd0_fit}
\end{figure}

Since the flavour fractions vary with \pt, the flavour fraction fits to the data are performed in bins of \pt\ of the two track-jets. For each jet \pT\ bin,
individual MC templates are used. The following jet-\pt bins are considered: \muonjet\ \pt bins = \{(0--100), (100--200), >200\}~\GeV\ and
\nonmuonjet\ \pt bins = \{(0--100), (100--200), (200--300), >300\}~\GeV. Figure~\ref{fig:Sd0_fit} shows an example of the flavour fraction fit to the \sdzero\ distributions of \muonjet\ and \nonmuonjet\ for one particular bin of the track-jet transverse momenta. The fit
uncertainty includes the statistical uncertainty of the templates and is evaluated using toy MC simulations. The flavour fraction corrections relative
to the simulated fractions vary between 0.7 and 1.7 in the jet \pt\ bins with a statistical uncertainty below 10\%.
 
After correcting for the observed flavour-pair fractions the level of agreement between data and MC simulation is evaluated in the selected event sample
before and after $b$-tagging is applied to the track-jets. The $70\%$ double-$b$-tagging working point is used.

\subsection{$b$-tagging results}
 
Since the flavour fractions are corrected in the MC simulation, differences between the data and predictions after the $b$-tagging can be
attributed to a difference between data and MC simulation in the dependence of the $b$-tagging performance on the \largeR\ jet topology,
in particular on the topology with two closely spaced track-$b$-jets.
 
\begin{figure}
\begin{center}
\includegraphics[width=0.44\textwidth]{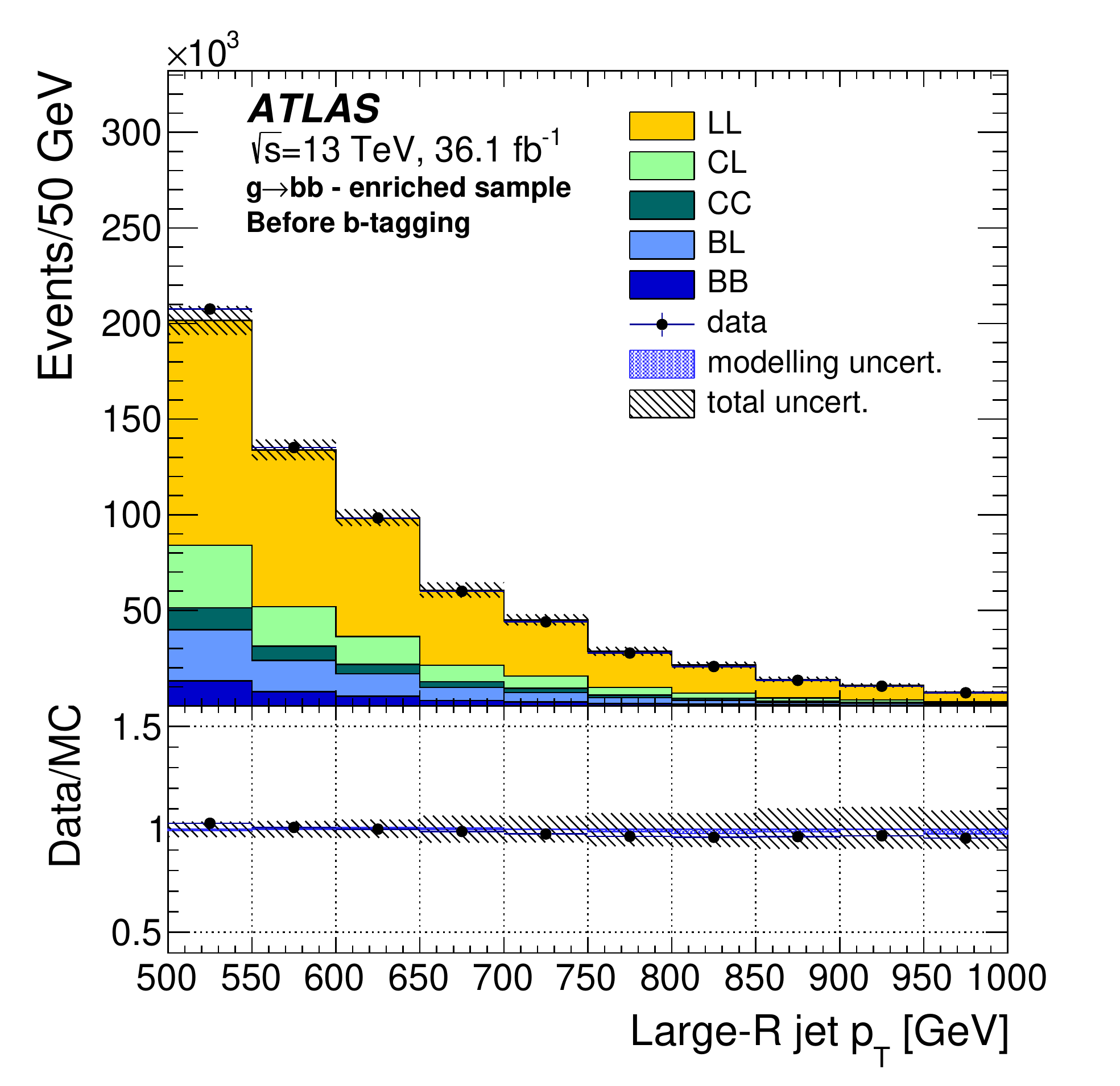}
\includegraphics[width=0.44\textwidth]{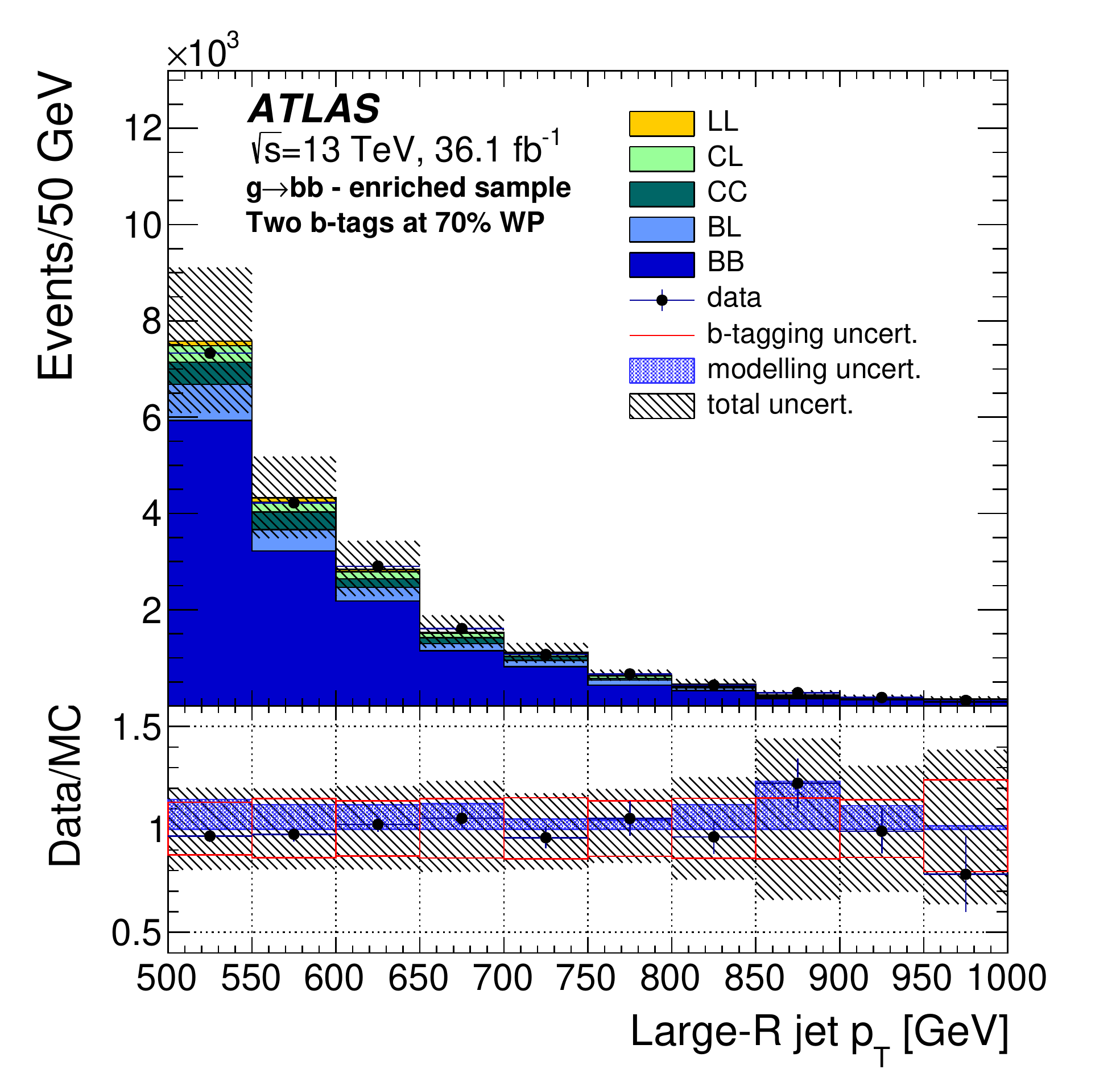}\\
\includegraphics[width=0.44\textwidth]{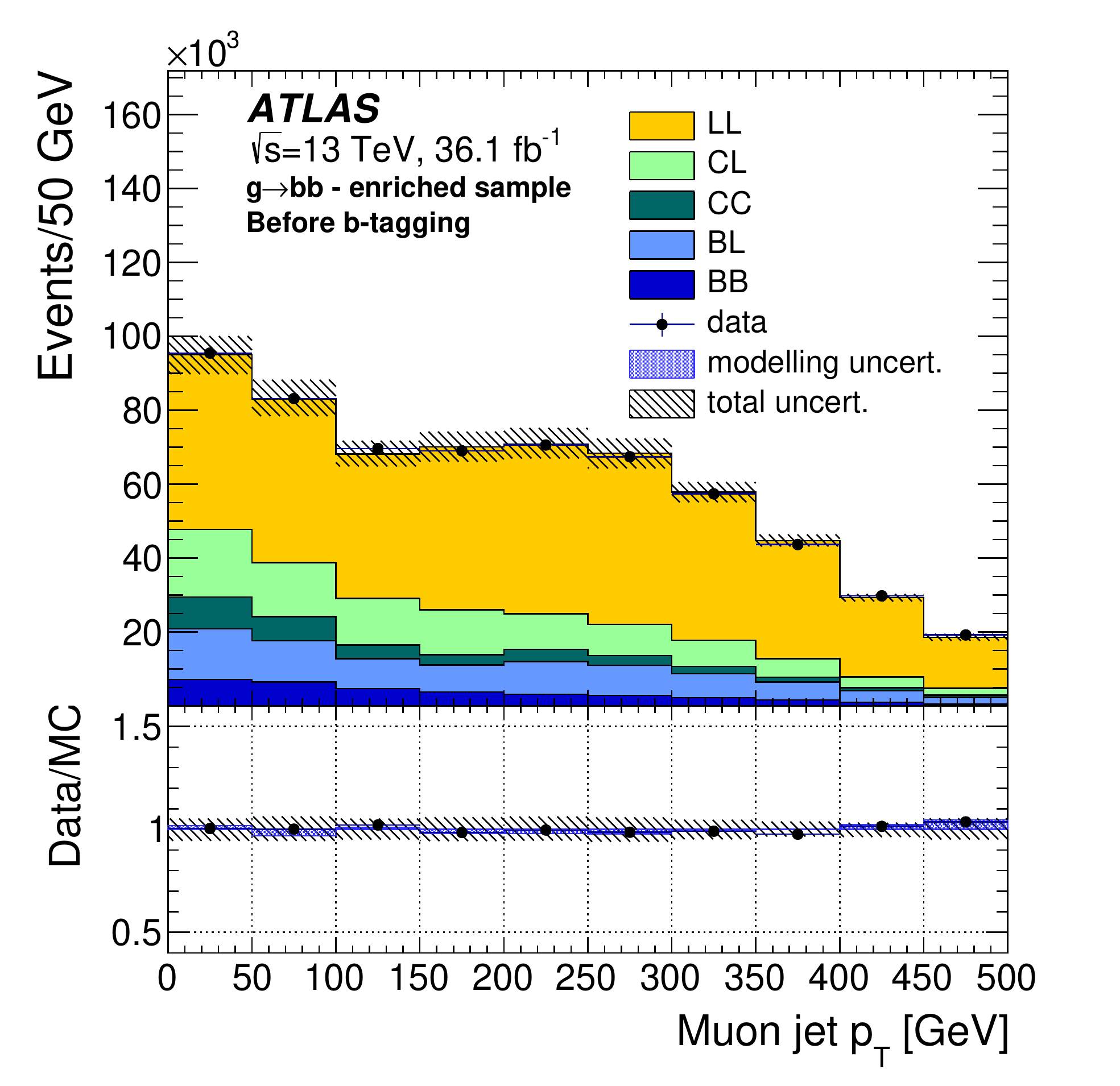}
\includegraphics[width=0.44\textwidth]{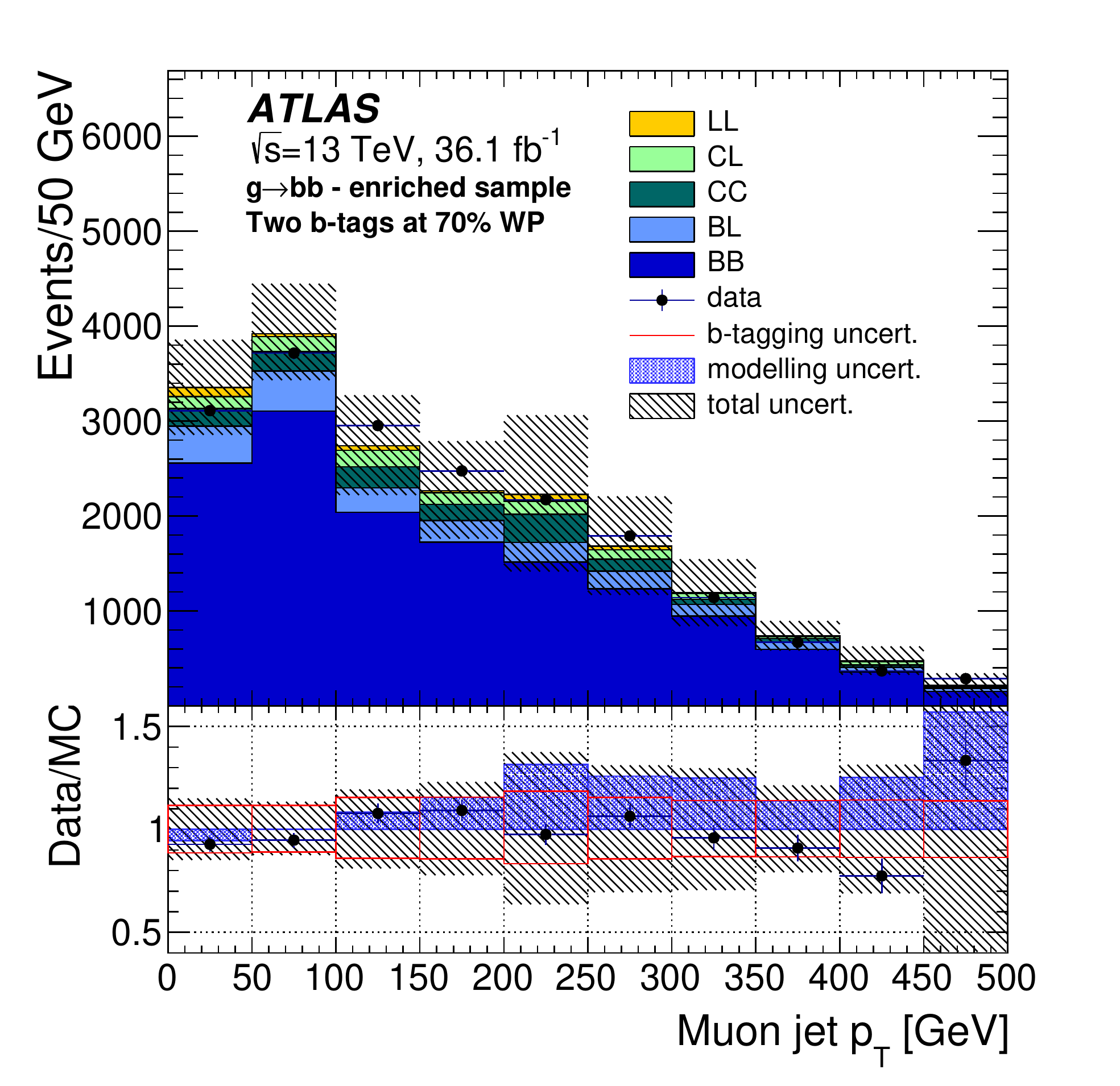} \\
\includegraphics[width=0.44\textwidth]{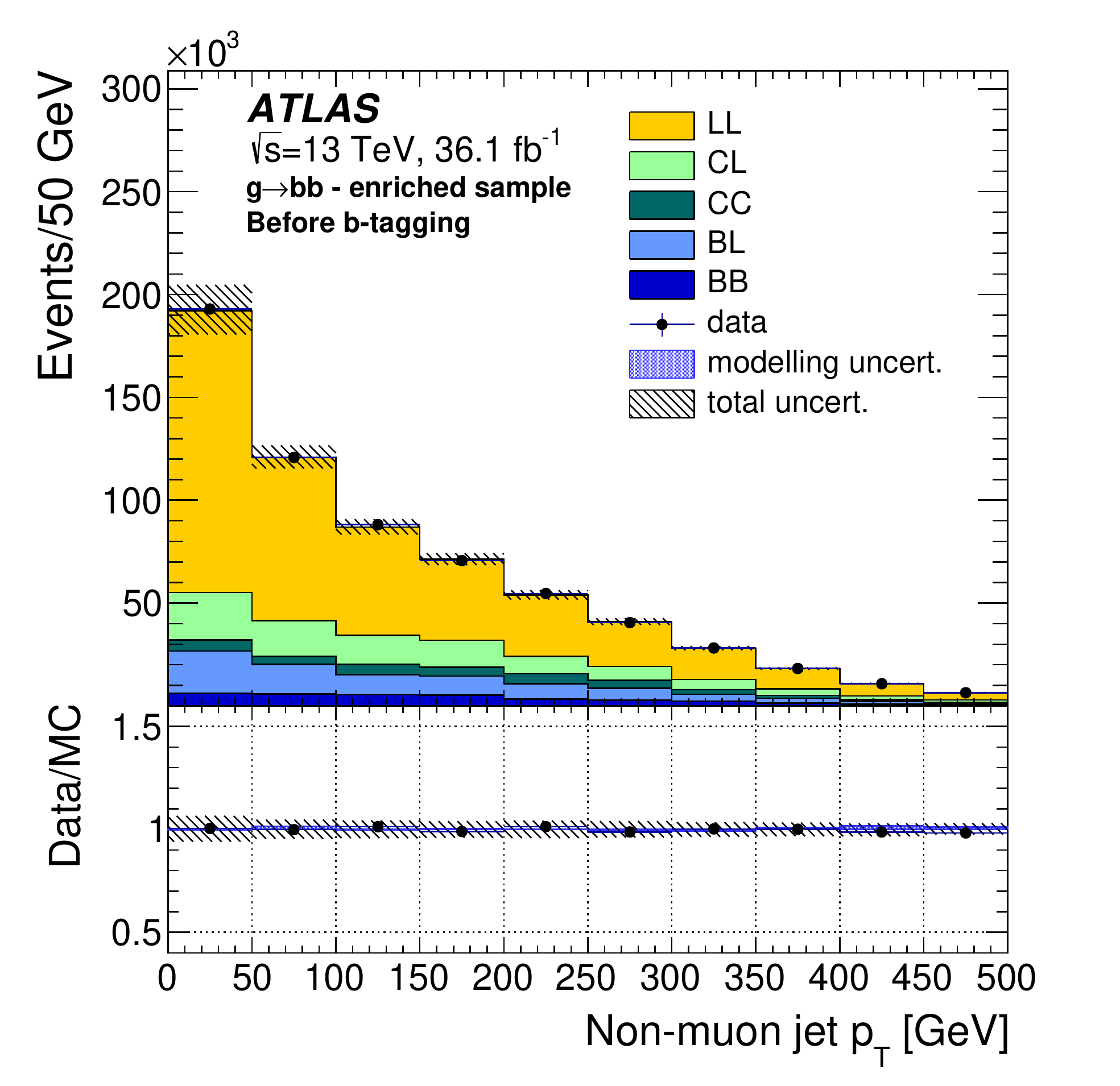}
\includegraphics[width=0.44\textwidth]{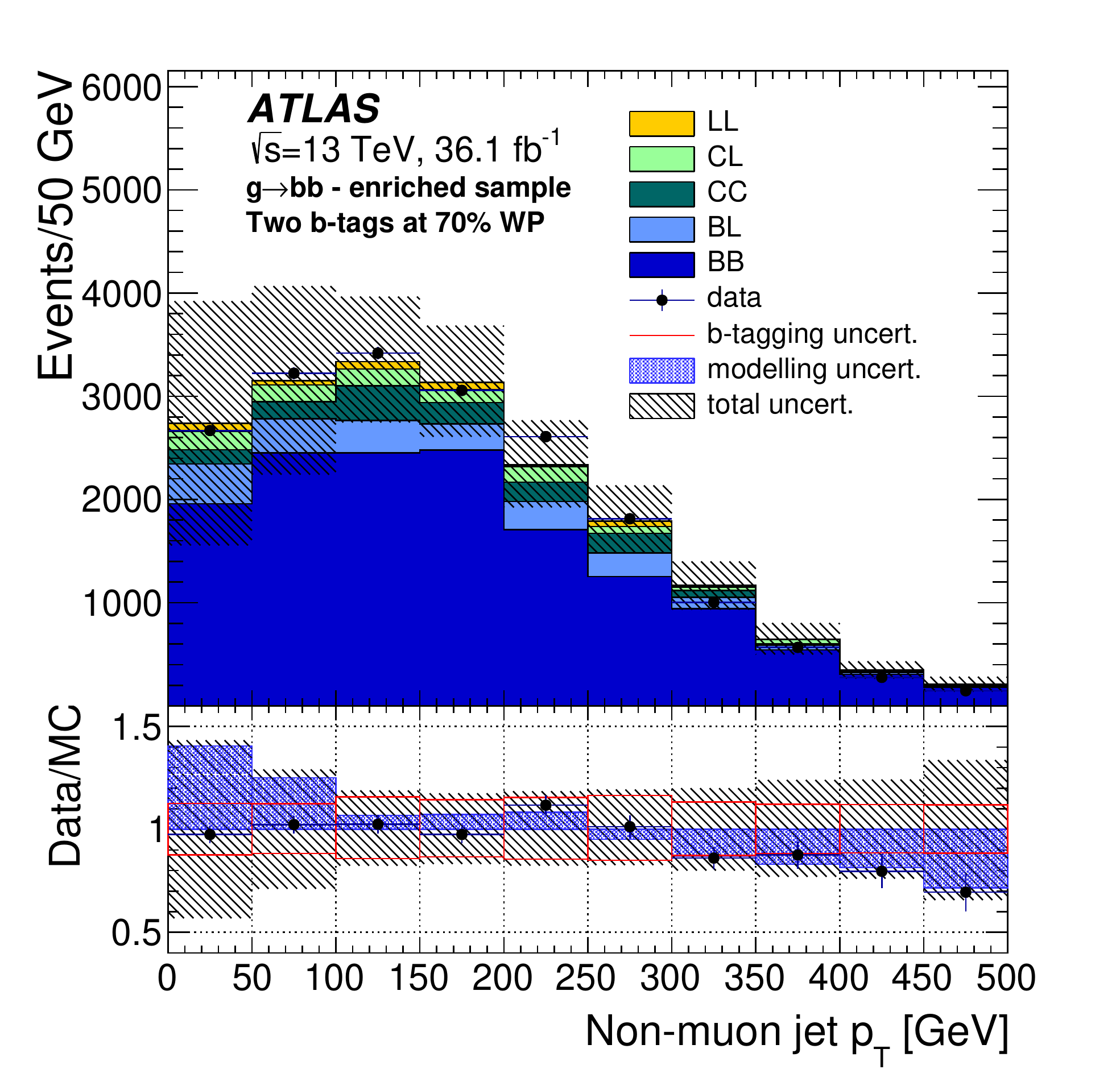} \\
\end{center}
\caption{Transverse momentum distributions of the \largeR\ jet (top), \muonjet\ (middle) and \nonmuonjet\ (bottom) before (left) and
after (right) double $b$-tagging. The flavour-tagging correction factors and the flavour-fit corrections have been applied. The two largest systematic
uncertainties, generator modelling and the $b$-tagging-related uncertainties, are shown as well. The total uncertainty includes all
systematic uncertainties listed in Section~\ref{sec:systematics} and the fit uncertainty summed in quadrature. }
\label{fig:jmupt}
\end{figure}

\begin{figure}
\begin{center}
\includegraphics[width=0.60\textwidth]{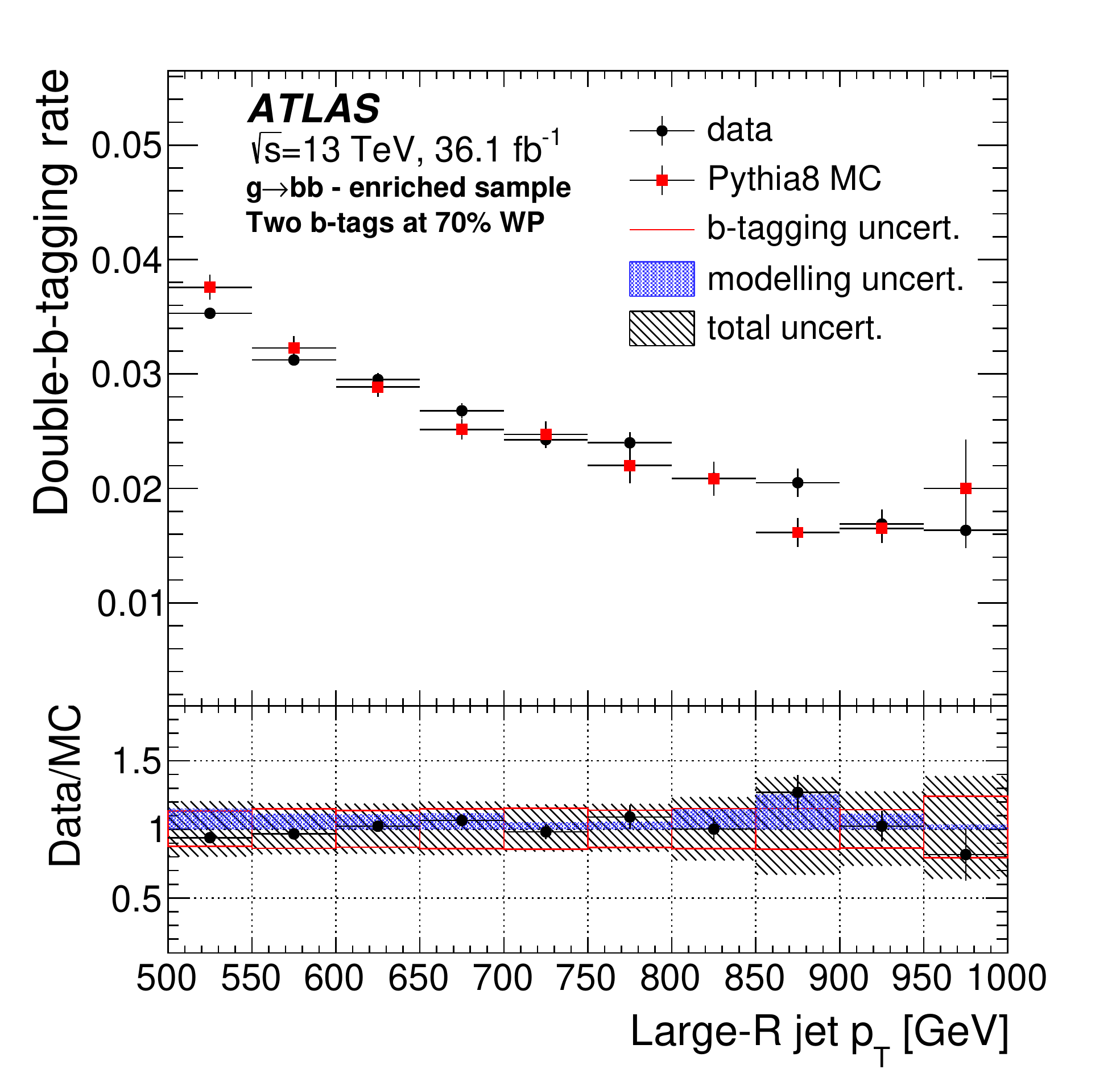}
\end{center}
\caption{Comparison of data and MC simulation double-$b$-tagging rates as a function of the \largeR\ jet \pt. The flavour-tagging correction
factors and the flavour-fit corrections have been applied. The two largest systematic
uncertainties, generator modelling and the $b$-tagging-related uncertainties, are shown as well. The total uncertainty includes all
systematic uncertainties listed in Section~\ref{sec:systematics} and the fit uncertainty summed in quadrature. The size of the flavour-fit uncertainty is below 1\%.}
\label{fig:btageff}
\end{figure}

Figure~\ref{fig:jmupt} shows the flavour-fit-corrected \pt\ spectrum of the \largeR\ jet as well the \muonjet\ and \nonmuonjet\ before and
after $b$-tagging. As seen in the ratio plots, there is good agreement within uncertainties between data and MC simulation. The shape
differences between data and MC simulations especially for the $j^\text{trk}_{\text{non-}\mu}$ transverse momentum can be partially
explained by the difference observed between \PythiaEight and \Herwigpp MC simulations. The double-$b$-tagging rate is defined as the
number of selected \largeR\ jets with at least two track-jets, two of which are $b$-tagged, divided by the number of all selected \largeR\
jets with at least two track-jets. Figure~\ref{fig:btageff} shows the double-$b$-tagging rate as a function of the \largeR\ jet \pt.
Data and MC simulation agree within the uncertainties. The performance of the double $b$-tagging applied to two track-jets seems not to depend
on the \largeR\ jet topology with two closely spaced track-$b$-jets, and the default $b$-tagging calibration described in Section~\ref{sec:systematics} can
be applied for this analysis.
 
\subsection{Jet substructure results}
 
\begin{figure}
\begin{center}
\includegraphics[width=0.44\textwidth]{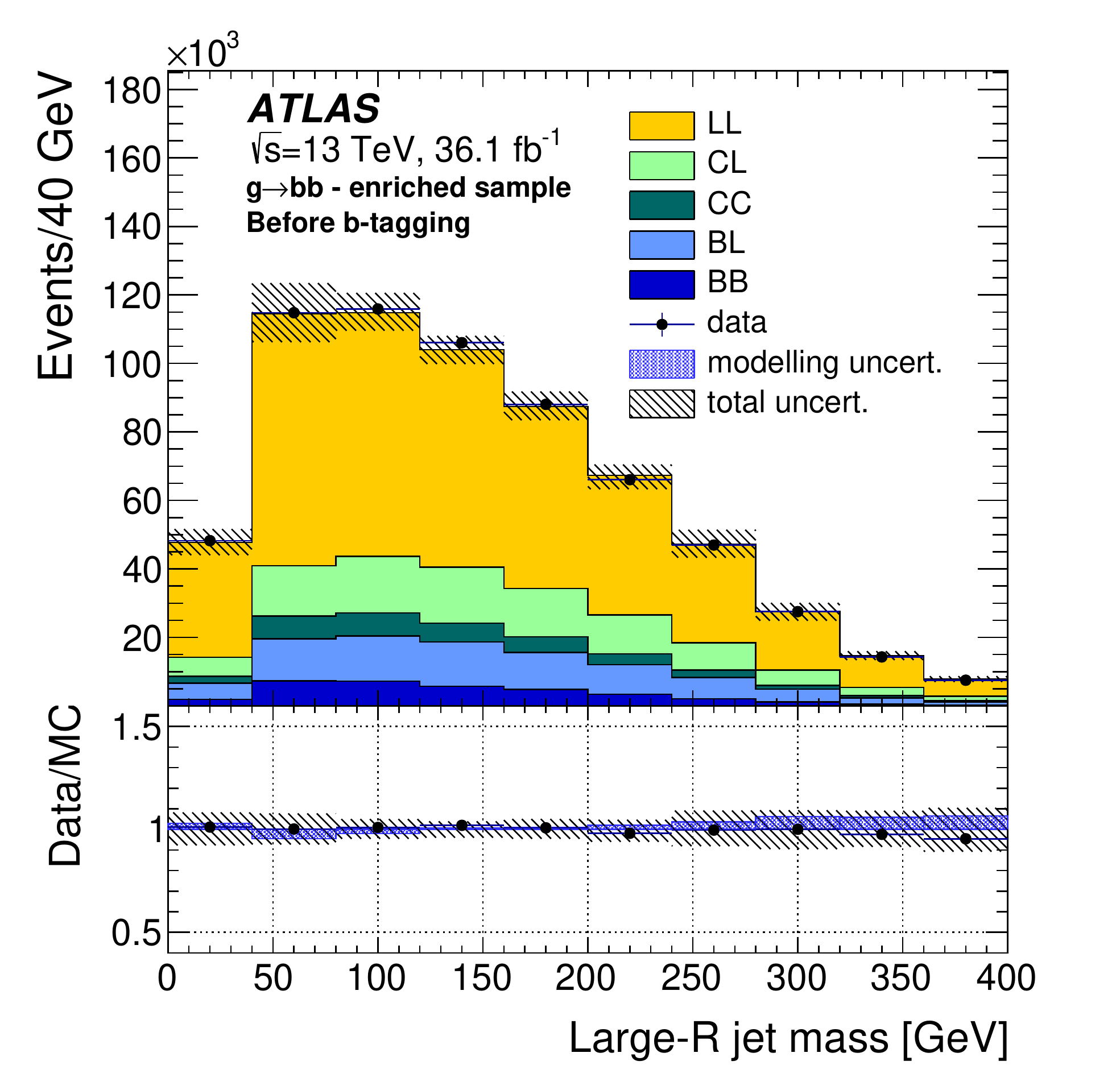}
\includegraphics[width=0.44\textwidth]{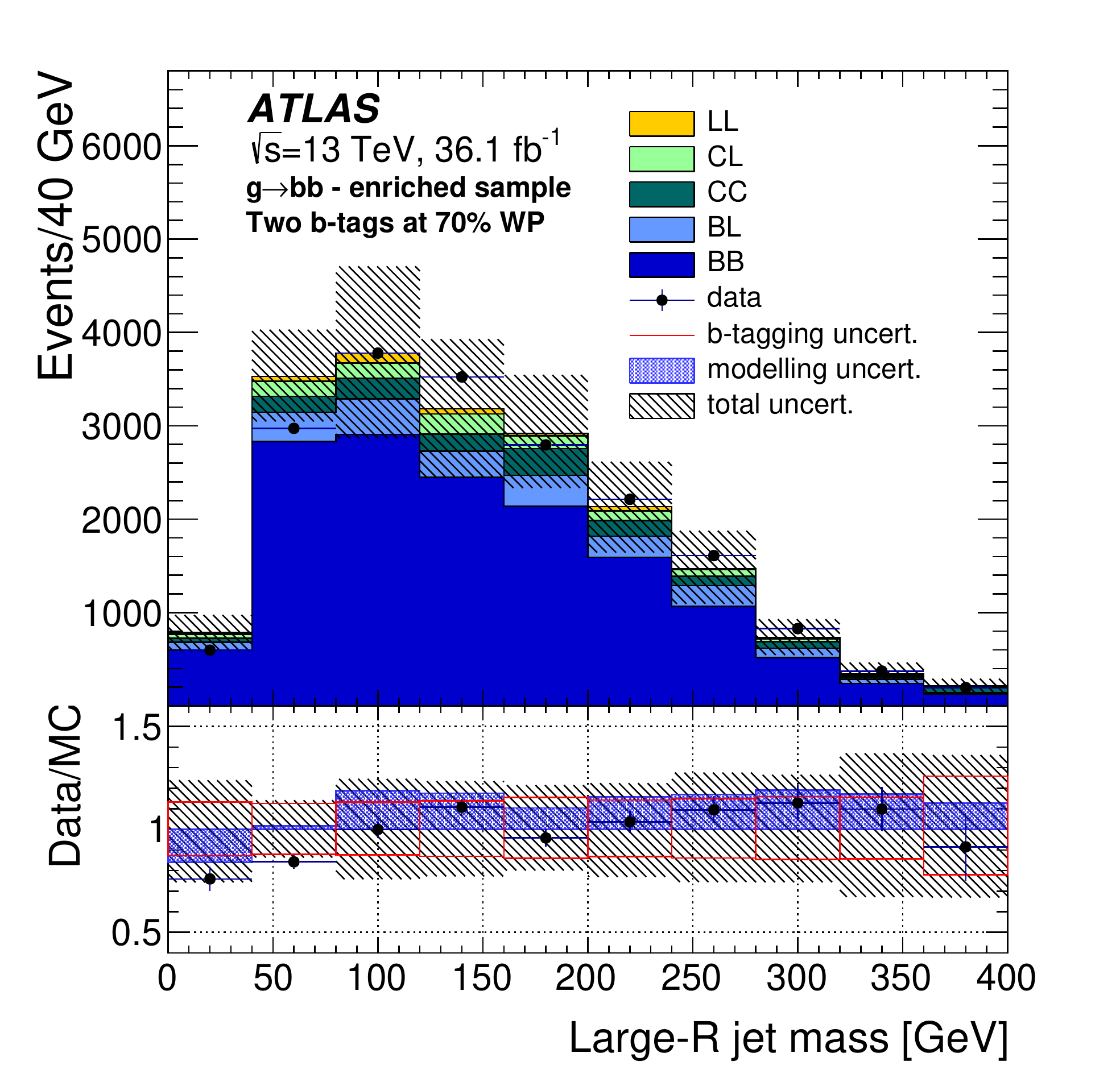}\\
\includegraphics[width=0.44\textwidth]{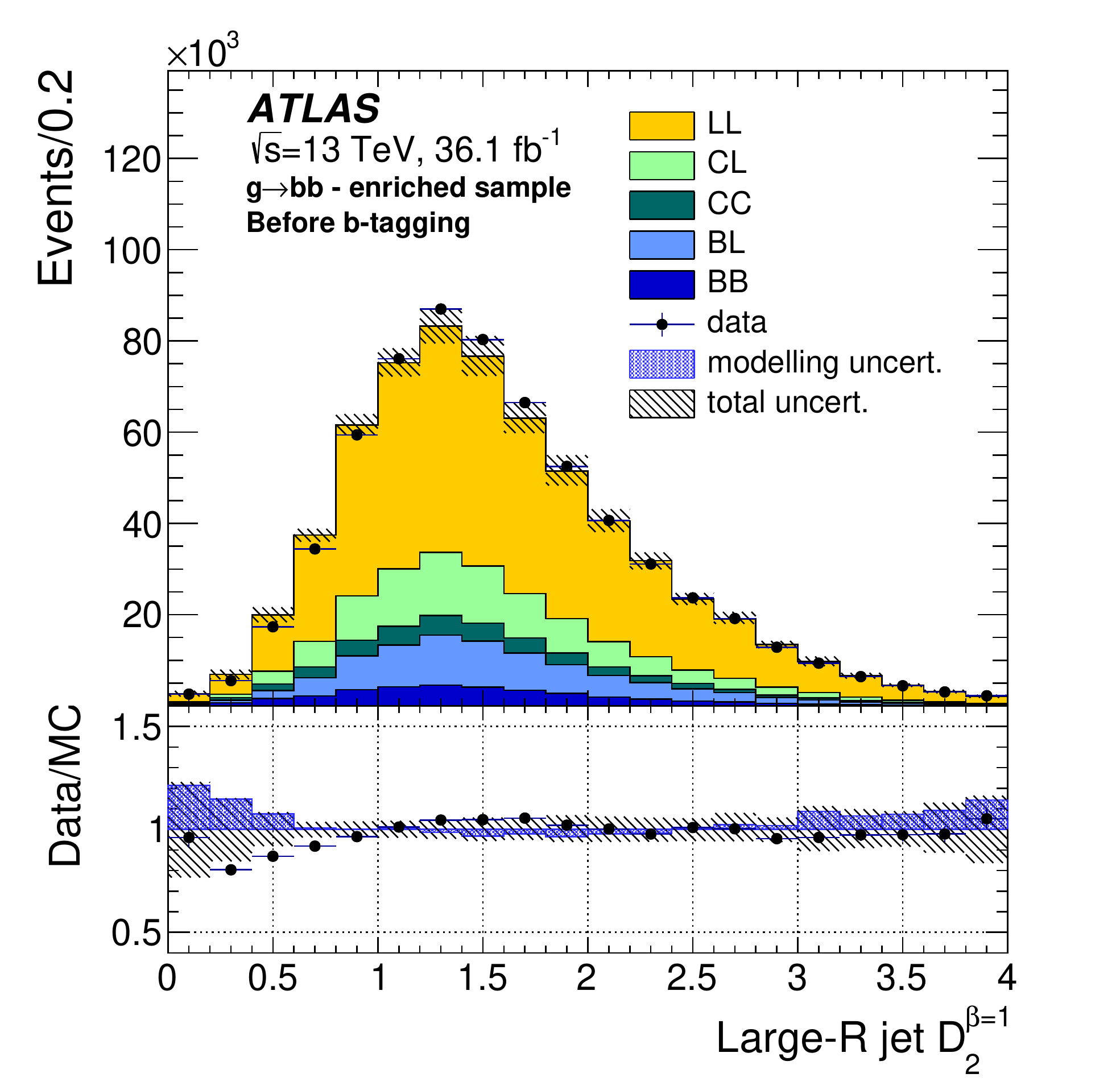}
\includegraphics[width=0.44\textwidth]{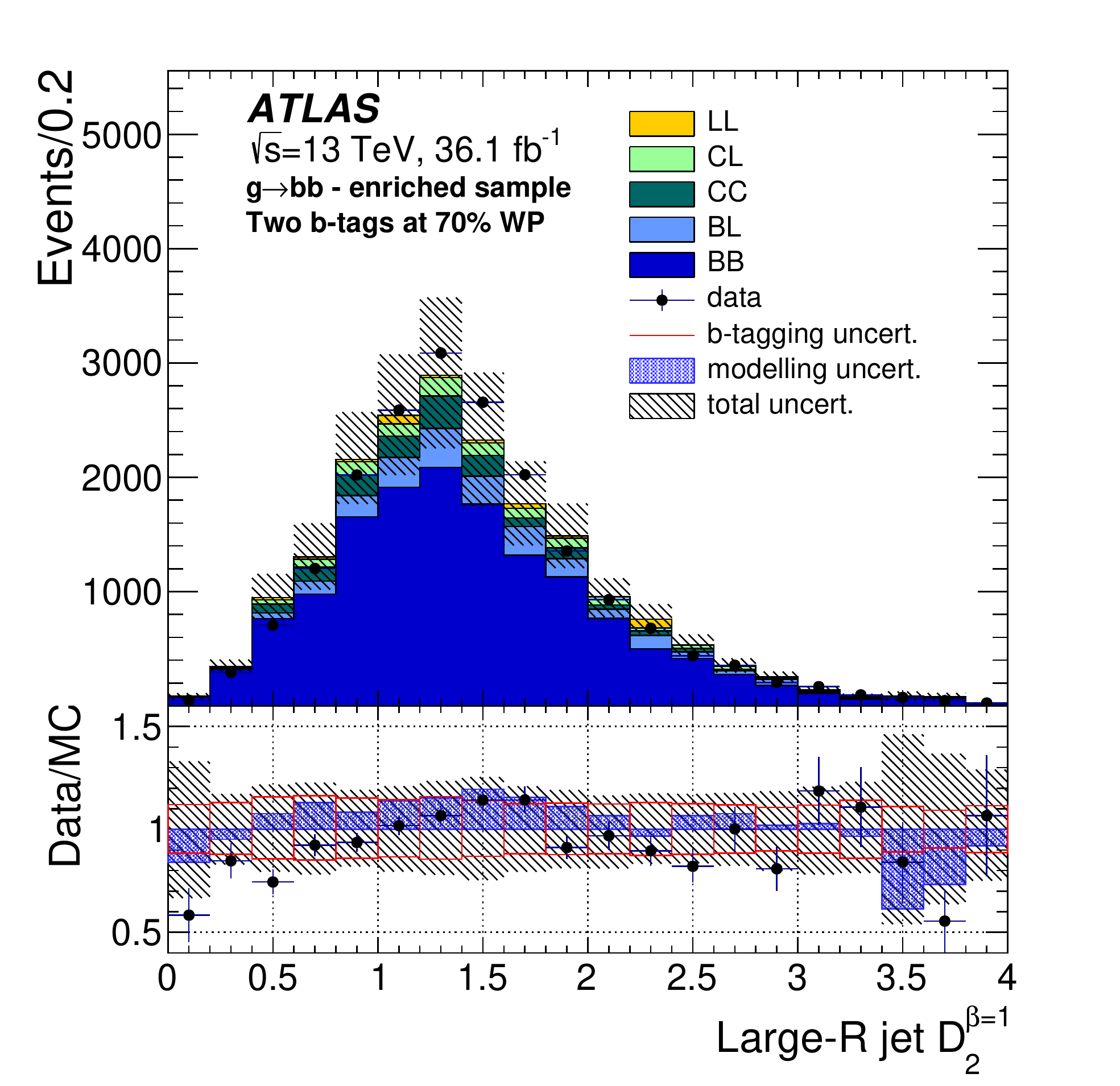} \\
\includegraphics[width=0.44\textwidth]{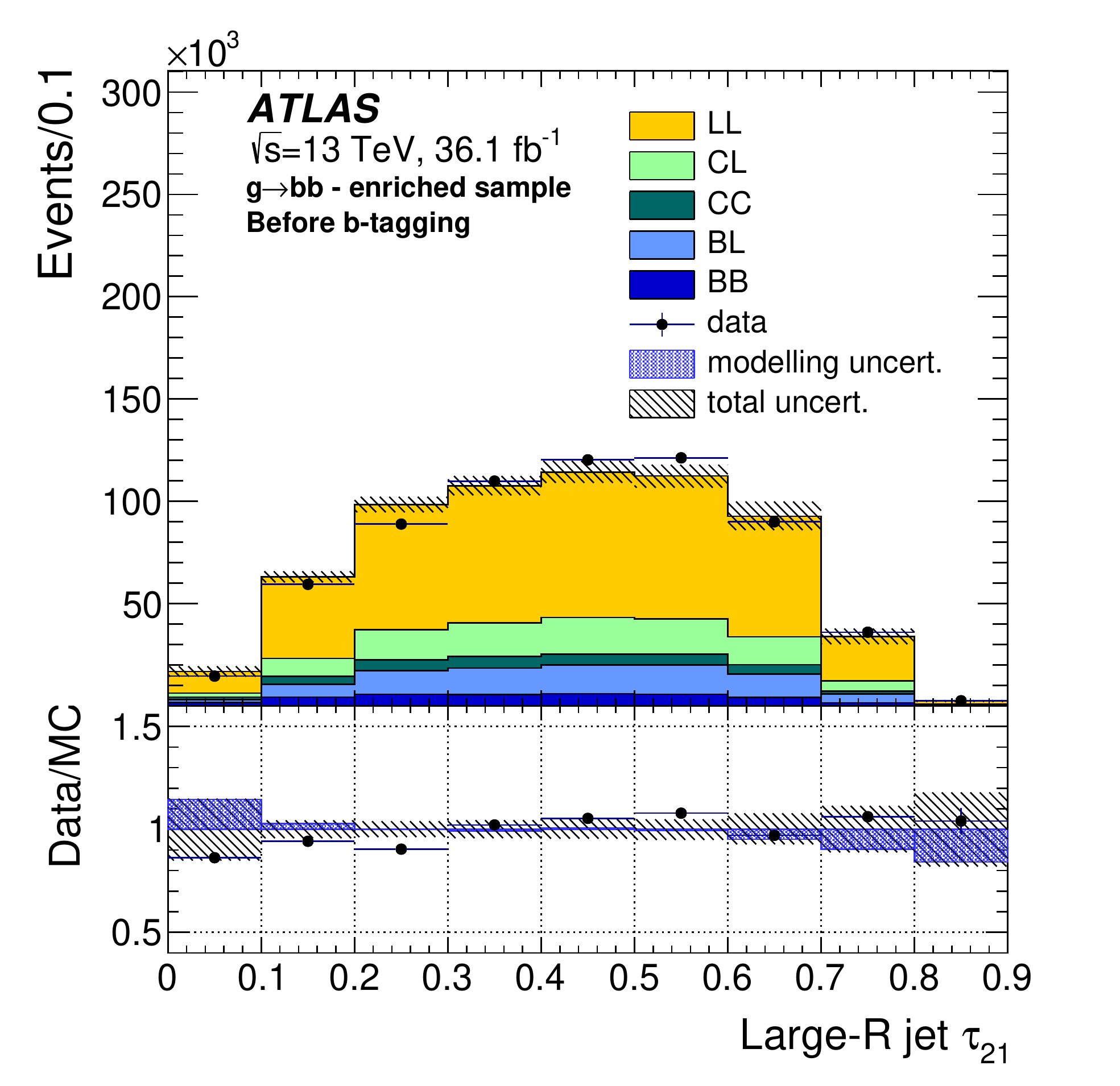}
\includegraphics[width=0.44\textwidth]{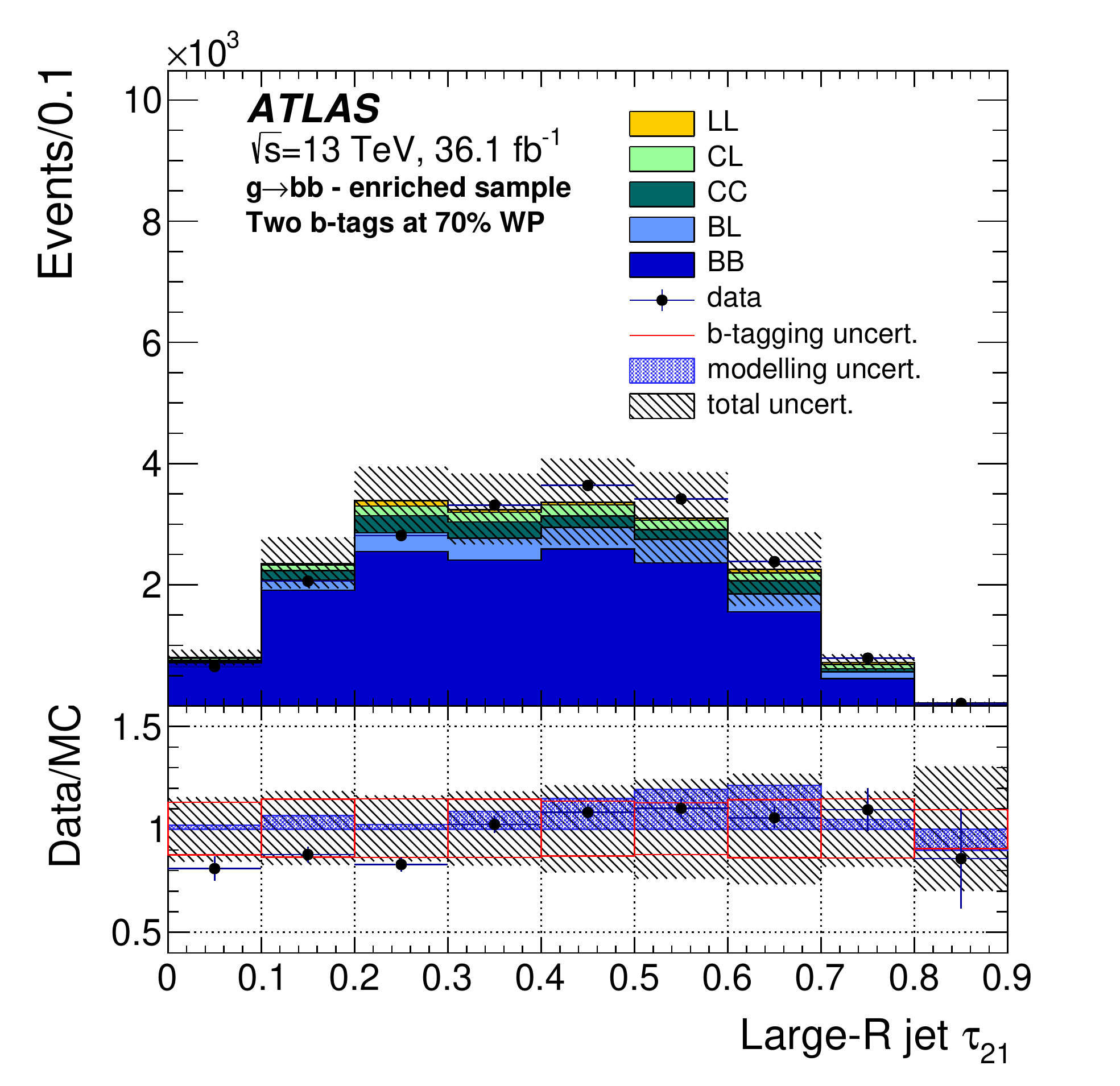} \\
\end{center}
\caption{Distributions of \largeR\ jet mass (top), \DTwoBetaOne\ (middle) and $\tau_{21}$ (bottom) before (left) and after (right) double $b$-tagging.
The flavour-tagging correction factors and the flavour-fit corrections have been applied. The two largest systematic
uncertainties, generator modelling and the $b$-tagging-related uncertainties, are shown as well.
The total uncertainty includes all systematic uncertainties listed in Section~\ref{sec:systematics} and the fit uncertainty summed in quadrature.
The \DTwoBetaOne\ and $\tau_{21}$ uncertainty bands include additional substructure variable uncertainties~\cite{JETM-2018-02}. }
\label{fig:gbb_substr}
\end{figure}

As possible variations of Higgs taggers may make use of the \largeR\ jet \pt, and substructure variables such as mass,
$n$-subjettiness, or \DTwoBetaOne, it is important to ensure that these variables are well modelled by MC simulations. The distributions
of kinematic and substructure variables are shown in Figure~\ref{fig:gbb_substr}, for double-$b$-tagged jets after the flavour-fit correction.
As seen in the ratio plots, there is acceptable agreement within uncertainties between data and MC simulations.
 
The relative impact of the systematic uncertainties on the yields of signal and background are presented in Table~\ref{tab:SysTable_yields_gbb}.
The dominant signal uncertainty is the modelling uncertainty followed by the $b$-tagging-related uncertainties. The $b$-tagging-related
uncertainties (misidentification of light-flavour jets and $c$-jets as $b$-jets) are dominant for background. The dominant uncertainties are shown separately in
Figure~\ref{fig:gbb_substr}. The difference in the shapes between data and MC simulations can be partially explained by the difference observed
between \PythiaEight and \Herwigpp MC simulations.

\begin{table}[h]
\caption{Relative impact of the systematic uncertainties on the yields of signal and the main background for the $g\rightarrow b\bar{b}$ analysis. Multiple independent components have been combined into groups of systematic uncertainties. `Jet scales' refers to the sum in quadrature of the jet energy, mass and substructure scale uncertainties.}
\medskip
\centering
\begin{tabular}{lS[table-format=3.2]S[table-format=3.2]}
\hline
Source                                                  & \text{Signal (B,B) [\%]} & \text{Background (non-B,non-B) [\%]} \\
\hline
Jet scales & 9.0 &  7.3\\
Jet energy resolution & 1.0  & 1.8\\
Jet mass resolution & 0.1 & 0.2 \\
JVT &  0.01 & 0.01\\
$b$-tagging related & 9.5 & 27\\
Modelling & 23  & 13\\
Fit statistics & 1.0 & 0.1\\
\hline
\end{tabular}
\label{tab:SysTable_yields_gbb}
\end{table}


\section{Modelling tests in $Z \rightarrow b\bar{b}$ data}
\label{sec:performance_in_Zbb}
 
As mentioned in the introduction, the $Z \to b \bar{b}$ process is a colour-singlet resonance with a mass close to the Higgs boson mass, so
kinematic properties of the $Z\rightarrow b\bar{b}$ and $H\rightarrow b\bar{b}$ events are expected to be similar. Events with one
double-$b$-tagged \largeR\ jet (`$Z \to b\bar{b}$ candidate jet') and a photon that are back-to-back are used for this study. The photon requirement
improves the signal-to-background ratio in comparison with the fully hadronic final state.
 
\subsection{Event selection}
 
Events are selected using a single-photon trigger with a transverse energy ($E_{\textrm T}$) threshold of 140~\GeV\ and  loose photon identification
requirements~\cite{TRIG-2016-01}. This trigger is non-prescaled for the entire data-taking period and is fully efficient for offline photons with
$E_{\textrm T}>175$~\GeV. The same primary vertex and jet-cleaning requirements are applied as for the $g \to b\bar{b}$ study, described in
Section~\ref{sec:performance_in_gbb}.
 
Exactly one photon and at least one \largeR\ jet are required to be present in the event. The \largeR\ jet is required to have $\pt > 200$~\GeV, $|\eta| < 2.0$,
and mass greater than 30~\GeV. A jet--photon overlap removal procedure is applied, removing photons within $\Delta R=1.0$ of the \largeR\ jet. The \largeR\ jet with
the highest \pt\ is chosen as the \Zbb candidate. The two highest-\pT\ track-jets that are associated with the \Zbb candidate are required to be identified as
$b$-jets using the 70\% working point.
 
\subsection{Background estimate}
 
The dominant SM background in this analysis is $\gamma$+jets with gluon-to-$b\bar{b}$ splitting. The contribution from the Standard Model
$t\bar{t}\gamma$ and $W\gamma$ processes is smaller than that from the \Gammajets\ process. Other background contributions such as jets faking photons,
electrons faking photons, and $t\bar{t}$ are found to be negligible. To extract the \Zbb and $\gamma$+jets normalisations, the \Zbb candidate jet mass
distribution is fitted to data. Both templates are taken from the MC simulation as described in Section~\ref{sec:samples}. The $t\bar{t}+\gamma$ and
$W(q\bar{q})+\gamma$ background contributions estimated from MC simulation are subtracted before the fit to data. The jet mass variable is used in the fit because
the difference between the shapes of the \Zbb and $\gamma$+jets templates is larger than for other substructure variables. The extracted normalisations
are applied to all other distributions.
 
\subsection{Jet substructure results}
 
\begin{figure}
\centering
\includegraphics[width=0.49\textwidth]{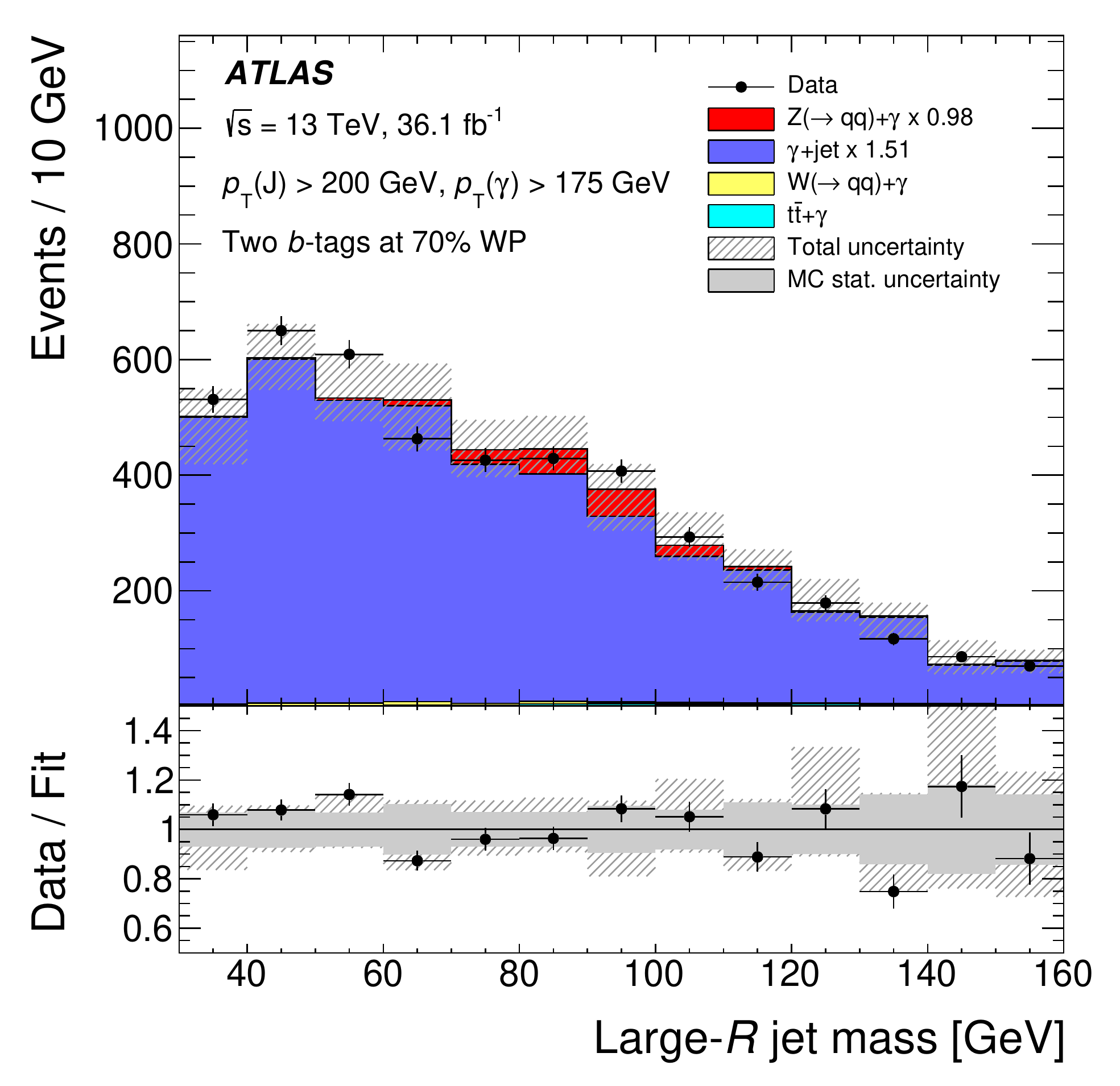}
\includegraphics[width=0.49\textwidth]{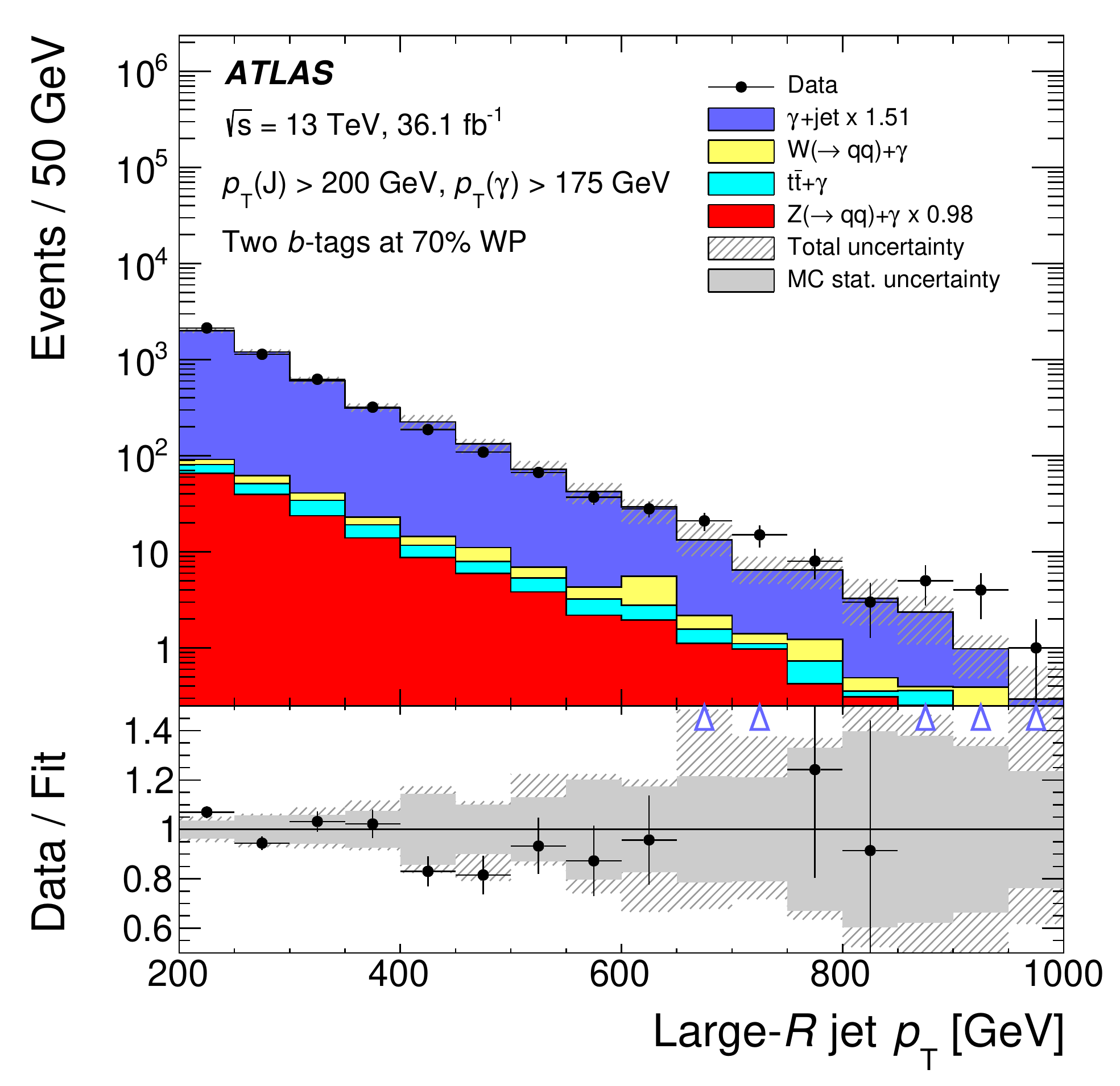} \\
\includegraphics[width=0.49\textwidth]{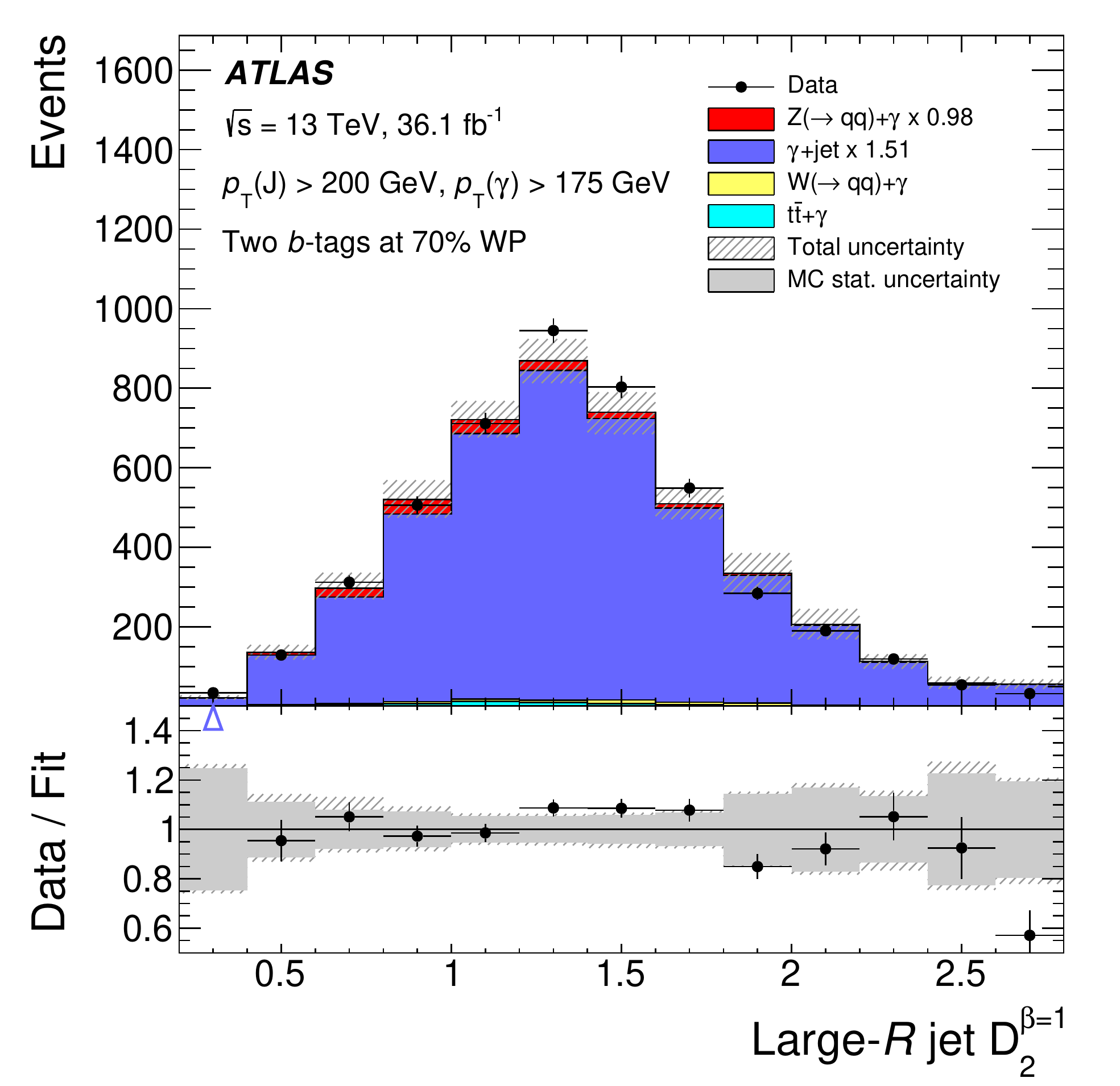}
\includegraphics[width=0.49\textwidth]{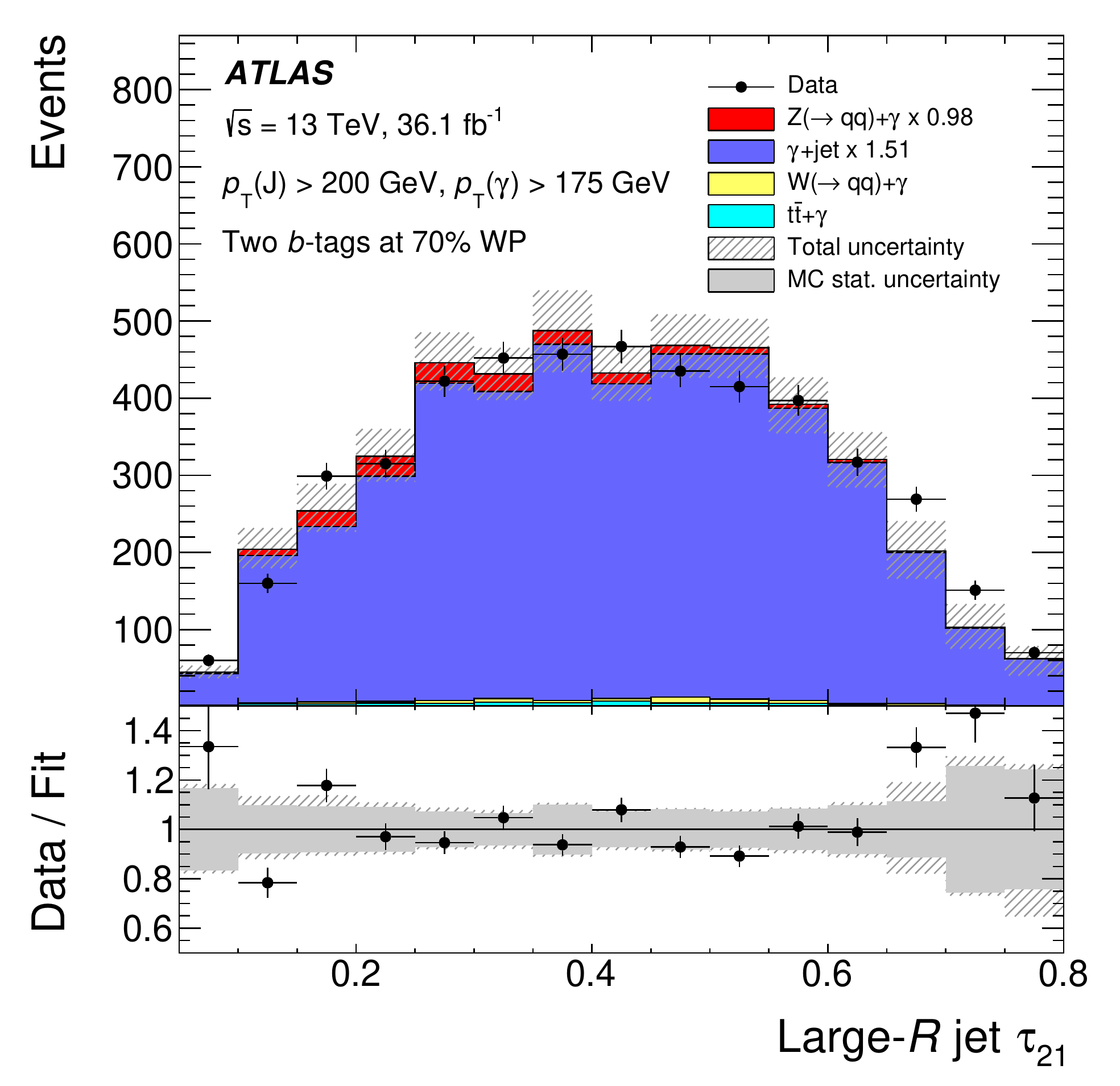}
\caption{Jet mass, \pt, $\DTwoBetaOne$ and $\tau_{21}$ distributions.
Events with two $b$-tagged track-jets are used.  The $\gamma$+ jets background and the \Zbb\ signal are normalised to data by applying
a scale factor of $1.51$ and $0.98$, respectively. The upward- or downward-pointing arrows indicate that the Data/Fit ratio is out of the histogram range for these bins. }
\label{fig:zjet_2tag_JSS}
\end{figure}

Figure~\ref{fig:zjet_2tag_JSS} shows the \Zbb candidate jet mass,  \pT, $\DTwoBetaOne$, and $\tau_{\textrm 21}$ distributions in data and MC simulation.
Systematic uncertainties summarised in Section~\ref{sec:systematics} are applied to the templates, and for each systematic variation the fit to data is performed. The fit
uncertainty and the contribution for each of the systematic uncertainties summed in quadrature are presented in Figure~\ref{fig:zjet_2tag_JSS}. The relative impact of the
systematic uncertainties on the \Zbb and $\gamma$+jets yields are presented in Table~\ref{tab:SysTable_yields}.
The observed data/MC discrepancies are covered by systematic uncertainties.

\begin{table}
\caption{Relative impact of the systematic uncertainties on the  \Zbb and $\gamma$+jets yields	. Multiple independent components have been combined into groups of systematic uncertainties. `Jet scales' refers to the sum in quadrature of the jet energy, mass and substructure scale uncertainties.}
\medskip
\centering
\begin{tabular}{lS[table-format=3.2]S[table-format=3.2]}
\hline
Source  &  \text{$Z\gamma$ [\%]} &  \text{$\gamma$+ jets [\%]} \\
\hline
Photon related & 1.3 & 0.1\\
$b$-tagging related & 23 & 1.4\\
Muon related & 0.9  & 0.1\\
Jet scales & 59 & 2.1 \\
Jet energy resolution & 25 & 1.0 \\
Jet mass resolution & 42 & 1.7 \\
$\gamma$+ jets modelling & 40 & 8.3 \\
$t\bar{t}+\gamma$ related & 4.0 & 0.2\\
$W\gamma$ related & 0.6 & 0.7 \\
Fit statistics & 37 & 2.1\\
 
\hline
 
\end{tabular}
\label{tab:SysTable_yields}
\end{table}

\begin{figure}
\centering
\includegraphics[width=0.45\textwidth]{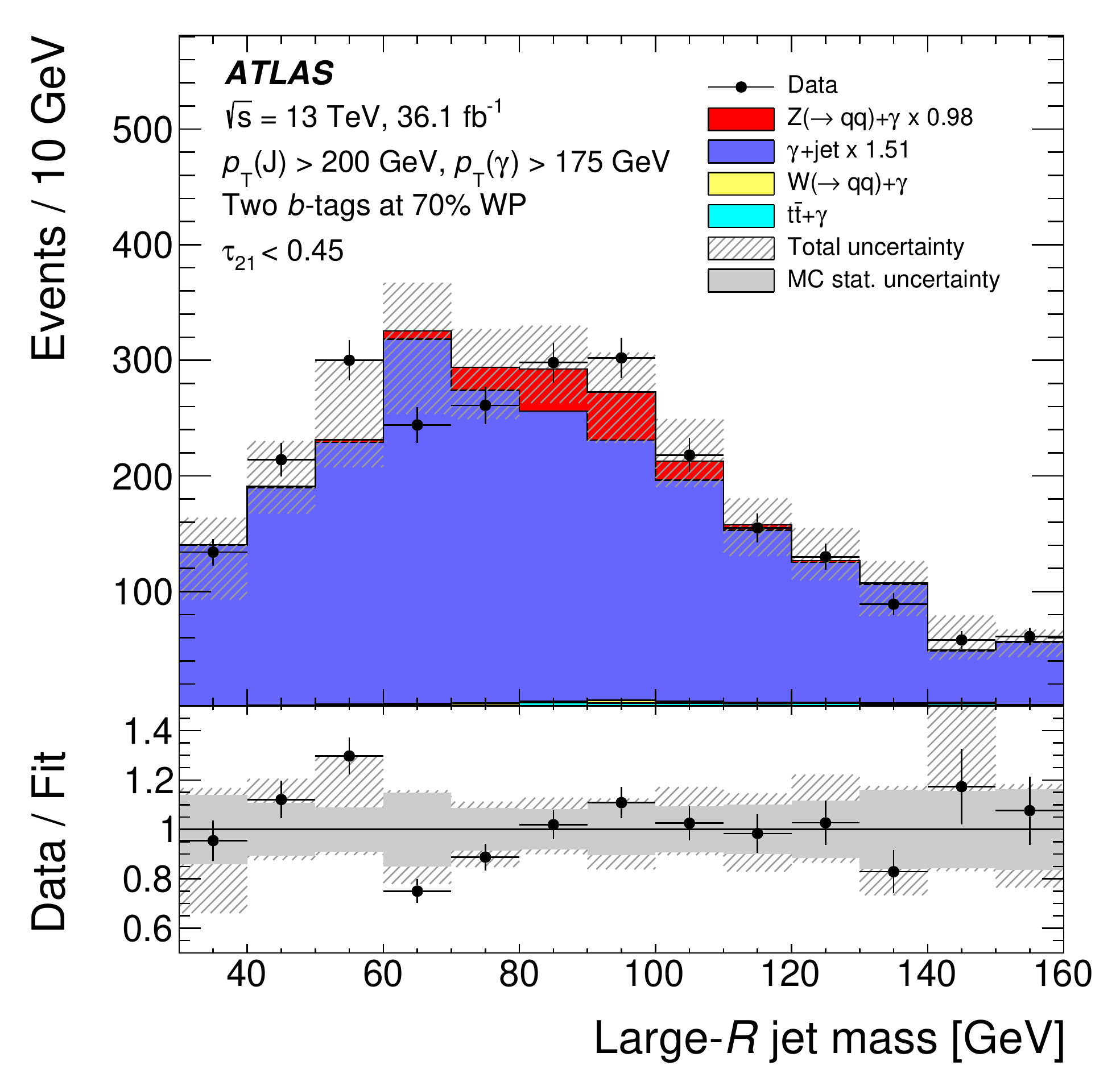}
\includegraphics[width=0.45\textwidth]{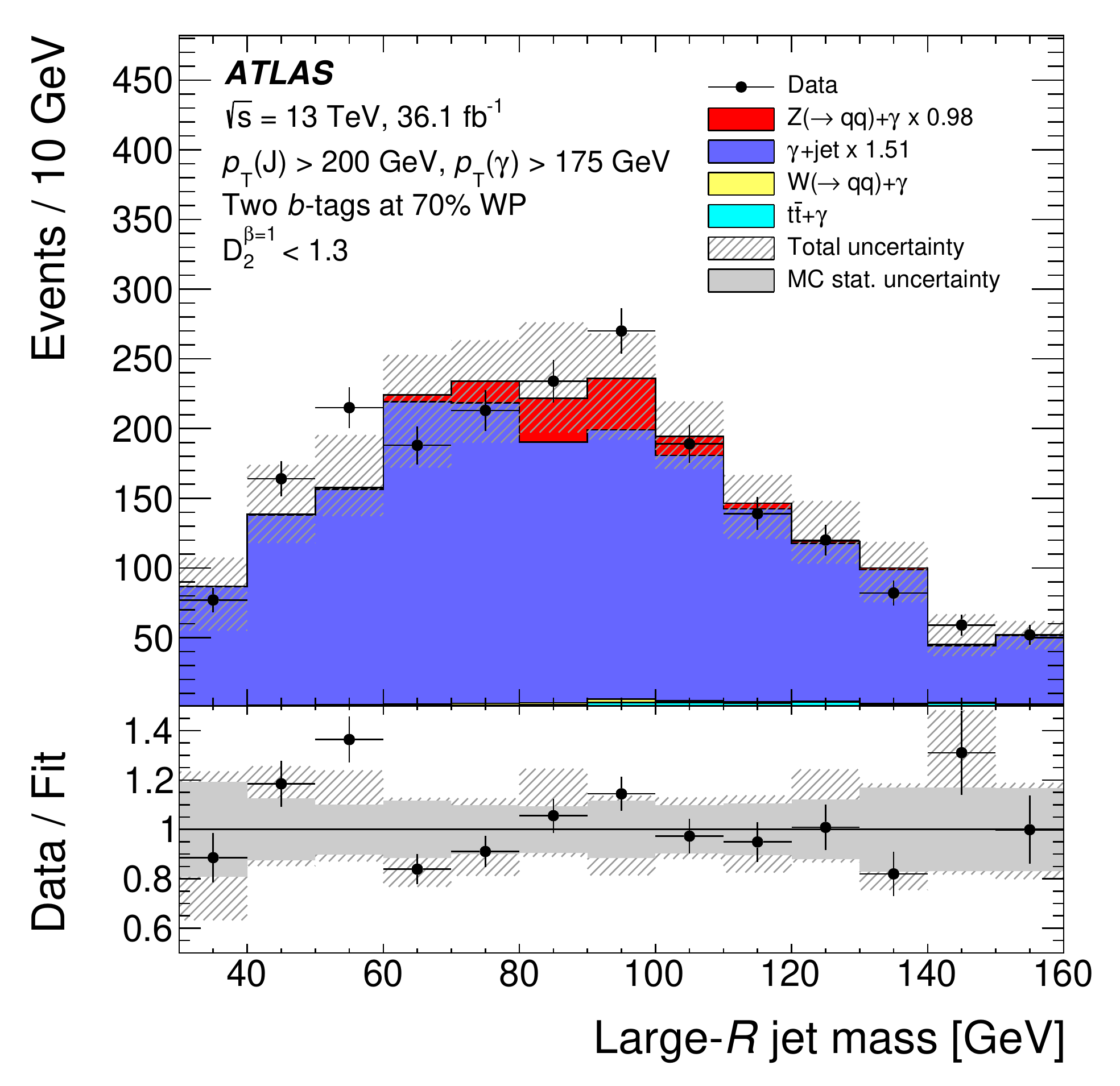}
\caption{\Zbb candidate jet mass after applying the $\tau_{21}<0.45$ (left) or $\DTwoBetaOne<1.3$ (right) requirement.
Events with two $b$-tagged track-jets are used. The $\gamma$+ jets background and the \Zbb\ signal are normalised to data by applying a scale factor of
$1.51$ and $0.98$, respectively.}
\label{fig:zjet_2tag_massafterJSS}
\end{figure}

\begin{figure}
\centering
\includegraphics[width=0.45\textwidth]{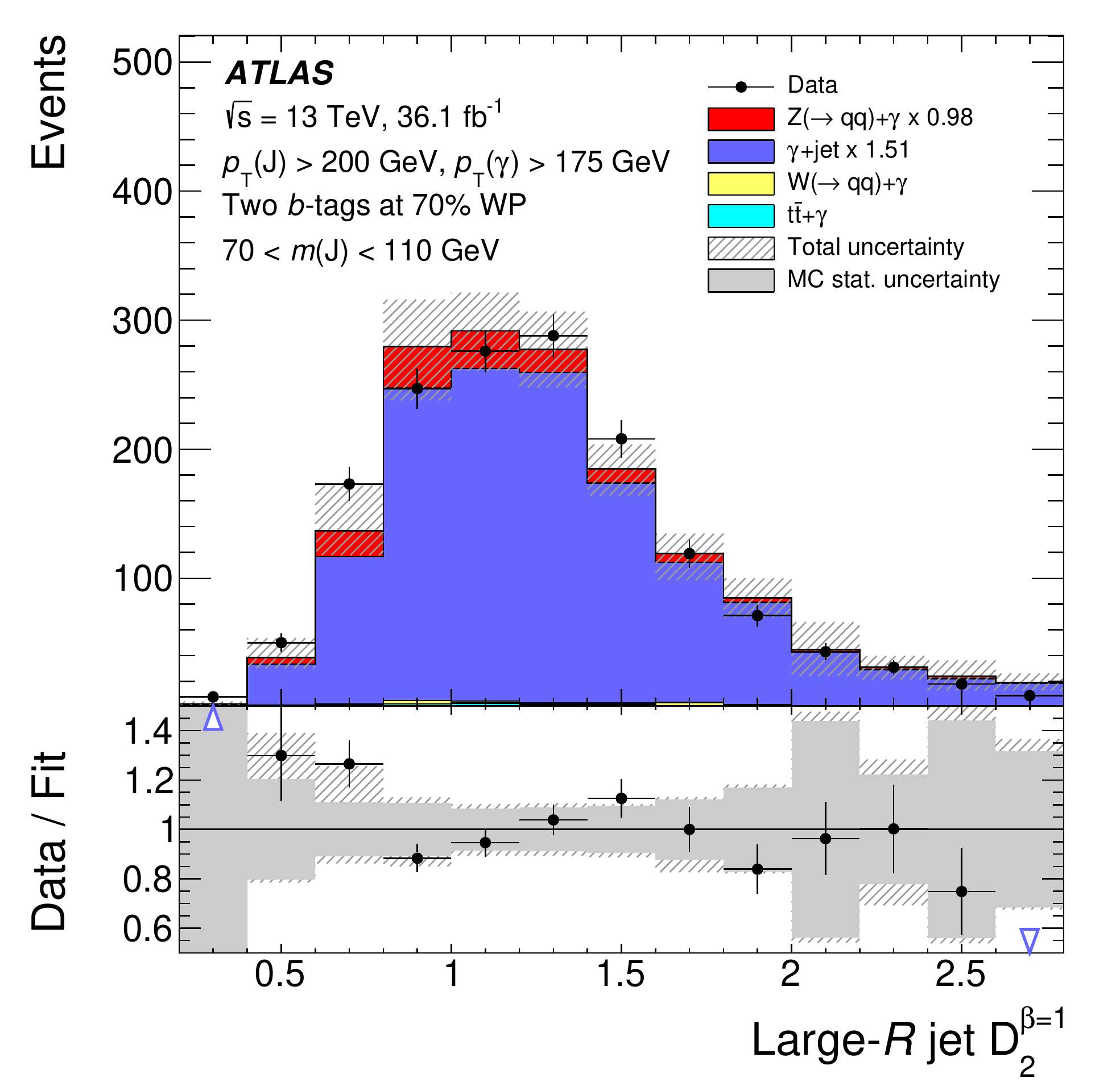}
\includegraphics[width=0.45\textwidth]{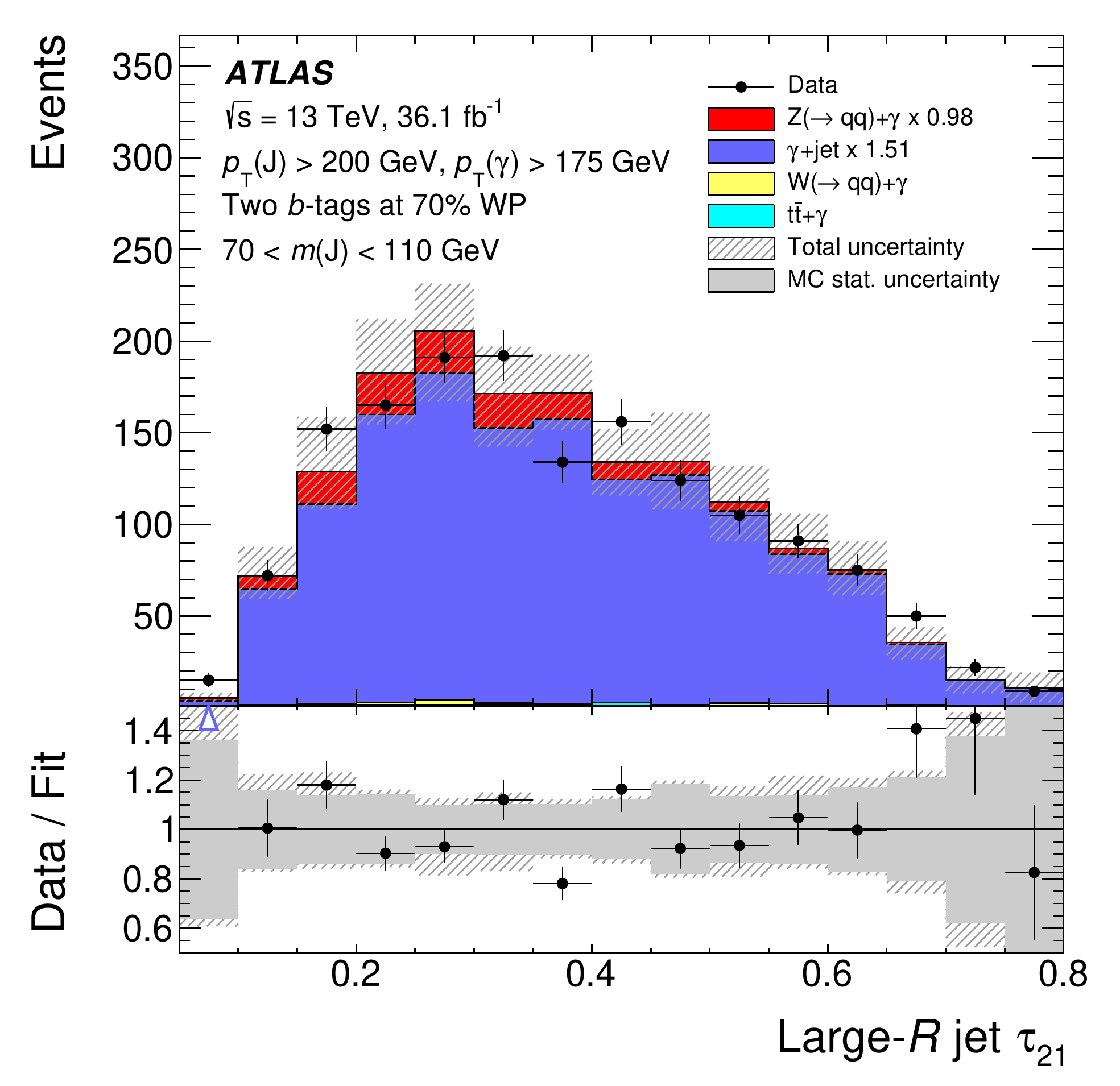}
\caption{\Zbb candidate \DTwoBetaOne\ and $\tau_{21}$ distributions after requiring the \Zbb candidate jet mass to be between 70 and 110~\GeV.
Events with two $b$-tagged track-jets are used. The $\gamma$+ jets background and the \Zbb signal are normalised to data by applying a scale factor of
$1.51$ and $0.98$, respectively. The upward- or downward-pointing arrows indicate that the Data/Fit ratio is out of the histogram range for these bins.}
\label{fig:zjet_2tag_JSS_aftermasscut}
\end{figure}

Further requirements on the jet substructure variables can improve the purity of the selection. Figure~\ref{fig:zjet_2tag_massafterJSS} shows the \Zbb candidate
jet mass after further selections: $\tau_{21}<0.45$ or $\DTwoBetaOne<1.3$. Figure~\ref{fig:zjet_2tag_JSS_aftermasscut} shows the \DTwoBetaOne\ and
$\tau_{21}$ distributions after requiring the \Zbb candidate jet mass to be between 70 and 110~\GeV.

The $Z(\to b \bar{b})\gamma$ process provides a unique possibility to validate the \HiggsJet\ tagging algorithm given the similarity of the $H \to b \bar{b}$
and $Z \to b \bar{b}$ processes. For the current integrated luminosity of 36~\ifb, the dominant uncertainties are the statistical and systematic uncertainties of the
jet scales and jet mass for the \Zbb process and the $\gamma$+jets modelling uncertainties. To reduce the dominant uncertainties, a larger dataset is needed.
Within the uncertainties the studied jet substructure variables are modelled well by the signal plus background MC simulations.
 
\FloatBarrier
 

\section{Conclusions}
\label{sec:conclusions}
 
Techniques to identify Higgs bosons at high transverse momenta decaying into bottom-quark pairs are described in this paper.
The identification is based on the $b$-tagging of $R = 0.2$ track-jets matched to the \HiggsJet\ and requirements placed on the \HiggsJet\ mass and other substructure variables. The modelling of the relevant input distributions is studied in 36~\ifb\ of 13~\TeV\ proton--proton collision data recorded by the ATLAS detector at the LHC in 2015 and 2016.
 
The choice of $b$-tagging working point for an analysis depends on the required background rejection rate and on the \HiggsJet\ \pT\
range relevant for the analysis. The double-$b$-tagging working points give the best background rejection for a large range of the \HiggsJet-tagging
efficiency but the efficiency decreases faster with increasing \HiggsJet\ \pT\ than it does for single-$b$-tagging working points. At high efficiencies
above $\sim 90\%$ ($\sim 55\%$) for \HiggsJet\ \pT above 250 (1000)~\GeV\ the single-$b$-tagging selection provides better background rejection.
 
Application of the Higgs boson mass window requirement improves the performance of the \HiggsJet\ tagger substantially. The multijet
background rejection improves by a factor of about five by adding a loose (corresponding to 80\% signal efficiency) mass window requirement
on top of the double-$b$-tagging requirement. The tight (corresponding to 68\% signal efficiency) mass window requirement leads to
an additional 30--50\% improvement in the multijet background rejection. The multijet background rejection has a weak dependence on
the jet \pT\ for both mass window requirements. The hadronic top-quark rejection depends strongly on the jet \pT. The rejection varies
between 60 and 230 for the loose mass window and double-$b$-tagging working points. The largest improvement in the top-quark
rejection for the tight mass window is about 70\% and corresponds to the high \pT\ and double-$b$-tagging working point.
 
The performance of the additional jet substructure variables depends on the chosen \HiggsJet\ tagger working point. The jet mass and other
substructure variables are often correlated and the double-$b$-tagging requirement enforces a two-prong structure. In general, the background rejection
is larger for the multijet background than for hadronically decaying top quarks but still below two for the individual variables and the loose
mass window working point. The $b$-tagging discriminant is very powerful but the jet substructure variables offer an alternative to the
$b$-tagging working points. Especially at high \HiggsJet\ \pT\ the efficiency to reconstruct two track-jets and the double-$b$-tagging
efficiency decrease quickly. A combination of several substructure variables using multivariate methods could potentially increase the gain in
performance in this phase space.
 
The modelling of representative \HiggsJet\ properties is tested in ATLAS data for $g \rightarrow b \bar{b}$ and
$Z (\rightarrow b \bar{b})\gamma$ event selections. Good modelling is observed given the size of the available data sample  and the systematic uncertainties.
In particular, the use of jet substructure variables is shown to improve the purity of the $Z (\rightarrow b \bar{b})\gamma$ event selection.

\clearpage
 
\section*{Acknowledgements}
 
 
We thank CERN for the very successful operation of the LHC, as well as the
support staff from our institutions without whom ATLAS could not be
operated efficiently.
 
We acknowledge the support of ANPCyT, Argentina; YerPhI, Armenia; ARC, Australia; BMWFW and FWF, Austria; ANAS, Azerbaijan; SSTC, Belarus; CNPq and FAPESP, Brazil; NSERC, NRC and CFI, Canada; CERN; CONICYT, Chile; CAS, MOST and NSFC, China; COLCIENCIAS, Colombia; MSMT CR, MPO CR and VSC CR, Czech Republic; DNRF and DNSRC, Denmark; IN2P3-CNRS, CEA-DRF/IRFU, France; SRNSFG, Georgia; BMBF, HGF, and MPG, Germany; GSRT, Greece; RGC, Hong Kong SAR, China; ISF and Benoziyo Center, Israel; INFN, Italy; MEXT and JSPS, Japan; CNRST, Morocco; NWO, Netherlands; RCN, Norway; MNiSW and NCN, Poland; FCT, Portugal; MNE/IFA, Romania; MES of Russia and NRC KI, Russian Federation; JINR; MESTD, Serbia; MSSR, Slovakia; ARRS and MIZ\v{S}, Slovenia; DST/NRF, South Africa; MINECO, Spain; SRC and Wallenberg Foundation, Sweden; SERI, SNSF and Cantons of Bern and Geneva, Switzerland; MOST, Taiwan; TAEK, Turkey; STFC, United Kingdom; DOE and NSF, United States of America. In addition, individual groups and members have received support from BCKDF, CANARIE, CRC and Compute Canada, Canada; COST, ERC, ERDF, Horizon 2020, and Marie Sk{\l}odowska-Curie Actions, European Union; Investissements d' Avenir Labex and Idex, ANR, France; DFG and AvH Foundation, Germany; Herakleitos, Thales and Aristeia programmes co-financed by EU-ESF and the Greek NSRF, Greece; BSF-NSF and GIF, Israel; CERCA Programme Generalitat de Catalunya, Spain; The Royal Society and Leverhulme Trust, United Kingdom.
 
The crucial computing support from all WLCG partners is acknowledged gratefully, in particular from CERN, the ATLAS Tier-1 facilities at TRIUMF (Canada), NDGF (Denmark, Norway, Sweden), CC-IN2P3 (France), KIT/GridKA (Germany), INFN-CNAF (Italy), NL-T1 (Netherlands), PIC (Spain), ASGC (Taiwan), RAL (UK) and BNL (USA), the Tier-2 facilities worldwide and large non-WLCG resource providers. Major contributors of computing resources are listed in Ref.~\cite{ATL-GEN-PUB-2016-002}.
 

\clearpage
 
\printbibliography
 
\clearpage
 
\begin{flushleft}
{\Large The ATLAS Collaboration}

\bigskip

G.~Aad$^\textrm{\scriptsize 101}$,    
B.~Abbott$^\textrm{\scriptsize 128}$,    
D.C.~Abbott$^\textrm{\scriptsize 102}$,    
O.~Abdinov$^\textrm{\scriptsize 13,*}$,    
A.~Abed~Abud$^\textrm{\scriptsize 70a,70b}$,    
K.~Abeling$^\textrm{\scriptsize 53}$,    
D.K.~Abhayasinghe$^\textrm{\scriptsize 93}$,    
S.H.~Abidi$^\textrm{\scriptsize 167}$,    
O.S.~AbouZeid$^\textrm{\scriptsize 40}$,    
N.L.~Abraham$^\textrm{\scriptsize 156}$
% ...
\bigskip
\\

% ...
$^{13}$Institute of Physics, Azerbaijan Academy of Sciences, Baku; Azerbaijan.\\
% ...
$^{40}$Niels Bohr Institute, University of Copenhagen, Copenhagen; Denmark.\\
% ...
$^{53}$II. Physikalisches Institut, Georg-August-Universit\"{a}t G\"ottingen, G\"ottingen; Germany.\\
% ...
$^{70}$$^{(a)}$INFN Sezione di Pavia;$^{(b)}$Dipartimento di Fisica, Universit\`a di Pavia, Pavia; Italy.\\
% ...
$^{93}$Department of Physics, Royal Holloway University of London, Egham; United Kingdom.\\
% ...
$^{101}$CPPM, Aix-Marseille Universit\'e, CNRS/IN2P3, Marseille; France.\\
$^{102}$Department of Physics, University of Massachusetts, Amherst MA; United States of America.\\
% ...
$^{128}$Homer L. Dodge Department of Physics and Astronomy, University of Oklahoma, Norman OK; United States of America.\\
% ...
$^{156}$Department of Physics and Astronomy, University of Sussex, Brighton; United Kingdom.\\
% ...
$^{167}$Department of Physics, University of Toronto, Toronto ON; Canada.\\
% ...
$^{a}$ Also at Borough of Manhattan Community College, City University of New York, New York NY; United States of America.\\
$^{b}$ Also at Centre for High Performance Computing, CSIR Campus, Rosebank, Cape Town; South Africa.\\
% ...
$^{*}$ Deceased

\end{flushleft}


\end{document}